\documentclass{ws-ijmpa}

\def\Vec#1{\mbox{\boldmath$#1$}}

\begin{document}

\markboth{Y.~Mizutani and T.~Inagaki}
{Non-Equilibrium Thermo Field Dynamics for 
Relativistic Complex Scalar and Dirac Fields}

%
\catchline{}{}{}{}{}
%

\title{NON-EQUILIBRIUM THERMO FIELD DYNAMICS FOR 
RELATIVISTIC COMPLEX SCALAR AND DIRAC FIELDS
}

\author{YUICHI MIZUTANI}

\address{Department of Physics, Hiroshima University \\
 Higashi-Hiroshima, Hiroshima 739-8526, Japan\\
mizutani@theo.phys.sci.hiroshima-u.ac.jp}

\author{TOMOHIRO INAGAKI}

\address{Information Media Center, Hiroshima University\\
Higashi-Hiroshima, Hiroshima 739-8521, Japan\\
inagaki@hiroshima-u.ac.jp}

\maketitle

\begin{history}
\received{Day Month Year}
\revised{Day Month Year}
\end{history}

\begin{abstract}
Relativistic quantum field theories for complex scalar and Dirac fields 
are investigated in non-equilibrium thermo field dynamics. 
The thermal vacuum is defined by the 
Bogoliubov transformed creation and annihilation operators. 
Two independent Bogoliubov parameters are introduced for a charged field. 
Its difference naturally induces the chemical potential. 
Time-dependent thermal Bogoliubov transformation generates 
the thermal counter terms.
We fix the terms by the self-consistency renormalization condition.
Evaluating the thermal self-energy under the self-consistency 
renormalization condition, 
we derive the quantum Boltzmann equations for the relativistic fields. 

\keywords{thermo field dynamics; out of equilibrium; relativistic system.}
\end{abstract}

\ccode{PACS numbers: 03.70.+k, 11.10.Wx}

\section{Introduction}
Much attention has been paid for 
thermal evolution of high energy phenomena, 
for example, hot and dense quark matter. 
It is expected that the time dependence of the 
system can be observed in a relativistic heavy ion 
collisions and cooling process of dense stars. 
In such a system a thermal quantum field theory, i.e. 
a quantum field theory to describe thermal dynamics, 
is necessary to investigate the non-equilibrium phenomena for relativistic fields. 
There are several theories, 
the thermo field dynamics (TFD), the Kadanoff-Baym formalism, 
the Langevin equation\cite{umezawa1,Kadanoff,lngvn1,ctp1,ctp2} and so on. 
In this paper we focus on thermo field dynamics 
to describe a non-equilibrium thermal system for relativistic charged fields. 

TFD has been proposed 
by Y.~Takahashi and H.~Umezawa\cite{umezawa1,finitetfd1,CQTFD0,arimitsu2}. 
In TFD the canonical formalism for a quantum field theory is 
established at finite temperature. 
The so-called thermal vacuum state is defined to calculate 
the thermal average of dynamical operators. It introduces the thermal 
Bogoliubov transformation. 
The fundamental formalism of TFD for relativistic complex scalar and 
Dirac fields is developed in a finite and homogeneous system, 
see for example Ref.~\refcite{henning1}. 
A propagator for the Dirac field has been discussed 
at finite chemical potential in Ref.~\refcite{df_prop1}. 
In Ref.~\refcite{csf_henning1} 
the complex scalar field has been applied to TFD 
in a spatially inhomogeneous system, 
and the thermal conductivity and the diffusion coefficient 
has been calculated for interacting pions in hot and dense nuclear matter. 

TFD has been extended to describe an out of equilibrium system\cite{netfd1}. 
In the non-equilibrium thermo field dynamics (NETFD) the thermal 
Bogoliubov transformation is generalized to depend on time variable. 
The time-dependent thermal Bogoliubov transformation induces the thermal 
counter term \cite{CQNETFD1,ThermC1,CQNETFD2}.
The term is fixed by imposing the self-consistency renormalization condition 
at the equal time limit\cite{ThermC1,CQNETFD2}. 
Then the quantum Boltzmann equation is derived from the renormalization 
condition. This procedure is applied to investigate the Bose-Einstein 
condensation in trapped cold atom 
systems\cite{ColdAtm1,ColdAtm2,ColdAtm3,ColdAtm4}. 
NETFD is also extended to an inhomogeneous system with 
diffusion\cite{inhomo1,inhomo2}. 

In Ref.~\refcite{CQNS1} NETFD has been applied to a neutral scalar field 
based on the canonical quantization. 
In NETFD the neutral scalar field 
depends on a thermal Bogoliubov parameter, 
which is equal to uncharged particle number density
under the self-consistency renormalization condition. 
The relativistic charged fields depend on 
two types of the Bogoliubov parameters, $n_{+}(t)$ and $n_{-}(t)$, 
which correspond to the particle and anti-particle number density
under the self-consistency renormalization conditions, respectively. 
The thermal counter term for the neutral scalar field fixes
one of the Bogoliubov parameters. An alternative condition or
assumption is necessary to fix all the Bogoliubov parameters.
In the case of the Dirac fermion it has been pointed out that
the positive and the negative frequency parts can not be 
described in an usual Lagrangian form 
and conjectured that 
the charge conjugation is broken in the framework of NETFD
\cite{NETFD_DF1}.

In this paper we approach relativistic complex scalar 
and Dirac fields at finite chemical potential 
based on the canonical quantization. 
In \S 2 we consider a scalar field in NETFD. 
Following the canonical formalism, we quantize the complex scalar 
field with finite chemical potential. 
The time-dependent Bogoliubov transformation 
modifies the unperturbed Hamiltonian and introduces the thermal counter term. 
In \S 3 we discuss the self-consistency renormalization condition 
for the complex scalar field. 
The time evolution equations are derived under the condition. 
In \S 4 a two loop quantum correction is calculated 
in a $(\phi^\dagger \phi)^2$ interaction model. 
Applying the self-consistency renormalization condition at the lowest order, 
we find that the time evolution equations for the Bogoliubov parameters 
coincide with the quantum Boltzmann equations for the charged Bose particles. 
In \S 5 we study a Dirac field in NETFD. 
The Dirac field can be quantized at finite chemical potential 
following a similar procedure with the complex scalar field. 
Because of the time-dependence of the thermal Bogoliubov transformation, 
the unperturbed Hamiltonian is modified 
and the thermal counter term appears. 
In \S 6 we extend the self-consistency renormalization condition to the Dirac field. 
The time evolution equations for the Dirac field 
are obtained from the condition. 
In \S 7 we apply the result to a Yukawa interaction model. 
One loop quantum corrections are calculated 
for the neutral scalar and the Dirac propagators.  
Imposing the self-consistency renormalization conditions on the scalar and Dirac fields, 
we reproduce the well-known quantum Boltzmann equations 
at the week coupling limit. 
Concluding remarks are given in \S 8. 

\section{Canonical Quantization for a Complex Scalar Field}

We start from the description of a complex scalar field at finite 
chemical potential in NETFD.
In this paper we confine ourselves to homogeneous processes. 
Thus the complex scalar field is expanded in terms of two independent 
operators, $a_p$ and $b_p$, where $p$ denotes the spatial momentum of 
a single Fourier mode. 
The creation and annihilation operators obey the commutation relations, 
\begin{eqnarray}
&& [a_{p}, a_{k}^\dagger ]
= (2\pi)^3 \delta^{(3)}(\Vec{p}-\Vec{k}),
  \label{cs_comm_a1} \\
&& [b_{p}, b_{k}^\dagger ]
= (2\pi)^3 \delta^{(3)}(\Vec{p}-\Vec{k}), 
\label{cs_comm_a2} \\
&& [a_{p}, b_{k}^\dagger ]
= [a_p^\dagger , b_k^\dagger]
=[a_p , b_k]
=[a_p^\dagger , b_k]
= 0. 
\end{eqnarray}

In TFD the tilde operators, $\tilde{a}_p$ and $\tilde{b}_p$, 
are introduced according to the tilde conjugation rules, 
\begin{eqnarray}
&& (A_1 A_2)\tilde{} = \tilde{A_1}\tilde{A_2},  \label{tilde1}\\
&& (c_1 A_1 + c_2 A_2)\tilde{} =c_1^\ast \tilde{A}_1 + c_2^\ast \tilde{A}_2, \label{tilde2} \\
&& (\tilde{A})\tilde{} =A,	\label{tilde3} \\
&& (A^\dagger)\tilde{} = \tilde{A}^\dagger, \label{tilde4}
\end{eqnarray}
where $A_1$ and $A_2$ show any operators, $c_1$ and $c_2$ are c-numbers. 
Following the tilde conjugation rules, we find that the tilde 
operators, $\tilde{a}_p$ and $\tilde{b}_p$, obey the commutation relations, 
\begin{eqnarray}
&& [\tilde{a}_{p}, \tilde{a}_{k}^\dagger ]
= (2\pi)^3 \delta^{(3)}(\Vec{p}-\Vec{k}),
  \label{cs_comm_ta1} \\
&&[\tilde{b}_{p}, \tilde{b}_{k}^\dagger ]
= (2\pi)^3 \delta^{(3)}(\Vec{p}-\Vec{k}), 
\label{cs_comm_ta2} \\
&& [\tilde{a}_{p}, \tilde{b}_{k}^\dagger ]
= [\tilde{a}_p^\dagger , \tilde{b}_k^\dagger]
=[\tilde{a}_p , \tilde{b}_k]
=[\tilde{a}_p^\dagger , \tilde{b}_k]
= 0. 
\end{eqnarray}
The ordinary operators commute with the tilde operators, 
\begin{eqnarray}
&& [ a_{p}, \tilde{a}_{k} ] 
= [ a_{p}, \tilde{a}_{k}^\dagger ] 
= [ a_{p}, \tilde{b}_{k}^\dagger ] 
= [ a_{p}, \tilde{b}_{k} ] = 0, \\
&& [ b_{p}, \tilde{b}_{k} ] 
= [ b_{p}, \tilde{b}_{k}^\dagger ] 
= [ b_{p}, \tilde{a}_{k} ] 
=[ b_{p}, \tilde{a}_{k}^\dagger ]
= 0. 
\end{eqnarray}
Time evolution of the field is described by the hat-Hamiltonian which is constructed by 
\begin{eqnarray}
\hat{H} = H-\tilde{H}, \label{hat-H}
\end{eqnarray}
where $H$ is an ordinary Hamiltonian and 
$\tilde{H}$ is the tilde conjugate Hamiltonian.

In TFD the thermal expectation value of a dynamical operator 
is given by the expectation value 
under the state so-called thermal vacuum state, 
$\langle \theta|$ and $|\theta\rangle$. 
For the complex scalar field the thermal vacuum state 
is defined by the two operators, $\xi_p$ and $\eta_p$, 
and their tilde conjugates, $\tilde{\xi}_p$ and $\tilde{\eta}_p$, 
which are the thermal Bogoliubov transformations 
of $a_p$, $b_p$, $\tilde{a}_p$ and $\tilde{b}_p$
with two independent thermal Bogoliubov parameters, $n_{\pm}(t;p)$, 
\begin{eqnarray}
&&\xi_{p}^\alpha (t) 
= B(n_{+}(t;p))^{\alpha \beta} a_{p}^{\beta} (t),\quad
\bar{\xi}_{p}^\alpha (t)
= \bar{a}_{p}^{\beta} (t) B^{-1}(n_{+}(t;p))^{\beta \alpha}, \label{csf_tbt1} \\
&&\eta_{p}^\alpha (t) 
= B(n_{-}(t;p))^{\alpha \beta} b_{p}^{\beta} (t),\quad
\bar{\eta}_{p}^\alpha (t)
= \bar{b}_{p}^{\beta} (t) B^{-1}(n_{-}(t;p))^{\beta \alpha}. \label{csf_tbt2}
\end{eqnarray}
Here and below the upper indices indicate the 
thermal doublet notation. 
In this notation creation and annihilation operators 
are described as 
\begin{eqnarray}
&& a_p^\alpha = 
\left(
   \begin{array}{c}
	a_p \\
	\tilde{a}_p^\dagger
   \end{array}
\right),\quad 
\bar{a}_p^\alpha =
\left(
   \begin{array}{cc}
	a_p^\dagger & -\tilde{a}_p
   \end{array}
\right).  \label{ns_tdn1}
\end{eqnarray}
The Bogoliubov transformation has a $2\times2$ matrix form. 

In the non-equilibrium and homogeneous system
the thermal Bogoliubov parameters, $n_{\pm}(t;p)$, 
depend on time, $t$, and the magnitude of the momentum, $p$. 
We set the thermal Bogoliubov matrices, $B$ and $B^{-1}$, as
\begin{eqnarray}
&& B(n_{\pm})=
\left(
   \begin{array}{cc}
	1+n_{\pm} & -n_{\pm} \\
	-1 & 1 \\
   \end{array}
\right),  \label{cs_bm1}  \\ 
&& B^{-1}(n_{\pm})=
\left(
   \begin{array}{cc}
	1 & n_{\pm} \\
	1 & 1+n_{\pm} \\
   \end{array}
\right). \label{cs_bm2}
\end{eqnarray}
This expression makes available to calculate quantum corrections
in the Dyson-Wick formalism\cite{FynPert1}. 

According to the transformed operators, $\xi_p, \eta_p, \tilde{\xi}_p$ 
and $\tilde{\eta}_p$, the thermal vacuum state is defined as
\begin{eqnarray}
&& \xi_{p}|\theta\rangle = \tilde{\xi}_{p}|\theta\rangle=
\eta_{p}|\theta\rangle = \tilde{\eta}_{p}|\theta\rangle=0, \label{csf_vs1} \\ 
&&\langle \theta| \xi_{p}^\dagger = \langle \theta| \tilde{\xi}_{p}^\dagger
=\langle \theta| \eta_{p}^\dagger = \langle \theta| \tilde{\eta}_{p}^\dagger =0.
\label{csf_vs2}
\end{eqnarray}
It should be noticed that the physical observables are constructed by the 
original operators, $a_{p}$ and $b_{p}$. 

In an equilibrium system the thermal vacuum state 
has a condensate nature structure 
with the thermal pairs, $a\tilde{a}$ and $b\tilde{b}$, \cite{umezawa1}
\begin{eqnarray}
&&|\theta\rangle = U_{B,a} (\chi_a) U_{B,b} (\chi_b)  |0 \rangle\rangle, \label{UB_1-1} \\
&&\langle \theta|= \langle \langle 0|U_{B,a}^{-1} (\chi_a) U_{B,b}^{-1} (\chi_b), 
\label{UB_1-2}
\end{eqnarray}
where $|0\rangle\rangle$ and $\langle\langle 0|$ 
represent the non-thermal vacuum states,
\begin{eqnarray}
&&a_p|0 \rangle \rangle = \tilde{a}_p|0 \rangle \rangle= 
b_p |0 \rangle \rangle = \tilde{b}_p |0 \rangle \rangle = 0, \label{UB_2-1} \\
&& \langle \langle 0|a_p^{\dagger}=\langle \langle 0|\tilde{a}_p^{\dagger}
=\langle \langle 0|b_p^{\dagger}=\langle \langle 0|\tilde{b}_p^{\dagger} = 0. \label{UB_2-2}
\end{eqnarray}
The transformation operators are described as
\begin{eqnarray}
&& U_{B,a} (\chi_a)=
{\rm exp}[i\chi_{a1} \hat{G}_{a1}] {\rm exp}[-i \chi_{a2} \hat{G}_{a2}]
{\rm exp}[i \chi_{a3} \hat{G}_{a1}]
{\rm exp}[i\chi_{a2} \hat{G}_{a2}]{\rm exp}[-i \chi_{a1} \hat{G}_{a1}], \nonumber \\
\label{UB_3-1} \\
&& U_{B,b}(\chi_b)=
{\rm exp}[i\chi_{b1} \hat{G}_{b1}] {\rm exp}[-i \chi_{b2} \hat{G}_{b2}]
{\rm exp}[i \chi_{b3} \hat{G}_{b1}]
{\rm exp}[i\chi_{b2} \hat{G}_{b2}]{\rm exp}[-i \chi_{b1} \hat{G}_{b1}], \nonumber \\
\label{UB_3-2}
\end{eqnarray}
with the generators, 
\begin{eqnarray}
&&\hat{G}_{a1}(k) = i (a_{k}^\dagger a_{k} + \tilde{a}_{k}^\dagger \tilde{a}_k), \quad
\hat{G}_{a2}(k) 
= i (a_{k}\tilde{a}_{k} + a_{k}^\dagger \tilde{a}_{k}^\dagger ), \label{UB_5-1} \\
&&\hat{G}_{b1}(k) = i (b_{k}^\dagger b_{k} + \tilde{b}_{k}^\dagger \tilde{b}_k), \quad
\hat{G}_{b2} (k)
= i (b_{k}\tilde{b}_{k} + b_{k}^\dagger \tilde{b}_{k}^\dagger ). \label{UB_5-2} 
\end{eqnarray}
We use an abbreviation,
\begin{eqnarray}
\chi_{a(b)i} \hat{G}_{a(b)j}=\int d^3 k \ \chi_{a(b)i} (k) \hat{G}_{a(b)j} (k). \label{UB_4}
\end{eqnarray}
The thermal vacuum state is classified 
according to the parameters, $\chi_{a1},\chi_{a2},\chi_{a3},\chi_{b1},\chi_{b2}$ 
and $\chi_{b3}$. 
From Eqs.~(\ref{UB_1-1}) and (\ref{UB_1-2}) 
we can find that the expectation value of operators 
under the thermal vacuum coincides with the thermal trace of the operators. 
By using the transformation operators, $U_{B,a}(\chi_a)$ and $U_{B,b}(\chi_b)$, 
the thermal Bogoliubov transformations are written as
\begin{eqnarray}
&&\xi_p^\alpha = U_{B,a}(\chi_a) a_p^\alpha U_{B,a}^{-1}(\chi_a), \quad
\bar{\xi}_p^\alpha =U_{B,a}(\chi_a) \bar{a}_p^\alpha U_{B,a}^{-1}(\chi_a), 
\label{UB_6-1} \\
&&\eta_p^\alpha = U_{B,b}(\chi_b) b_p^\alpha U_{B,b}^{-1}(\chi_b), \quad
\bar{\eta}_p^\alpha = U_{B,b}(\chi_b) \bar{b}_p^\alpha U_{B,b}^{-1}(\chi_b). 
\label{UB_6-2}
\end{eqnarray}
The thermal Bogoliubov parameters, $n_{+}(k)$ and $n_{-}(k)$, 
can be represented as a function of $\chi_{ai}(k)$ and $\chi_{bi}(k)$, respectively. 
However, the operators, $U_{B,a}$ and $U_{B,b}$, 
contain divergent coefficients in field theories 
with an infinite degree of freedom. 
These operators are mathematically ill-defined. 
Here we employ well-defined expressions (\ref{csf_tbt1}) and (\ref{csf_tbt2}) 
and quantize the fields.

The Bogoliubov transformation keeps the commutation relations. Thus
the transformed operators, $\xi_p$ and $\eta_p$, 
and their tilde conjugates, $\tilde{\xi}_p$ and $\tilde{\eta}_p$, 
satisfy the commutation relations, 
\begin{eqnarray}
&& [\xi_{p}, \xi_{k}^\dagger ]
= [\tilde{\xi}_{p}, \tilde{\xi}_{k}^\dagger ]
= (2\pi)^3 \delta^{(3)}(\Vec{p}-\Vec{k}),
  \label{cs_cr:xi1} \\
&& [\eta_{p}, \eta_{k}^\dagger ]
= [\tilde{\eta}_{p}, \tilde{\eta}_{k}^\dagger ]
= (2\pi)^3 \delta^{(3)}(\Vec{p}-\Vec{k}),  \label{cs_cr:xi2} 
\end{eqnarray}
with all other commutators vanish. 
The scalar field is quantized under these commutation relations.

We develop the quantum field theory for the complex scalar field 
with finite chemical potential according to NETFD which has been 
proposed by H.~Umezawa and Y.~Yamanaka\cite{CQNETFD1}. 
Below we work in the interaction picture. 
To guarantee the time independence of the thermal vacuum 
we assume that the complex scalar field is decomposed as 
a Hermitian form in terms of the transformed operators,
\begin{eqnarray}
&&  \phi_{\xi} (x) \equiv \int \frac{d^3\Vec{p}}{(2\pi)^3}
 \frac{1}{\sqrt{2\omega_p}} 
\left\{ \xi_{p}(t_x) 
 e^{ i \Vec{p} \cdot \Vec{x}} 
+ \eta_{p}^\dagger (t_x) 
 e^{ -i \Vec{p} \cdot \Vec{x}} \right\}, \label{phic_xi1} \\
&&  \phi_{\xi}^\dagger (x) \equiv \int \frac{d^3\Vec{p}}{(2\pi)^3}
 \frac{1}{\sqrt{2\omega_p}} 
\left\{ \xi_{p}^\dagger (t_x) 
 e^{ -i \Vec{p} \cdot \Vec{x}} 
+ \eta_{p} (t_x) 
 e^{ i \Vec{p} \cdot \Vec{x}} \right\}, \label{phic_xi2}
\end{eqnarray}
with
\begin{eqnarray}
&&\xi_p (t_x) = \xi_p e^{-i \omega_{+,p} t_x},~~ 
\xi_p^\dagger (t_x) = \xi_p^\dagger e^{ i \omega_{+,p} t_x}, \label{csf_xi1} \\
&&\eta_p (t_x) = \eta_p e^{-i \omega_{-,p} t_x},~~
\eta_p^\dagger (t_x) = \eta_p^\dagger e^{i \omega_{-,p} t_x}, \label{csf_eta1}
\end{eqnarray}
and
\begin{eqnarray}
&& \omega_{+,p} \equiv \omega_p - \mu, \label{csf_exi_1} \\
&& \omega_{-,p} \equiv \omega_p + \mu, \label{csf_exi_2}
\end{eqnarray}
where $\omega_p$ is the relativistic energy eigenvalue, 
$\sqrt[]{\Vec{p}^2+m^2}$, and $\mu$ denotes the chemical potential. 
The energy eigenvalue, $\omega_p$, 
generally depends on time through the screening 
effects and so on. In this paper we focus on a system with a 
time independent energy eigenvalue, for simplicity. 
The chemical potential is introduced 
according to the conserved charge in an equilibrium system. 
See appendix A. 

The tilde conjugate field is defined by the tilde conjugation rules, 
(\ref{tilde1})-(\ref{tilde3}). 
\begin{eqnarray}
&&  \tilde{\phi}_{\xi} (x) = \int \frac{d^3\Vec{p}}{(2\pi)^3}
 \frac{1}{\sqrt{2\omega_p}} 
\Bigl\{ \tilde{\xi}_{p}(t_x) 
 e^{ -i \Vec{p} \cdot \Vec{x}} 
+ \tilde{\eta}_{p}^\dagger (t_x) 
 e^{ i \Vec{p} \cdot \Vec{x}} \Bigr\}, \label{phic_xi3} \\
&&  \tilde{\phi}_{\xi}^\dagger (x) = \int \frac{d^3\Vec{p}}{(2\pi)^3}
 \frac{1}{\sqrt{2\omega_p}} 
\Bigl\{ \tilde{\xi}_{p}^\dagger (t_x) 
 e^{ i \Vec{p} \cdot \Vec{x}} 
+ \tilde{\eta}_{p} (t_x) 
 e^{ -i \Vec{p} \cdot \Vec{x}}\Bigr\}. \label{phic_xi4} 
\end{eqnarray}
Thus the time dependent operators 
can be combined as the thermal doublet notation,
\begin{eqnarray}
&&\xi_{p}^\alpha (t_x)
 = \xi_{p}^\alpha e^{-i \omega_{+,p} t_x},~~
\bar{\xi}_{p}^\alpha (t_x)
 =\bar{\xi}_{p}^{\alpha} e^{i \omega_{+,p} t_x}, \label{csf_e_xi-1} \\
&& \eta_{p}^\alpha (t_x)
 = \eta_{p}^\alpha e^{-i \omega_{-,p} t_x},~~ 
\bar{\eta}_{p}^\alpha (t_x)
 =\bar{\eta}_{p}^{\alpha} e^{i \omega_{-,p} t_x}. \label{csf_e_xi-2}
\end{eqnarray}

The complex scalar field is written in the thermal doublet notation, 
\begin{eqnarray}
\phi_{\xi}^{\alpha}(x) &\equiv& \int \frac{d^3\Vec{p}}{(2\pi)^3}
 \frac{1}{\sqrt{2\omega_p}} 
\left\{ 
\left(
  \begin{array}{c}
	\xi_{p}(t_x) \\
	\tilde{\xi}_{p}^\dagger (t_x) 
  \end{array}
\right) 
 e^{ i \Vec{p} \cdot \Vec{x}} 
+ 
\left(
  \begin{array}{c}
	\eta_{p}^\dagger (t_x) \\
	\tilde{\eta}_{p}(t_x) 
  \end{array}
\right)
 e^{ -i \Vec{p} \cdot \Vec{x}} \right\} \nonumber \\
&=& \int \frac{d^3\Vec{p}}{(2\pi)^3}
 \frac{1}{\sqrt{2\omega_p}} 
\left\{ \xi_{p}^{\alpha}(t_x) 
 e^{ i \Vec{p} \cdot \Vec{x}} 
+ (\tau_3 \bar{\eta}_{p} (t_x)^T)^\alpha 
 e^{ -i \Vec{p} \cdot \Vec{x}} \right\},   \label{csf_phixi1} \\
\bar{\phi}_{\xi}^{\alpha}(x) &\equiv& \int \frac{d^3\Vec{p}}{(2\pi)^3} 
\frac{1}{\sqrt{2\omega_p}} 
\Bigl\{
\left(
  \begin{array}{cc}
	\xi_{p}^\dagger (t_x) & - \tilde{\xi}_{p}(t_x)
  \end{array}
\right)
 e^{ -i \Vec{p} \cdot \Vec{x}} \nonumber \\
&& \qquad
+ \left(
  \begin{array}{cc}
	\eta_{p}(t_x) & - \tilde{\eta}_{p}^\dagger (t_x)
  \end{array}
\right)
 e^{ i \Vec{p} \cdot \Vec{x}} \Bigr\} \nonumber \\
&=& \int \frac{d^3\Vec{p}}{(2\pi)^3} 
\frac{1}{\sqrt{2\omega_p}} 
\left\{ \bar{\xi}_{p}^{\alpha}(t_x) 
 e^{ -i \Vec{p} \cdot \Vec{x}} 
+ (\eta_{p} (t_x)^{T} \tau_3)^\alpha 
 e^{ i \Vec{p} \cdot \Vec{x}} \right\},  \label{csf_phixi2}
\end{eqnarray}
where $\tau_3^{\alpha \beta}$ is the third Pauli matrix acting on the thermal indices. 
The canonical conjugate field is given by 
\begin{eqnarray}
&&  \pi_{\xi}^\alpha(x) = (-i)\int \frac{d^3\Vec{p}}{(2\pi)^3}
 \sqrt{\frac{\omega_p}{2}} 
\left\{ \xi_{p}^\alpha (t_x) 
 e^{ i \Vec{p} \cdot \Vec{x}} 
- (\tau_3 \bar{\eta}_{p}(t_x)^T)^\alpha 
 e^{ -i \Vec{p} \cdot \Vec{x}} \right\},  
 \label{csf_pixi1} \\
&&  \bar{\pi}_{\xi}^\alpha(x) = (-i)\int \frac{d^3\Vec{p}}{(2\pi)^3} 
 \sqrt{\frac{\omega_p}{2}} 
\left\{ -\bar{\xi}_{p}^\alpha (t_x) 
 e^{ -i \Vec{p} \cdot \Vec{x}} 
+ (\eta_{p} (t_x)^{T} \tau_3)^\alpha 
 e^{ i \Vec{p} \cdot \Vec{x}} \right\}. \label{csf_pixi2}
\end{eqnarray}
The canonical commutation relations are satisfied at the equal time,
\begin{eqnarray}
 [\phi_{\xi}^\alpha (t,\Vec{x}),\bar{\pi}_{\xi}^\beta (t,\Vec{y})]
 =  [ \bar{\phi}_{\xi}^\alpha (t,\Vec{x}),\pi_{\xi}^\beta (t,\Vec{y}) ]
 = i \delta^{(3)}(\Vec{x}-\Vec{y}) \delta^{\alpha \beta}.
\end{eqnarray}

Next we decompose the complex scalar field 
in terms of the original operators, $a_p$ and $b_p$. 
Differentiating the operators, $\xi_{p}^{\alpha}(t_x)$ and $\bar{\xi}_{p}^{\alpha}(t_x)$, 
with respect to the time variable, we obtain 
\begin{eqnarray}
&& \partial_{t_x} \xi_{p}^\alpha (t_x)
 = -i \omega_{+,p} \xi_{p}^\alpha (t_x),  \label{csf_t-evo_xi1} \\
&& \partial_{t_x} \bar{\xi}_{p}^\alpha (t_x)
 = i  \omega_{+,p} \bar{\xi}_{p}^\alpha (t_x). \label{csf_t-evo_xi2} 
\end{eqnarray}
The thermal Bogoliubov transformation (\ref{csf_tbt1}) 
gives the time evolution equations 
for the operators, $a_p^{\alpha}(t_x)$ and $\bar{a}_p^{\alpha}(t_x)$, 
\begin{eqnarray}
&&\partial_{t_x}a_{p}^{\alpha}(t_x)
=-i \{ \omega_{+,p} -i \dot{n}_{+}(t_x;p)T_0 \}^{\alpha \beta} a_{p}^{\beta}(t_x), 
 \label{csf_t-evo_a1} \\
&&\partial_{t_x}\bar{a}_{p}^\alpha (t_x) 
= \bar{a}_{p}^\beta (t_x)
 i \{ \omega_{+,p} -i \dot{n}_{+}(t_x;p)T_0 \}^{\beta \alpha}, 
 \label{csf_t-evo_a2} 
\end{eqnarray}
where the matrix, $T_0^{\alpha \beta}$, is given by 
\begin{eqnarray}
T_0^{\alpha \beta} =
\left(
   \begin{array}{cc}
	1 & -1 \\
	1 & -1 \\
   \end{array}
\right). \label{t_0}
\end{eqnarray}
It should be noted that the matrix, $T_0^{\alpha \beta}$, satisfies, 
$T_0^{\alpha \beta}T_0^{\beta \gamma} = 0$. 
Eqs.~(\ref{csf_t-evo_a1}) and (\ref{csf_t-evo_a2}) indicate that 
the energy eigenvalue for the operator, $a_p(t_x)$, 
depends on the time derivative 
of the thermal Bogoliubov parameter, $\dot{n}_{+}(t;p)$. 

Hence, we define the time dependent operators, 
$a_{p}^\alpha (t_x)$ and $\bar{a}_{p}^\alpha (t_x)$, by
\begin{eqnarray}
&& a_{p}^\alpha (t_x) = 
 {\rm exp} \left\{ -i\int_{-\infty}^{t_x} dt_s \Omega_{+,p}(t_s) 
 \right\}^{\alpha \beta}
 a_{p}^\beta, \label{cs_a1} \\
&& \bar{a}_{p}^\alpha (t_x) = \bar{a}_{p}^{\beta}\ 
 {\rm exp} \left\{ i \int_{-\infty}^{t_x} dt_s \Omega_{+,p} (t_s)
  \right\}^{\beta \alpha}, \label{cs_a2} 
\end{eqnarray}
where we write
\begin{eqnarray}
&&\Omega_{+,p}^{\alpha \beta} (t_s)
 \equiv \omega_{+,p} \delta^{\alpha \beta}
  -i \dot{n}_{+}(t_s;p) T_0^{\alpha \beta}. \label{csf_ea1} 
\end{eqnarray}

The time dependence of the operators, 
$b_p^{\alpha}(t_x)$ and $\bar{b}_p^{\alpha}(t_x)$, 
is obtained from Eq.~(\ref{csf_e_xi-2}). 
Time derivatives of $\eta_p^{\alpha}(t_x)$ and $\bar{\eta}_p^{\alpha}(t_x)$ give 
\begin{eqnarray}
&& \partial_{t_x} \eta_{p}^\alpha (t_x)
 = -i \omega_{-,p} \eta_{p}^\alpha (t_x), \label{csf_t-evo_eta1} \\
&& \partial_{t_x} \bar{\eta}_{p}^\alpha (t_x)
 = i  \omega_{-,p} \bar{\eta}_{p}^\alpha (t_x). \label{csf_t-evo_eta2}
\end{eqnarray}
Applying the thermal Bogoliubov transformation, 
we find the time evolution equations 
for $b_p^{\alpha}(t_x)$ and $\bar{b}_p^{\alpha}(t_x)$,
\begin{eqnarray}
&&\partial_{t_x} b_{p}^{\alpha}(t_x)
=-i \{ \omega_{-,p} -i \dot{n}_{-} (t_x;p) T_0 \}^{\alpha \beta} 
 b_{p}^{\beta}(t_x), 
\label{csf_t-evo_b1} \\
&&\partial_{t_x}\bar{b}_{p}^\alpha (t_x) 
= \bar{b}_{p}^\beta (t_x)
 i \{ \omega_{-,p} -i \dot{n}_{-}(t_x;p) T_0 \}^{\beta \alpha}.  
 \label{csf_t-evo_b2}
\end{eqnarray}
Thus the time dependent operators, $b_{p}^\alpha (t_x)$ and 
$\bar{b}_{p}^\alpha (t_x)$, are described as 
\begin{eqnarray}
&& b_{p}^\alpha (t_x) = 
 {\rm exp} \left\{ -i\int_{-\infty}^{t_x} dt_s \Omega_{-,p}(t_s) 
 \right\}^{\alpha \beta}
b_{p}^\beta,  \label{cs_b1} \\
&& \bar{b}_{p}^\alpha (t_x) = \bar{b}_{p}^{\beta}\ 
 {\rm exp} \left\{ i \int_{-\infty}^{t_x} dt_s \Omega_{-,p} (t_s)
  \right\}^{\beta \alpha}, \label{cs_b2} 
\end{eqnarray}
where the energy eigenvalue of the operators, 
$\Omega_{-,p}^{\alpha\beta}(t_s)$, is given by
\begin{eqnarray}
&&\Omega_{-,p}^{\alpha \beta} (t_s)
 \equiv \omega_{-,p} \delta^{\alpha \beta}
  -i \dot{n}_{-}(t_s;p) T_0^{\alpha \beta}. \label{csf_eb1} 
\end{eqnarray}

Therefore the operators, $a_{p}$ and $b_{p}$, are organized into 
the positive and the negative frequency parts of the complex scalar field, 
\begin{eqnarray}
&&  \phi_{a}^\alpha(x) \equiv \int \frac{d^3\Vec{p}}{(2\pi)^3}
 \frac{1}{\sqrt{2\omega_p}} 
\left\{ 
a_{p}^\alpha (t_x) 
e^{ i \Vec{p} \cdot \Vec{x}} 
+ (\tau_3 \bar{b}_{p}(t_x)^T)^\alpha 
e^{ -i \Vec{p} \cdot \Vec{x}} \right\}, \label{csf_phia1} \\
&&  \bar{\phi}_{a}^\alpha (x) \equiv \int \frac{d^3\Vec{p}}{(2\pi)^3} 
\frac{1}{\sqrt{2\omega_p}} 
\left\{ \bar{a}_{p}^\alpha (t_x) 
e^{ -i \Vec{p} \cdot \Vec{x}} 
+ ( b_{p}(t_x)^{T} \tau_3)^\alpha 
e^{ i \Vec{p} \cdot \Vec{x}} \right\}. \label{csf_phia2}
\end{eqnarray}
Since the energy eigenvalues, 
$\Omega_{\pm,p}^{\alpha \beta} (t_s)$, are not Hermitian, 
the complex scalar field (\ref{csf_phia1}) and (\ref{csf_phia2}) 
are not invariant under the time-reversal transformation. 

We define the canonical conjugate field, 
\begin{eqnarray}
&&  \pi_{a}^{\alpha} (x) \equiv (-i) \int \frac{d^3\Vec{p}}{(2\pi)^3}
 \sqrt{\frac{\omega_p}{2}} 
\left\{ 
a_{p}^\alpha (t_x) 
e^{ i \Vec{p} \cdot \Vec{x}} 
- ( \tau_3 \bar{b}_{p}(t_x)^T )^\alpha 
e^{ -i \Vec{p} \cdot \Vec{x}} \right\}, \label{csf_pia1}  \\
&&  \bar{\pi}_{a}^{\alpha}(x) \equiv (-i) \int \frac{d^3\Vec{p}}{(2\pi)^3} 
\sqrt{\frac{\omega_p}{2}} 
\left\{ -\bar{a}_{p}^\alpha (t_x) 
e^{ -i \Vec{p} \cdot \Vec{x}} 
+ ( b_{p}(t_x)^T \tau_3)^\alpha 
e^{ i \Vec{p} \cdot \Vec{x}} \right\}, \label{csf_pia2}
\end{eqnarray}
to satisfy the equal-time canonical commutation relations, 
\begin{eqnarray}
&& [\phi_{a}^{\alpha}(t,\Vec{x}),\bar{\pi}_{a}^{\beta}(t,\Vec{y})]
= [ \bar{\phi}_{a}^{\alpha}(t,\Vec{x}),\pi_{a}^{\beta}(t,\Vec{y}) ]
 = i \delta^{(3)}(\Vec{x}-\Vec{y}) \delta^{\alpha \beta}. \label{csf_phia_comm1}
\end{eqnarray}
Thus the complex scalar field is constructed by
$a_p$ and $b_p$ with an ordinary canonical commutation relations. 

The unperturbed hat-Hamiltonian for the complex scalar field, 
$\phi_a^{\alpha}$ and $\bar{\phi}_a^{\alpha}$, 
should be defined to derive the equations of motion. 
The time derivatives of Eqs.~(\ref{csf_phia1})-(\ref{csf_pia2}) 
give the equations of motion,
\begin{eqnarray}
&& (\partial_{t_x} - i \mu )\phi_{a}^{\alpha}(x) 
 = \pi_{a}^{\alpha}(x) \nonumber \\
&&\quad - \dot{n}_{+}(t_x;|\nabla_x|) T_0^{\alpha \beta}\phi_{a,+}^{\beta}(x)
+ \dot{n}_{-}(t_x;|\nabla_x| ) T_0^{\alpha\beta}\phi_{a,-}^{\beta}(t_x), 
 \label{csf_EOMa1} \\ 
&& (\partial_{t_x} + i \mu )\bar{\phi}_{a}^{\alpha}(x) 
 = \bar{\pi}_{a}^{\alpha}(x) \nonumber \\
&&\quad + \dot{n}_{+}(t_x;|\nabla_x|) \bar{\phi}_{a,-}^{\beta}(x)T_0^{\beta \alpha}
- \dot{n}_{-}(t_x;|\nabla_x| ) \bar{\phi}_{a,+}^{\beta}(t_x) T_0^{\beta \alpha}, 
 \label{csf_EOMa2} \\
&& (\partial_{t_x} -i \mu ) \pi_{a}^{\alpha}(x)= - (-\nabla_{x}^2 + m^2) \phi_{a}^{\alpha}(x) 
  \nonumber \\
&&\quad - \dot{n}_{+}(t_x;|\nabla_x|) T_{0}^{\alpha\beta} \pi_{a,+}^{\beta}(x)
+ \dot{n}_{-}(t_x;|\nabla_x|) T_{0}^{\alpha\beta} \pi_{a,-}^{\beta}(x), 
 \label{csf_EOMa3} \\
&& (\partial_{t_x} + i \mu ) \bar{\pi}_{a}^{\alpha}(x)= 
  - (-\nabla_{x}^2 + m^2) \bar{\phi}_{a}^{\alpha}(x) 
\nonumber \\
&&\quad  + \dot{n}_{+}(t_x;|\nabla_x|) \bar{\pi}_{a,-}^{\beta}(x) T_{0}^{\beta\alpha} 
- \dot{n}_{-}(t_x;|\nabla_x|) \bar{\pi}_{a,+}^{\beta}(x) T_{0}^{\beta\alpha}, 
 \label{csf_EOMa4}
\end{eqnarray}
where $\phi_{a,\pm}^{\alpha}$ and $\bar{\phi}_{a,\pm}^{\alpha}$ denote the positive 
and the negative frequency parts of the complex scalar field, 
(\ref{csf_phia1}) and (\ref{csf_phia2}). $\pi_{a,\pm}^{\alpha}$ 
and $\bar{\pi}_{a,\pm}^{\alpha}$ represent 
its canonical conjugate (\ref{csf_pia1}) and (\ref{csf_pia2}). 
The terms proportional to $\dot{n}_{\pm}(t_x;|\nabla_x|)$ appear from the time 
dependence of the thermal Bogoliubov transformation.

The unperturbed hat-Hamiltonian, $\hat{H}_0$, 
for the complex scalar field in an equilibrium system 
consists with the kinetic term and the conserved charge as
\begin{eqnarray}
&&\hat{H}_0 = 
 \int d^3\Vec{x} \bigl[ \bar{\pi}_a^{\alpha}(x) \pi_a^{\alpha} (x)
 + \bar{\phi}_a^{\alpha}(x) (-\nabla_x^2 + m^2) \phi_a^{\alpha}(x) \nonumber \\
&&\quad -i\mu \left\{ \bar{\phi}_{a}^{\alpha}(x)\pi_{a}^{\alpha}(x) 
- \bar{\pi}_{a}^{\alpha}(x) \phi_{a}^{\alpha}(x) \right\} \bigr], \label{csf_Ha0}
\end{eqnarray}
see Appendix A. 
In NETFD the time dependent Bogoliubov parameters 
induce the thermal counter term. 

As a simple extension of the thermal counter term 
for the neutral scalar field we set
\begin{eqnarray}
&&\hat{Q}_n = \int d^3{\Vec{x}} \Bigl\{  \bar{\pi}_a^{\alpha}(x)
 i \frac{\dot{\bar{n}}_{c}(t_x;|\nabla_x|)}{\hat{\omega}_{\nabla_x}} 
T_0^{\alpha \beta} \pi_{a}^{\beta}(x) \nonumber \\
&&\quad + \bar{\phi}_a^{\alpha}(x) 
 i \frac{\dot{\bar{n}}_{c}(t_x;|\nabla_x|)}{\hat{\omega}_{\nabla_x}} T_0^{\alpha \beta} 
 (-\nabla_x^2 + m^2)
 \phi_{a}^{\beta} (x) \Bigr\},	\label{csf_Q_1}
\end{eqnarray}
where $\hat{\omega}_{\nabla_x}$ means $\sqrt[]{\nabla_x^2 + m^2}$ 
and $\dot{\bar{n}}_{c}(t_x;|\nabla_x|)$ is 
a coefficient which consists of the Bogoliubov parameters.

The time dependence of one of the Bogoliubov parameters 
is fixed by the self-consistency renormalization condition 
with the thermal counter term, $\hat{Q}_n$\cite{CQNS1}. 
However, the complex scalar field depends on 
two Bogoliubov parameters, $n_{+}(t;|\nabla_x|)$ and $n_{-}(t;|\nabla_x|)$. 
To fix all the Bogoliubov parameters, 
we introduce an additional thermal counter term. 
The second line in Eq.~(\ref{csf_Ha0}) appears for the charged 
scalar field, i.e. no such term for the neutral field. 
We introduce an additional counter term for the second line in Eq.~(\ref{csf_Ha0}),
\begin{eqnarray}
&&\hat{Q}_{c} 
= -\frac{1}{2} \int d^3{\Vec{x}} \Bigl\{  
\bar{\phi}_{a}^{\alpha}(x) \dot{\mu}_{c}(t_x;|\nabla_x|) T_0^{\alpha\beta} \pi_{a}^{\beta}(t_x)
\nonumber \\
&&\quad - \bar{\pi}_{a}^{\alpha}(x) \dot{\mu}_{c}(t_x;|\nabla_x|) T_0^{\alpha\beta}
 \phi_{a}^{\beta}(x)
\Bigr\},	\label{csf_Q_2}
\end{eqnarray}
where $\dot{\mu}_{c}(t_x;|\nabla_x|)$ is 
an coefficient which consists of the Bogoliubov parameters. 
Thus the thermal counter term for the complex scalar field is given by
\begin{eqnarray}
&&\hat{Q} = \hat{Q}_{n} + \hat{Q}_{c} \nonumber \\
&&=\int d^3{\Vec{x}} \Biggl[  \bar{\pi}_a^{\alpha}(x)
 i \frac{\dot{\bar{n}}_{c}(t_x;|\nabla_x|)}{\hat{\omega}_{\nabla_x}} 
T_0^{\alpha \beta} \pi_{a}^{\beta}(x) \nonumber \\
&&\quad + \bar{\phi}_a^{\alpha}(x) 
 i \frac{\dot{\bar{n}}_{c}(t_x;|\nabla_x|)}{\hat{\omega}_{\nabla_x}} T_0^{\alpha \beta} 
 (-\nabla_x^2 + m^2)
 \phi_{a}^{\beta} (x)  \nonumber \\
&& -\frac{1}{2} \Bigl\{  
\bar{\phi}_{a}^{\alpha}(x) \dot{\mu}_{c}(t_x;|\nabla_x|) T_0^{\alpha\beta} \pi_{a}^{\beta}(t_x)
\nonumber \\
&&\quad - \bar{\pi}_{a}^{\alpha}(x) \dot{\mu}_{c}(t_x;|\nabla_x|) T_0^{\alpha\beta}
 \phi_{a}^{\beta}(x)
\Bigr\} \Biggr].	\label{csf_Q}
\end{eqnarray}

The unperturbed hat-Hamiltonian 
for the non-equilibrium complex scalar field, $\hat{H}_Q$, 
is represented as
\begin{eqnarray}
&&\hat{H}_Q = \hat{H}_0 - \hat{Q} \nonumber \\
&& = \int d^3\Vec{x} \Biggl[ 
 \bar{\pi}_a^{\alpha}(x)
 \left( 1 - i \frac{\dot{\bar{n}}_{c}(t_x;|\nabla_x|)}{\hat{\omega}_{\nabla_x}} 
T_0 \right)^{\alpha \beta} \pi_{a}^{\beta}(x)
 \nonumber \\
&&\quad + \bar{\phi}_a^{\alpha}(x) \left( 1 
- i \frac{\dot{\bar{n}}_{c}(t_x;|\nabla_x|)}{\hat{\omega}_{\nabla_x}} 
T_0 \right)^{\alpha \beta} 
 (-\nabla_x^2 + m^2)
 \phi_{a}^{\beta} (x) \nonumber \\
&&\quad + \bar{\phi}_{a}^{\alpha}(x) 
\left( -i \mu + \frac{1}{2}\dot{\mu}_{c}(t_x;|\nabla_x|) T_0 \right)^{\alpha \beta} 
\pi_{a}^{\beta}(x) \nonumber \\
&&\quad - \bar{\pi}_{a}^{\alpha}(x) 
\left( -i \mu + \frac{1}{2} \dot{\mu}_{c}(t_x;|\nabla_x|) T_0 \right)^{\alpha \beta}
\phi_{a}^{\beta}(x)
\Biggr].	\label{csf_HQ}
\end{eqnarray}

The time dependence of the field is described
by the Heisenberg equations of motion.
Calculating the commutator between the
unperturbed hat-Hamiltonian and the field,
we obtain
\begin{eqnarray}
&&\!\!\!\!\!\!\!\! (\partial_{t_x} - i \mu )\phi_{a}^{\alpha}(x) 
 = \left( 1-i \frac{\dot{\bar{n}}_{c}(t_x;|\nabla_x|)}
{\hat{\omega}_{\nabla_x}}T_0 \right)^{\alpha \beta} \pi_{a}^{\beta}(x) 
 - \frac{1}{2} \dot{\mu}_{c}(t_x;|\nabla_x|) T_0^{\alpha \beta}\phi_{a}^{\beta}(x), 
\label{csf_EOMa2-1}  \\ 
&&\!\!\!\!\!\!\!\! (\partial_{t_x} + i \mu )\bar{\phi}_{a}^{\alpha}(x) 
 = \bar{\pi}_{a}^{\beta}(x) 
\left( 1-i \frac{\dot{\bar{n}}_{c}(t_x;|\overleftarrow{\nabla}_x|)}
{\hat{\omega}_{\overleftarrow{\nabla}_x}}T_0 \right)^{\beta \alpha}  
+ \frac{1}{2} \dot{\mu}_{c}(t_x;|\nabla_x|) \bar{\phi}_{a}^{\beta}(x)T_0^{\beta \alpha}, 
 \label{csf_EOMa2-2} \\
&&\!\!\!\!\!\!\!\!  (\partial_{t_x} -i \mu ) \pi_{a}^{\alpha}(x) \nonumber \\
&&\!\!\!\!\!\!\!\! = - \left( 1-i \frac{\dot{\bar{n}}_{c}(t_x;|\nabla_x|)}
{\hat{\omega}_{\nabla_x}}T_0 \right)^{\alpha \beta} 
(-\nabla_{x}^2 + m^2) \phi_{a}^{\beta}(x) 
- \frac{1}{2} \dot{\mu}_{c}(t_x;|\nabla_x|) T_{0}^{\alpha\beta} \pi_{a}^{\beta}(x), 
 \label{csf_EOMa2-3} 
\end{eqnarray}

\begin{eqnarray}
&&\!\!\!\!\!\!\!\! (\partial_{t_x} + i \mu ) \bar{\pi}_{a}^{\alpha}(x) \nonumber \\
&&\!\!\!\!\!\!\!\! = 
- \bar{\phi}_{a}^{\beta}(x) 
\left( 1-i \frac{\dot{\bar{n}}_{c}(t_x;|\overleftarrow{\nabla}_x|)}
{\hat{\omega}_{\overleftarrow{\nabla}_x}}T_0 \right)^{\beta \alpha} 
(-\overleftarrow{\nabla}_{x}^2 + m^2)  
+ \frac{1}{2} \dot{\mu}_{c}(t_x;|\nabla_x|) 
\bar{\pi}_{a}^{\beta}(x) T_{0}^{\beta\alpha}. \label{csf_EOMa2-4}
\end{eqnarray}
These equations should coincide with Eqs.~(\ref{csf_EOMa1})-(\ref{csf_EOMa4}). 
This condition fixes the coefficient in the thermal counter term, 
\begin{eqnarray}
&&\dot{\bar{n}}_{c}(t_x;|\nabla_x|) 
= \frac{\dot{n}_{+}(t_x;|\nabla_x|) + \dot{n}_{-}(t_x;|\nabla_x|)}{2}, \label{csf_nc_1} \\
&&\dot{\mu}_{c}(t_x;|\nabla_x|) = \dot{n}_{+}(t_x;|\nabla_x|) - \dot{n}_{-}(t_x;|\nabla_x|).
\label{csf_nu_1}
\end{eqnarray}
We notice that the chemical potential, $\dot{\mu}_{c}(t_x;|\nabla_x|)$ is given
by the difference of the Bogoliubov parameters for the positive and the 
negative frequency parts. 

\section{Self-Consistency Condition for the Complex Scalar Field}

In this section we discuss the 
self-consistency renormalization condition for the complex scalar field.
For this purpose we evaluate the scalar propagator in NETFD.
The time evolution for the field is given by the 
hat-Hamiltonian, $\hat{H}$. It is divided into two parts, 
\begin{eqnarray}
\hat{H} = \hat{H}_0 + \hat{H}_{int}, \label{cs_H1}
\end{eqnarray}
where $\hat{H}_0$ and $\hat{H}_{int}$ represent 
the hat-Hamiltonian for a free complex scalar field 
and the interaction part in equilibrium system, respectively. 

Since the unperturbed hat-Hamiltonian for the 
field, $\phi_{a}$, is not $\hat{H}_0$ but $\hat{H}_Q$
in non-equilibrium systems, we rearrange the 
hat-Hamiltonian (\ref{cs_H1}) as
\begin{eqnarray}
\hat{H} = \hat{H}_Q + \hat{H}_I, \label{cs_H2}
\end{eqnarray}
where $\hat{H}_I$ is the interaction hat-Hamiltonian in NETFD,
\begin{eqnarray}
\hat{H}_I = \hat{H}_{int} + \hat{Q}. \label{cs_HI1}
\end{eqnarray}
The perturbative calculation should be accomplished by the unperturbed 
hat-Hamiltonian, $\hat{H}_Q$, and 
the interaction hat-Hamiltonian, $\hat{H}_I$. 
As is shown in Ref.~\refcite{umezawa1}, 
the interaction hat-Hamiltonian (\ref{cs_HI1}) satisfies the condition, 
\begin{eqnarray}
\langle \theta|\hat{H}_I=0. \label{cs_HI_c1}
\end{eqnarray}

In the interaction picture the full scalar propagator is given by
\begin{eqnarray}
D_{H}^{\alpha\beta}(t_x, t_y,\Vec{x}-\Vec{y})
= \langle \theta| T[\phi_a^{\alpha} (x) \bar{\phi}_a^{\beta} (y)
u(\infty,-\infty)]|\theta\rangle,
\label{csf_full_pro1}
\end{eqnarray}
where $T$ denotes time-ordering operator
and $u(t,t^{\prime})$ is the time evolution operator,
\begin{eqnarray}
u(t, t^{\prime}) = {\exp} \left( -i\int_{t^{\prime}}^{t} dt_s \hat{H}_I (t_s) \right). 
\label{u1}
\end{eqnarray} 
From Eq.~(\ref{cs_HI_c1}) we find that 
the thermal vacuum, $\langle \theta|$, 
is annihilated by the interaction hat-Hamiltonian and satisfies, 
\begin{eqnarray}
\langle \theta|u(t,t^\prime)=\langle \theta|, \label{cs_bra_vac1}
\end{eqnarray}
which is necessary to adopt the Feynman diagram procedure\cite{umezawa1,FynPert1}. 

Applying the thermal Bogoliubov transformation 
with Eq.~(\ref{cs_bra_vac1}), we rewrite 
the thermal full propagator (\ref{csf_full_pro1}) as 
\begin{eqnarray}
&& D_{H}^{\alpha \beta}(t_{x}, t_{y},\Vec{x}-\Vec{y}) \nonumber \\
&& = B^{-1}(n_{+}(t_x;|\nabla_{x}|)) 
 \left(
   \begin{array}{cc}
	d_{c,1}^{11}(x,y) & d_{c,1}^{12}(x,y) \\
	0 & d_{c,1}^{22}(x,y)
   \end{array}
\right) B(n_{+}(t_y;|\overleftarrow{\nabla}_{y}|)) \nonumber \\
&& + B^{-1}(n_{+}(t_x;|\nabla_{x}|)) 
 \left(
   \begin{array}{cc}
		d_{c,2}^{11}(x,y) & d_{c,2}^{12}(x,y) \\
		d_{c,2}^{21}(x,y) & 0
   \end{array}
\right) 
 B^{-1}(n_{-}(t_y;|\overleftarrow{\nabla}_{y}|))^T \tau_3  \nonumber \\
&& + \tau_3 B(n_{-}(t_x;|\nabla_{x}|))^T  
 \left(
   \begin{array}{cc}
	0 & d_{c,3}^{12}(x,y) \\
	d_{c,3}^{21}(x,y) & d_{c,3}^{22}(x,y)
   \end{array}
\right) 
 B(n_{+}(t_y;|\overleftarrow{\nabla}_{y}|)) \nonumber \\
&& +  \tau_3 B(n_{-}(t_x;|\nabla_{x}|))^T 
 \left(
   \begin{array}{cc}
	d_{c,4}^{11} (x,y) & 0 \\
	d_{c,4}^{21} (x,y) & d_4^{22}(x,y)
   \end{array}
\right) 
   B^{-1}(n_{-}(t_y;|\overleftarrow{\nabla}_{y}|))^T \tau_3, 
\label{csf_p2}
\end{eqnarray}
with
\begin{eqnarray}
&& d_{c,1}^{\gamma_1 \gamma_2}(x,y) 
=  \theta (t_x-t_y) \langle \theta| \phi_{\xi,+}^{\gamma_1}(x)
 u(t_x,t_y) \bar{\phi}_{\xi,-}^{\gamma_2}(y)u(t_y,-\infty)|\theta\rangle  \nonumber \\
&&\quad + \theta (t_y-t_x) \langle \theta| \bar{\phi}_{\xi,-}^{\gamma_2}(y)
 u(t_y,t_x) \phi_{\xi,+}^{\gamma_1}(x)u(t_x,-\infty)|\theta\rangle, 
\label{cs_d1} \\
&& d_{c,2}^{\gamma_1 \gamma_2}(x,y)
=\theta (t_x-t_y) \langle \theta| \phi_{\xi,+}^{\gamma_1}(x)
  u(t_x,t_y) \{\bar{\phi}_{\xi,+}(y)\tau_3\}^{\gamma_2}
u(t_y,-\infty)|\theta\rangle	\nonumber \\
&&\quad + \theta (t_y-t_x) \langle \theta| \{ \bar{\phi}_{\xi,+}(y) \tau_3 \}^{\gamma_2} 
 u(t_y,t_x) \phi_{\xi,+}^{\gamma_1}(x) u(t_x,-\infty) |\theta \rangle, 
\label{cs_d2} \\
&& d_{c,3}^{\gamma_1 \gamma_2}(x,y)
=\theta (t_x-t_y) \langle \theta| \{ \tau_3 \phi_{\xi,-}(x) \}^{\gamma_1} 
 u(t_x,t_y) \bar{\phi}_{\xi,-}^{\gamma_2}(y) u(t_y,-\infty) |\theta \rangle \nonumber \\
&&\quad + \theta (t_y-t_x)\langle \theta| \bar{\phi}_{\xi,-}^{\gamma_2}(y)
  u(t_y,t_x) \{ \tau_3 \phi_{\xi,-}(x) \}^{\gamma_1}u(t_x,-\infty)|\theta\rangle, 
\label{cs_d3} \\
&& d_{c,4}^{\gamma_1 \gamma_2}(x,y)
=\theta (t_x-t_y) 
\langle \theta| \{ \tau_3 \phi_{\xi,-}(x)\}^{\gamma_1}
 u(t_x,t_y) \{ \bar{\phi}_{\xi,+}(y)\tau_3\}^{\gamma_2}
 u(t_y,-\infty) |\theta\rangle 
\nonumber \\
&&\quad + \theta (t_y-t_x) \langle \theta| \{ \bar{\phi}_{\xi,+}(y) \tau_3 \}^{\gamma_2}
 u(t_y,t_x) \{ \tau_3 \phi_{\xi,-}(x) \}^{\gamma_1}
 u(t_x,-\infty) |\theta\rangle, 
\label{cs_d4}
\end{eqnarray}
where $\phi_{\xi,\pm}^{\alpha}$ and $\bar{\phi}_{\xi,\pm}^{\alpha}$ represent 
the positive and the negative frequency parts of 
the complex scalar field (\ref{csf_phixi1}) and (\ref{csf_phixi2}).  
Only the diagonal elements in $d_{c,1}^{\gamma_1 \gamma_2}$ 
and $d_{c,4}^{\gamma_1 \gamma_2}$ have non-vanishing values 
in the free propagator (\ref{csf_app1_3}). 
Radiative corrections generally induce all the elements 
in Eqs.~(\ref{cs_d1})-(\ref{cs_d4}).

The time evolution operator $u(t,t^{\prime})$ is 
expanded perturbatively in terms of the thermal counter term, 
$\hat{Q}$ and the interaction part $\hat{H}_{int}$.
First we evaluate the contribution in the leading order
of the thermal counter term, $\hat{Q}$, at the tree level 
with respect to $\hat{H}_{int}$.
Performing the thermal Bogoliubov transformations in 
Eq.~(\ref{csf_Q}), we rewrite the thermal counter term, $\hat{Q}$, as
\begin{eqnarray}
\hat{Q}= -i \int \frac{d^3\Vec{p}}{(2\pi)^3} 
\left\{ \dot{n}_{+}(t_x;p)
\tilde{\xi}_{p}^\dagger \xi_{p}^\dagger 
+ \dot{n}_{-}(t_x;p) \tilde{\eta}_{p}^{\dagger} \eta_{p}^{\dagger}  \right\}.
\label{cs_Q_xi}
\end{eqnarray}
Substituting Eq.~(\ref{u1}) and Eq.~(\ref{cs_Q_xi}) into Eq.~(\ref{csf_p2}), 
we obtain in the leading order of the thermal counter term, $\hat{Q}$,
\begin{eqnarray}
&& \langle \theta| T[\phi_a^\alpha (x) \bar{\phi}_a^\beta (y)
u(\infty,-\infty)]|\theta\rangle 
- \langle \theta| T[\phi_a^\alpha (x) \bar{\phi}_a^\beta (y)]|\theta\rangle 
\nonumber \\
&&=  \int \frac{d^3 \Vec{p}}{(2\pi)^3}
\frac{1}{2\omega_p}e^{-i\omega_{+,p} (t_x-t_y)}
e^{i \Vec{p}\cdot (\Vec{x}-\Vec{y})} \nonumber \\
&& ~ \times \Biggl( \theta (t_x-t_y)\int_{-\infty}^{t_y} dt_s 
  \dot{n}_{+}(t_s;p)  
 + \theta (t_y-t_x)\int_{-\infty}^{t_x} dt_s 
 \dot{n}_{+}(t_s;p) 
 \Biggr) \nonumber \\
&& ~ \times B^{-1}(n_{+}(t_x;p)) 
\left(
   \begin{array}{cc}
	0 & 1 \\
	0 & 0 \\
   \end{array}
\right) B(n_{+}(t_y;p)) \nonumber \\
&& +  \int \frac{d^3 \Vec{p}}{(2\pi)^3}
\frac{1}{2\omega_p}e^{ i \omega_{-,p} (t_x-t_y)} 
e^{-i \Vec{p}\cdot (\Vec{x}-\Vec{y})} \nonumber \\
&& ~ \times \Biggl( \theta (t_x-t_y)\int_{-\infty}^{t_y} dt_s 
  \dot{n}_{-}(t_s;p)  
 + \theta (t_y-t_x)\int_{-\infty}^{t_x} dt_s 
  \dot{n}_{-}(t_s;p) 
 \Biggr) \nonumber \\
&& ~ \times \tau_3 B(n_{-} (t_x;p))^T  
\left(
   \begin{array}{cc}
	0 & 0 \\
	1 & 0 \\
   \end{array}
\right) 
 B^{-1}(n_{-}(t_y;p))^T \tau_3. 
\label{csf_p:1}
\end{eqnarray}
These counter terms contribute 
$d_{c,1}^{12}$ and $d_{c,4}^{21}$ in Eq.~(\ref{csf_p2}). 

Next we evaluate the contribution from the interaction, $\hat{H}_{int}$.
The contribution to the perturbed propagator is represented as
\begin{eqnarray}
&& \int d^4 z_1 d^4 z_2 D_{0}^{\alpha\gamma_1}(t_x,t_{z_1},\Vec{x}-\Vec{z}_1) 
i\Sigma^{\gamma_1 \gamma_2}(t_{z_1}, t_{z_2}, \Vec{z}_1-\Vec{z}_2)
D_{0}^{\gamma_2 \beta}(t_{z_2},t_y,\Vec{z}_2 - \Vec{y}) \nonumber \\
&& = \int dt_{z_1} dt_{z_2} \bigl[ B^{-1}(n_{+}(t_x;|\nabla_{x}|))^{\alpha \gamma_1} 
\delta \Sigma_{B,1}^{\gamma_1 \gamma_2}( x, z_1, z_2, y) 
B(n_{+}(t_y;|\overleftarrow{\nabla}_{y}|))^{\gamma_2 \beta} \nonumber \\
&& + B^{-1}(n_{+}(t_x;|\nabla_{x}|))^{\alpha \gamma_1} 
\delta \Sigma_{B,2}^{\gamma_1 \gamma_2}( x, z_1, z_2, y) 
\{ B^{-1}(n_{-}(t_y;|\overleftarrow{\nabla}_{y}|))^T \tau_3\}^{\gamma_2 \beta} 
\nonumber \\
&& + \{ \tau_3 B(n_{-}(t_x;|\nabla_{x}|))^T \}^{\alpha \gamma_1} 
\delta \Sigma_{B,3}^{\gamma_1 \gamma_2}( x, z_1, z_2, y)
 B(n_{+}(t_y;|\overleftarrow{\nabla}_{y}|))^{\gamma_2 \beta} \nonumber \\
&& +  \{ \tau_3 B(n_{-}(t_x;|\nabla_{x}|))^T \}^{\alpha \gamma_1} 
\delta \Sigma_{B,4}^{\gamma_1 \gamma_2}( x, z_1, z_2, y)
 \{ B^{-1}(n_{-}(t_y;|\overleftarrow{\nabla}_{y}|))^T \tau_3 \}^{\gamma_2 \beta} \bigr], 
\label{csf_p:2}
\end{eqnarray}
where $D_{0}^{\alpha\beta}$ represents the free propagator 
(\ref{csf_app1_3}) and 
\begin{eqnarray}
&& \delta \Sigma_{B,1}^{\gamma_1 \gamma_2} ( x, z_1, z_2, y)  \\
&& =\left(
   \begin{array}{c}
	D_{0,R}^{11}(x-z_1) i\Sigma_R (t_{z_1}, t_{z_2},\Vec{z}_1-\Vec{z}_2)
 D_{0,R}^{11}(z_2 - y)
 \quad \delta \Sigma_{B,1}^{12} (x,z_1,z_2,y) \\
  \qquad \qquad 0 \qquad \qquad
  D_{0,R}^{22}(x - z_1) i\Sigma_A (t_{z_1}, t_{z_2},\Vec{z}_1-\Vec{z}_2)
 D_{0,R}^{22}(z_2 - y) 
   \end{array}
\right), \nonumber \\
&& \delta \Sigma_{B,2}^{\gamma_1 \gamma_2}( x, z_1, z_2, y) \\
&& =\left(
   \begin{array}{c}
	  \delta \Sigma_{B,2}^{11} ( x, z_1,z_2,y) \quad
	  -D_{0,R}^{11}(x - z_1)i\Sigma_R (t_{z_1}, t_{z_2},\Vec{z}_1-\Vec{z}_2)
 D_{0,A}^{22}( z_2 - y) \\
	  -D_{0,R}^{22}(x - z_1)i\Sigma_A (t_{z_1}, t_{z_2}, \Vec{z}_1-\Vec{z}_2)
 D_{0,A}^{11}(z_2 - y)
	  \qquad \qquad 0 \qquad \qquad
   \end{array}
\right), \nonumber \\
&& \delta \Sigma_{B,3}^{\gamma_1 \gamma_2} ( x, z_1, z_2, y) \\
&& = \left(
   \begin{array}{c}
	  \qquad \qquad 0 \qquad \qquad
	  -D_{0,A}^{11}(x - z_1) i\Sigma_A (t_{z_1}, t_{z_2},\Vec{z}_1-\Vec{z}_2)
D_{0,R}^{22}(z_2 - y) \\
	  -D_{0,A}^{22}(x - z_1) i\Sigma_R (t_{z_1}, t_{z_2},\Vec{z}_1-\Vec{z}_2)
D_{0,R}^{11}(z_2 - y)
	  \quad \delta \Sigma_{B,3}^{44} (x,z_1,z_2,y)
   \end{array}
\right),  \nonumber 
\end{eqnarray}
\begin{eqnarray}
&& \delta \Sigma_{B,4}^{\gamma_1 \gamma_2}( x, z_1, z_2, y) \\
&& =\left(
   \begin{array}{c}
	D_{0,A}^{11}(x - z_1) i\Sigma_A (t_{z_1}, t_{z_2},\Vec{z}_1-\Vec{z}_2)
 D_{0,A}^{11}( z_2 - y)
  \qquad \qquad 0 \qquad \qquad \\
  \delta \Sigma_{B,4}^{21} (x, z_1, z_2,y)  \quad 
  D_{0,A}^{22}(x - z_1) i\Sigma_R (t_{z_1}, t_{z_2},\Vec{z}_1-\Vec{z}_2)
 D_{0,A}^{22}( z_2 - y)    
   \end{array}
\right), \nonumber
\end{eqnarray}
with
\begin{eqnarray}
&& \delta \Sigma_{B,1}^{12}( x, z_1, z_2, y)  \nonumber \\ 
&&= D_{0,R}^{11}(x - z_1)
 \bigl\{ i\Sigma^{12} (t_{z_1}, t_{z_2},\Vec{z}_1-\Vec{z}_2) 
 + i \Sigma_R(t_{z_1}, t_{z_2},\Vec{z}_1-\Vec{z}_2) 
 n_{+}(t_{z_2};|\overleftarrow{\nabla}_{z_2}|) \nonumber \\
&&\quad - n_{+}(t_{z_1};|\nabla_{z_1}|) i \Sigma_A (t_{z_1}, t_{z_2},\Vec{z}_1-\Vec{z}_2) \bigr\}
 D_{0,R}^{22}(z_2 - y), \label{csf_SEND1} \\
&& \delta \Sigma_{B,2}^{11} ( x, z_1, z_2, y) \nonumber \\
&& = D_{0,R}^{11}( x - z_1)
 \bigl\{ i\Sigma^{11} (t_{z_1}, t_{z_2},\Vec{z}_1-\Vec{z}_2) 
 + i \Sigma_R(t_{z_1}, t_{z_2},\Vec{z}_1-\Vec{z}_2) 
n_{-}(t_{z_2};|\overleftarrow{\nabla}_{z_2}|) \nonumber \\
&&\quad + n_{+}(t_{z_1};|\nabla_{z_1}|) i \Sigma_A(t_{z_1}, t_{z_2},\Vec{z}_1-\Vec{z}_2) 
\bigr\} D_{0,A}^{11}( z_2 - y), \label{csf_SEND2} \\
&& \delta \Sigma_{B,3}^{44} ( x, z_1, z_2, y) \nonumber \\
&& = -D_{0,A}^{22}( x - z_1)
 \bigl\{ i\Sigma^{22} (t_{z_1}, t_{z_2},\Vec{z}_1-\Vec{z}_2) 
 + i \Sigma_R(t_{z_1}, t_{z_2},\Vec{z}_1-\Vec{z}_2) 
n_{+}(t_{z_2};|\overleftarrow{\nabla}_{z_2}|) \nonumber \\
&&\quad + n_{-}(t_{z_1};|\nabla_{z_1}|) i \Sigma_A (t_{z_1}, t_{z_2},\Vec{z}_1-\Vec{z}_2) 
\bigr\} D_{0,R}^{22}( z_2 - y), \label{csf_SEND3} \\
&& \delta \Sigma_{B,4}^{21} (x,z_1,z_2,y) \nonumber \\
&& = D_{0,A}^{22}(x - z_1)
 \bigl\{ -i\Sigma^{21} (t_{z_1}, t_{z_2},\Vec{z}_1-\Vec{z}_2) 
- i \Sigma_R(t_{z_1}, t_{z_2},\Vec{z}_1-\Vec{z}_2) 
n_{-}(t_{z_2};|\overleftarrow{\nabla}_{z_2}|) \nonumber \\
&&\quad + n_{-}(t_{z_1};|\nabla_{z_1}|) i \Sigma_A(t_{z_1}, t_{z_2},\Vec{z}_1-\Vec{z}_2) \bigr\}
D_{0,A}^{11}(z_2 - y). \label{csf_SEND4}
\end{eqnarray}
The self-energy, $\Sigma^{\alpha\beta}$, has a $2 \times 2$ matrix form 
in the thermal doublet notation. 
The retarded and the advanced parts of the self-energy, 
$\Sigma_R$ and $\Sigma_A$, are defined by
\begin{eqnarray}
\Sigma_R \equiv \Sigma^{11}+\Sigma^{12} = \Sigma^{21}+\Sigma^{22},
\quad
\Sigma_A \equiv  \Sigma^{11}-\Sigma^{21} = \Sigma^{22}-\Sigma^{12}.
\label{SelfERA}
\end{eqnarray}
The first and the last terms in the right-hand side of 
Eq.~(\ref{csf_p:2}) have the same Bogoliubov transformation
structure as the first and the last terms in the 
right-hand side of Eq.~(\ref{csf_p:1}), respectively.

In Ref.~\refcite{SCRC2} the self-consistency renormalization 
condition has been proposed for a non-relativistic field by imposing 
$\langle \theta|\xi_{full,p}(t_x) \tilde{\xi}_{full,k}(t_x)|\theta\rangle=0$, 
where the subscript "$full$" denotes the perturbed operator 
with all order of radiative corrections. 
In Ref.~\refcite{CQNS1} it has been shown that 
the self-consistency renormalization condition 
induces the quantum Boltzmann equation 
for a neutral relativistic scalar field. 
The self-consistency renormalization condition is naturally
generalized for the complex scalar field, 
$\langle \theta|\xi_{full,p}(t_x) \tilde{\xi}_{full,k}(t_x)|\theta\rangle=0$ 
and $\langle \theta|\tilde{\eta}_{full,p}(t_x) \eta_{full,k}(t_x)|\theta\rangle=0$. 
The conditions fix the terms, $d_{c,1}^{12}$ and $d_{c,4}^{21}$ 
in Eq.~(\ref{csf_p2}).\footnote{
Substituting the fields (\ref{csf_phixi1})-(\ref{csf_phixi2}) and taking the equal time limit, 
we obtain
\begin{eqnarray*}
\lim_{t_x\rightarrow t_y}
d_{c,1}^{12}(x,y)&=& -\int \frac{d^3 \Vec{p}}{(2\pi)^3} \frac{d^3 \Vec{k}}{(2\pi)^3}
\frac{1}{\sqrt[]{2\omega_p}} \frac{1}{\sqrt[]{2\omega_k}} 
e^{ i \Vec{p} \cdot \Vec{x}} e^{ -i \Vec{k} \cdot \Vec{y}} 
\langle \theta|  T[\xi_p(t_x) \tilde{\xi}_k(t_x)u(\infty, -\infty)] |\theta\rangle,\\
\lim_{t_x\rightarrow t_y}
d_{c,4}^{21}(x,y)
&=& -\int \frac{d^3 \Vec{p}}{(2\pi)^3} \frac{d^3 \Vec{k}}{(2\pi)^3}
\frac{1}{\sqrt[]{2\omega_p}} \frac{1}{\sqrt[]{2\omega_k}} 
e^{ -i \Vec{p} \cdot \Vec{x}} e^{ i \Vec{k} \cdot \Vec{y}} 
\langle \theta| T[ \tilde{\eta}_p (t_x) \eta_k (t_x)u(\infty, -\infty)] |\theta\rangle.
\end{eqnarray*}
}
Thus we obtain integral equations for $\dot{n}_{\pm}(t;p)$,
\begin{eqnarray}
&& \int_{-\infty}^{t_x} dt_s \int \frac{d^3 \Vec{p}}{(2\pi)^3} 
e^{i \Vec{p} \cdot (\Vec{x} - \Vec{y})} 
\frac{1}{2\omega_p} \dot{n}_{+}(t_s;p) 
+\lim_{t_x\rightarrow t_y}\int dt_{z_1}dt_{z_2}
 \delta \Sigma_{B,1}^{12} (x, z_1, z_2, y) = 0, \nonumber \\
\label{csf_Self-C1} \\
&& \int_{-\infty}^{t_x} dt_s \int \frac{d^3 \Vec{p}}{(2\pi)^3} 
e^{-i\Vec{p} \cdot (\Vec{x}-\Vec{y})}
\frac{1}{2\omega_p} \dot{n}_{-}(t_s;p) 
 +\lim_{t_x\rightarrow t_y}\int dt_{z_1}dt_{z_2}
 \delta \Sigma_{B,4}^{21} (x, z_1, z_2, y)=0. \nonumber \\
\label{csf_Self-C2}
\end{eqnarray}
After the spatial Fourier transformation 
the self-consistency renormalization conditions are
rewritten in the t-representation.
Acting the time derivative operator, $\partial_{t_x}$, on
Eqs.~(\ref{csf_Self-C1}) and (\ref{csf_Self-C2}), we find
\begin{eqnarray}
\dot{n}_{+}(t_x;p) =
-2 \omega_p \partial_{t_x} \biggl\{ \lim_{t_x\rightarrow t_y}\int dt_{z_1}dt_{z_2}
 \delta \Sigma_{B,1}^{12} (t_x,t_{z_1},t_{z_2},t_y;\Vec{p}) \biggr\}, \label{csf_Boltz_eq1} \\
\dot{n}_{-}(t_x;p) = 
-2 \omega_p \partial_{t_x} \biggl\{ \lim_{t_x\rightarrow t_y}\int dt_{z_1}dt_{z_2}
 \delta \Sigma_{B,4}^{21} (t_x,t_{z_1},t_{z_2},t_y;\Vec{p})\biggr\}, \label{csf_Boltz_eq2}
\end{eqnarray}
where $\delta \Sigma_{B,1}$ and $\delta \Sigma_{B,4}$ are given by
\begin{eqnarray}
&& \delta \Sigma_{B,1}^{12} (t_x, t_{z_1}, t_{z_2}, t_y;\Vec{p})  \nonumber \\ 
&&= D_{0,R}^{11}(t_x - t_{z_1};\Vec{p})
 \Bigl\{ i\Sigma^{12} (t_{z_1}, t_{z_2};\Vec{p}) 
 + n_{+}(t_{z_2};p) \ i \Sigma_R(t_{z_1}, t_{z_2};\Vec{p}) \nonumber \\
&&\quad - n_{+}(t_{z_1};p) \ i \Sigma_A (t_{z_1}, t_{z_2};\Vec{p}) \Bigr\}
 D_{0,R}^{22}(t_{z_2} - t_{y};\Vec{p}), \label{csf_SEND1_t-rep} \\
&& \delta \Sigma_{B,4}^{21} (t_x,t_{z_1},t_{z_2},t_y;\Vec{p}) \nonumber \\
&& = D_{0,A}^{22}(t_x - t_{z_1};\Vec{p})
 \Bigl\{ -i\Sigma^{21} (t_{z_1}, t_{z_2};\Vec{p}) 
- n_{-}(t_{z_2};p) \ i \Sigma_R(t_{z_1}, t_{z_2};\Vec{p}) 
 \nonumber \\
&&\quad + n_{-}(t_{z_1};p) \ i \Sigma_A(t_{z_1}, t_{z_2};\Vec{p}) \Bigr\}
D_{0,A}^{11}(t_{z_2} - t_y;\Vec{p}). \label{csf_SEND4_t-rep}
\end{eqnarray}
The equations (\ref{csf_Boltz_eq1}) and (\ref{csf_Boltz_eq2})
describe the time evolution of the thermal Bogoliubov parameters, $n_{\pm}(t_x;p)$. 
In a similar manner developed in Ref.~\refcite{CQNS1} it can be shown that
the Bogoliubov parameters, $n_{\pm}(t_x;p)$, 
for the complex scalar field correspond to the particle and anti-particle 
number densities 
under the self-consistency renormalization conditions 
(\ref{csf_Self-C1}) and (\ref{csf_Self-C2}).\cite{umezawa1,pertN1,pertN2} 
Thus Eqs.~(\ref{csf_Boltz_eq1}) and (\ref{csf_Boltz_eq2}) 
give the quantum Boltzmann equations for the charged Bosons. 

\section{Boltzmann Equation for the Complex Scalar Field}

The time evolution equations for the Bogoliubov parameters 
(\ref{csf_Boltz_eq1}) and (\ref{csf_Boltz_eq2}) 
represent the transport equations for the number densities. 
In this section 
we discuss the validity of the time evolution equation. 
For this purpose we perturbatively evaluate 
Eqs.~(\ref{csf_Boltz_eq1}) and (\ref{csf_Boltz_eq2}) 
for interacting complex scalar field. 
One of the simplest interactions is a four-point 
interaction between complex scalars. 
We start from the Hamiltonian 
with a $\lambda(\phi_a^{\dagger}\phi_a)^2$ interaction,
\begin{eqnarray}
H&=& \int d^3\Vec{x} \Bigl[ \left\{ \pi_{a}^{\dagger} (x) \pi_{a}(x) 
 + \phi_{a}^{\dagger}(x) (-\nabla_x^2 + m^2) \phi_{a}(x) \right\} \nonumber \\
&&  -i\mu \left\{ \phi_{a}^{\dagger}(x)\pi_{a}(x) 
- \pi_{a}^{\dagger}(x) \phi_{a}(x) \right\}
 + \frac{\lambda}{4}\phi_{a}^{\dagger}(x)^2 \phi_{a}(x)^2 \Bigr]. 
\label{csf_mdl1}
\end{eqnarray}
In NETFD the Hamiltonian (\ref{csf_mdl1}) is extended to the 
hat-Hamiltonian which is defined by
\begin{eqnarray}
\hat{H} &=& \int d^3\Vec{x} \Bigl[ \left\{
\bar{\pi}_a^{\alpha} (x) \pi_{a}^{\alpha} (x) 
 + \bar{\phi}_{a}^{\alpha}(x) (-\nabla_x^2 + m^2) 
\phi_{a}^{\alpha}(x) \right\} \nonumber \\
&&-i\mu \left\{ \bar{\phi}_{a}^{\alpha}(x)\pi_{a}^{\alpha}(x) 
- \bar{\pi}_{a}^{\alpha}(x) \phi_{a}^{\alpha}(x) \right\}
+ \sum_{\gamma=1}^2 \frac{\lambda^{\gamma}}{4} 
\left\{ (\bar{\phi}_{a}(x) \tau_3)^{\gamma} \phi_{a}^{\gamma}(x) \right\}^2
 \Bigr].
\end{eqnarray}

Employing the Feynman rules, we calculate the off-diagonal 
elements (\ref{csf_SEND1}) and (\ref{csf_SEND4}) in the self-energy. 
In the thermal doublet notation the propagator 
(\ref{csf_app1_5}) is assigned for an internal line 
as $D_{0}^{\alpha\beta}\tau_3^{\beta\gamma}$. 
For a vertex we assign, $(-i)\lambda^{\alpha}$, which is given by
\begin{eqnarray}
(-i) \lambda^{\alpha} = (-i) \lambda
\left(
   \begin{array}{c}
	1 \\
	-1 \\
   \end{array}
\right ).
\label{csf_mdl_c1}
\end{eqnarray}

\begin{figure}[pb]
\centerline{\psfig{file=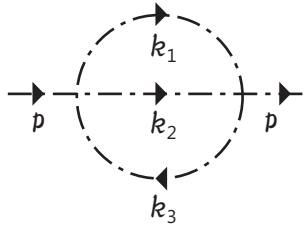}}
\vspace*{8pt}
	\caption{2-loop thermal self-energy 
in $\lambda (\phi_a^{\dagger})^2\phi_a^2$ interaction model}
	\label{csf_figTSE2}
\end{figure}

Since the radiative correction for the propagator has no 
momentum transfer to the internal line at the one-loop level,
the self-energy has a diagonal form up to the one-loop level.
The time evolution of the thermal Bogoliubov parameter
can be induced from the 2-loop self-energy illustrated in 
Fig.~\ref{csf_figTSE2}. We compute the diagram and find 
\begin{eqnarray}
&&i\Sigma^{\alpha \beta}(t_{z_1}, t_{z_2};\Vec{p}) 
= \frac{1}{2}\int \frac{d^3 \Vec{k}_1}{(2\pi)^3}
	\frac{d^3 \Vec{k}_2}{(2\pi)^3} \frac{d^3 \Vec{k}_3}{(2\pi)^3}
	(2\pi)^3 \delta^{(3)}(\Vec{p} - \Vec{k}_1 - \Vec{k}_2 + \Vec{k}_3 ) 
 \tau_3^{\alpha \gamma} (-i) \lambda^{\gamma} \nonumber \\
&& \times
\{ D_{0} (t_{z_1}, t_{z_2};\Vec{k}_1) \tau_3 \}^{\gamma \beta}
\{ D_{0} (t_{z_1}, t_{z_2};\Vec{k}_2) \tau_3 \}^{\gamma \beta}
(-i) \lambda^{\beta}
\{ D_{0} (t_{z_2}, t_{z_1};\Vec{k}_3) \tau_3 \}^{\beta \gamma}. \label{csf_SE1}
\end{eqnarray}
After the contraction of the index "$\gamma$" 
Eq.~(\ref{csf_SE1}) reads\footnote {A simpler formalism is developed 
to calculate the self-energy in Refs.~\refcite{FynRule1} and \refcite{FynRule2}. } 
\begin{eqnarray}
&&\!\!\!\!\!\! i\Sigma^{\alpha \beta}(t_{z_1}, t_{z_2};\Vec{p})
=-\frac{\lambda^2}{2} \sum^{2}_{i_1 =1}\sum^{2}_{i_2 =1}\sum^{2}_{i_3 =1}
\int \frac{d^3\Vec{k}_1}{(2\pi)^3} \frac{d^3\Vec{k}_2}{(2\pi)^3} 
\frac{d^3\Vec{k}_3}{(2\pi)^3} 
\frac{1}{8\omega_{k_1}\omega_{k_2}\omega_{k_3}} \nonumber \\
&&\!\!\!\!\!\! \times	(2\pi)^3 \delta^{(3)}(\Vec{p} - \Vec{k}_1 - \Vec{k}_2 + \Vec{k}_3 ) 
e^{-i(E_{C,k_1,i_1}+E_{C,k_2,i_2}-E_{C,k_3,i_3})(t_{z_1}-t_{z_2})} \nonumber \\
&&\!\!\!\!\!\! \times \Biggl[ \theta( t_{z_1}-t_{z_2} ) 
 \left(
   \begin{array}{cc}
	F_{1}(t_{z_2};k_1,k_2,k_3,i_1,i_2,i_3)
 & -F_{2}(t_{z_2};k_1,k_2,k_3,i_1,i_2,i_3) \\
	F_{1}(t_{z_2};k_1,k_2,k_3,i_1,i_2,i_3)
 & -F_{2}(t_{z_2};k_1,k_2,k_3,i_1,i_2,i_3) \\
   \end{array}
\right) \nonumber \\
&&\!\!\! + \theta( t_{z_2}-t_{z_1} ) 
\left(
   \begin{array}{cc}
	F_{2}(t_{z_1};k_1,k_2,k_3,i_1,i_2,i_3)
 & -F_{2}(t_{z_1};k_1,k_2,k_3,i_1,i_2,i_3) \\
	F_{1}(t_{z_1};k_1,k_2,k_3,i_1,i_2,i_3)
 & -F_{1}(t_{z_1};k_1,k_2,k_3,i_1,i_2,i_3) \\
   \end{array}
\right) 
\Biggr], \label{csf_SE2}
\end{eqnarray}
where we write as
\begin{eqnarray}
&& F_{1}(t;k_1,k_2,k_3,i_1,i_2,i_3) 
\equiv \bar{f}_{i_1}(t; k_1) \bar{f}_{i_2}(t;k_2) f_{i_3}(t;k_3), \\
&& F_{2}(t;k_1,k_2,k_3,i_1,i_2,i_3)
\equiv f_{i_1}(t;k_1) f_{i_2}(t;k_2) \bar{f}_{i_3}(t;k_3),
\end{eqnarray}
and 
\begin{eqnarray}
&& E_{C,q,1} \equiv \omega_{+,q}=\omega_q - \mu,
\quad E_{C,q,2} \equiv -\omega_{-,q}=-\omega_q - \mu, \label{csf_se_n_1} \\
&& f_{1}(t;q) \equiv n_{+}(t;q),\quad f_{2}(t;q) \equiv 1+n_{-}(t;q), \\
&& \bar{f}_{1}(t; q) \equiv 1+n_{+}(t;q),\quad \bar{f}_{2}(t;q) \equiv n_{-}(t;q).
\end{eqnarray}

We substitute the self-energy (\ref{csf_SE2}) into 
Eqs.~(\ref{csf_Boltz_eq1}) and (\ref{csf_Boltz_eq2}) 
and integrate over the time variables. 
Taking the equal time limit, we obtain
\begin{eqnarray}
&& \lim_{t_x\rightarrow t_y}\int dt_{z_1} dt_{z_2} 
\delta \Sigma_{B,1}^{12} (t_x,t_{z_1},t_{z_2},t_y;\Vec{p}) \nonumber \\
&&= \int_{-\infty}^{t_x} dt_s \frac{1}{2\omega_p}
\frac{\lambda^2}{2} \sum^{2}_{i_1 =1}\sum^{2}_{i_2 =1}\sum^{2}_{i_3 =1}
 \int \frac{d^3\Vec{k}_1}{(2{\pi})^3} \frac{d^3\Vec{k}_2}{(2{\pi})^3} 
\frac{d^3\Vec{k}_3}{(2{\pi})^3}
	\frac{1}{8\omega_{p} \omega_{k_1} \omega_{k_2} \omega_{k_3} }
 \nonumber \\
&&\times \frac{{\rm sin} 
\bigl\{ (\omega_{+,p}-E_{C,k_1,i_1}-E_{C,k_2,i_2}+E_{C,k_3,i_3}) (t_x-t_s)
  \bigr\}}
 { \omega_{+,p}-E_{C,k_1,i_1} - E_{C,k_2,i_2} + E_{C,k_3,i_3} }	
 (2\pi)^3 \delta^{(3)}(\Vec{p} - \Vec{k}_1 - \Vec{k}_2 + \Vec{k}_3 ) \nonumber \\
&&\times \bigl\{ n_{+}(t_s;p)
	 \bar{f}_{i_1}(t_s; k_1) \bar{f}_{i_2}(t_s;k_2) f_{i_3}(t_s;k_3) \nonumber \\
&&\quad \quad - (1+n_{+}(t_s;p)) f_{i_1}(t_s; k_1) f_{i_2}(t_s; k_2) \bar{f}_{i_3}(t_s;k_3) \bigr\}, 
	\label{df_off_dia_1} 
\end{eqnarray}
\begin{eqnarray}
&& \lim_{t_x\rightarrow t_y}\int dt_{z_1}dt_{z_2} 
\delta \Sigma_{B,4}^{21} (t_x,t_{z_1},t_{z_2},t_y;\Vec{p}) \nonumber \\
&&= \int_{-\infty}^{t_x}dt_s \frac{1}{2\omega_p}
\frac{\lambda^2}{2} \sum^{2}_{i_1 =1}\sum^{2}_{i_2 =1}\sum^{2}_{i_3 =1}
 \int \frac{d^3\Vec{k}_1}{(2{\pi})^3} \frac{d^3\Vec{k}_2}{(2{\pi})^3} 
\frac{d^3\Vec{k}_3}{(2{\pi})^3}
	\frac{1}{8\omega_{p} \omega_{k_1} \omega_{k_2} \omega_{k_3} }
 \nonumber \\
&&\times \frac{{\rm sin} 
\bigl\{ (\omega_{-,p}+E_{C,k_1,i_1}+E_{C,k_2,i_2}-E_{C,k_3,i_3}) (t_x-t_s) \bigr\}}
 { \omega_{-,p}+E_{C,k_1,i_1}+E_{C,k_2,i_2} - E_{C,k_3,i_3} }	
 (2\pi)^3 \delta^{(3)}(\Vec{p} - \Vec{k}_1 - \Vec{k}_2 + \Vec{k}_3 ) \nonumber \\
&&\times \bigl\{ n_{-}(t_s;p)
	 f_{i_1}(t_s; k_1) f_{i_2}(t_s;k_2) \bar{f}_{i_3}(t_s;k_3) \nonumber \\
&&\quad \quad 
- (1+n_{-}(t_s;p)) \bar{f}_{i_1}(t_s;k_1) \bar{f}_{i_2}(t_s;k_2) f_{i_3}(t_s;k_3) \bigr\}. 
	\label{df_off_dia_2}
\end{eqnarray}
Though 
the chemical potential seems to contribute the frequency and 
the amplitude of the time oscillation, 
Eq.~(\ref{df_off_dia_2}) is independent on $\mu$
because of the relationship (\ref{csf_se_n_1}).

Thus we find the time evolution equations (\ref{csf_Boltz_eq1}) and 
(\ref{csf_Boltz_eq2}) for the thermal Bogoliubov parameters of the complex scalar field,
\begin{eqnarray}
&&\dot{n}_{+}(t_x;p) = 
 -\frac{\lambda^2}{2} \sum^{2}_{i_1 =1}\sum^{2}_{i_2 =1}\sum^{2}_{i_3 =1}
 \int_{-\infty}^{t_x} dt_s \int \frac{d^3\Vec{k}_1}{(2{\pi})^3} \frac{d^3\Vec{k}_2}{(2{\pi})^3}
 \frac{d^3\Vec{k}_3}{(2{\pi})^3} 
	\frac{1}{8\omega_{p} \omega_{k_1} \omega_{k_2} \omega_{k_3} }	\nonumber \\
&&\times {\cos} \bigl\{ (\omega_{p} - E_{k_1,i_1} - E_{k_2,i_2} + E_{k_3,i_3}) (t_x-t_s) \bigr\}	
 (2\pi)^3 \delta^{(3)}(\Vec{p} - \Vec{k}_1 - \Vec{k}_2 + \Vec{k}_3 )
 \nonumber \\
&&\times \bigl\{ n_{+}(t_s;p)
	 \bar{f}_{i_1}(t_s;k_1) \bar{f}_{i_2}(t_s;k_2) f_{i_3}(t_s;k_3) \nonumber \\
&&\quad\quad  
- (1+n_{+}(t_s;p)) f_{i_1}(t_s;k_1) f_{i_2}(t_s;k_2) \bar{f}_{i_3}(t_s;k_3) \bigr\}, 
	\label{csf_Boltz_mdl1} \\
&&\dot{n}_{-}(t_x;p) = 
 -\frac{\lambda^2}{2} \sum^{2}_{i_1 =1}\sum^{2}_{i_2 =1}\sum^{2}_{i_3 =1}
 \int_{-\infty}^{t_x} dt_s \int \frac{d^3\Vec{k}_1}{(2{\pi})^3} \frac{d^3\Vec{k}_2}{(2{\pi})^3}
 \frac{d^3\Vec{k}_3}{(2{\pi})^3} 
	\frac{1}{8\omega_{p} \omega_{k_1} \omega_{k_2} \omega_{k_3} }	\nonumber \\
&&\times {\cos} \bigl\{ (\omega_{p} + E_{k_1,i_1} + E_{k_2,i_2} - E_{k_3,i_3}) (t_x-t_s) \bigr\}	
 (2\pi)^3 \delta^{(3)}(\Vec{p} - \Vec{k}_1 - \Vec{k}_2 + \Vec{k}_3 )
 \nonumber \\
&&\times \bigl\{ n_{-}(t_s;p)
	 f_{i_1}(t_s;k_1) f_{i_2}(t_s;k_2) \bar{f}_{i_3}(t_s;k_3) \nonumber \\
&&\quad\quad  
- (1+n_{-}(t_s;p)) \bar{f}_{i_1}(t_s;k_1) \bar{f}_{i_2}(t_s;k_2) f_{i_3}(t_s;k_3) \bigr\},
	\label{csf_Boltz_mdl2}
\end{eqnarray}
where we use
\begin{eqnarray}
&& E_{q,1} \equiv \omega_{q},\ \ E_{q,2} \equiv -\omega_{q}. 
\end{eqnarray}
These equations describe the thermal evolution of the Bogoliubov parameters
for the complex scalar field with the $\lambda(\phi_a^{\dagger}\phi_a)^2$ interaction. 
It has the consistent statistical structure with the quantum Boltzmann equations 
for the two body charged Bose particle scattering process. 

\section{Canonical Quantization for a Dirac Field}

We quantize the Dirac field in a similar manner to that 
used for the complex scalar field developed in the previous sections. 
Because it is easy to distinguish, 
we use the same expressions for the Dirac field with those for the scalar field, 
for example, the creation and annihilation operators, mass and so on. 

We review the framework of NETFD for the Dirac field, simply. 
Ordinary and tilde operators for the Dirac field 
obey the anticommutation rules, 
\begin{eqnarray}
&& \{ a_{p}^r, a_{k}^{r\dagger} \}
= \{ \tilde{a}_{p}^r, \tilde{a}_{k}^{r\dagger} \}
= (2\pi)^3 \delta^{(3)}(\Vec{p}-\Vec{k}) \delta^{rs}, \label{df_cr:a1} \\
&& \{ b_{p}^r, b_{k}^{s\dagger} \}
= \{ \tilde{b}_{p}^r, \tilde{b}_{k}^{s\dagger} \}
= (2\pi)^3 \delta^{(3)}(\Vec{p}-\Vec{k}) \delta^{rs}, \label{df_cr:a2} 
\end{eqnarray}
where the subscripts $r$ and $s$ represent two-component spinors. 
All other anticommutators vanish. 
We use the tilde conjugation rules given in Eqs.~(\ref{tilde1})-(\ref{tilde4}). 
For the Dirac field it is convenient to use the thermal doublet notation defined by
\begin{eqnarray}
&& a_p^{s,\alpha} = 
\left(
   \begin{array}{c}
	a_p^s \\
	i \tilde{a}_p^{s\dagger}
   \end{array}
\right) ,\quad
\bar{a}_p^{s,\alpha} =
\left(
   \begin{array}{cc}
	a_p^{s\dagger} & -i \tilde{a}_p^s
   \end{array}
\right).
\end{eqnarray}
It should be noted that the definition is different from Eq.~(\ref{ns_tdn1}). 

Time-dependent thermal Bogoliubov transformations 
are introduced for each operator 
with two independent thermal Bogoliubov parameters, $n_{\pm}(t;p)$, 
\begin{eqnarray}
&&\xi_{p}^{s,\alpha} (t) 
= B(n_{+}(t;p))^{\alpha \beta} a_{p}^{s,\beta} (t),~~ 
\bar{\xi}_{p}^{s,\alpha} (t)
= \bar{a}_{p}^{s,\beta} (t) B^{-1}(n_{+}(t;p))^{\beta \alpha}, 
 \label{df_tbt1} \\
&&\eta_{p}^{s,\alpha} (t) 
= B(n_{-}(t;p))^{\alpha \beta} b_{p}^{s,\beta} (t),~~ 
\bar{\eta}_{p}^{s,\alpha} (t)
= \bar{b}_{p}^{s,\beta} (t) B^{-1}(n_{-}(t;p))^{\beta \alpha},
 \label{df_tbt2}
\end{eqnarray}
where the thermal Bogoliubov matrices are given by 
\begin{eqnarray}
&& B(n_{\pm})=
\left(
   \begin{array}{cc}
	1-n_{\pm} & n_{\pm} \\
	-1 & 1 \\
   \end{array}
\right), \label{df_tbm1} \\ 
&& B^{-1}(n_{\pm})=
\left(
   \begin{array}{cc}
	1 & -n_{\pm} \\
	1 & 1 - n_{\pm}  \\
   \end{array}
\right). \label{df_tbm2}
\end{eqnarray}
Here the thermal Bogoliubov parameters are 
assumed to be independent on the spinor index, $s$. 
Thus the thermal Bogoliubov parameters depend only on 
time, $t$, and the absolute value of the momentum, $p$, 
in the homogeneous and out of equilibrium system. 
The transformed operators, $\xi_p^s$ and $\eta_p^s$, 
and their tilde conjugates, $\tilde{\xi}_p^s$ and $\tilde{\eta}_p^s$, 
obey the anticommutation relations, 
\begin{eqnarray}
&& \{ \xi_{p}^r, \xi_{k}^{r\dagger} \}
= \{ \tilde{\xi}_{p}^r, \tilde{\xi}_{k}^{r\dagger} \}
= (2\pi)^3 \delta^{(3)}(\Vec{p}-\Vec{k}) \delta^{rs},
  \label{df_cr:xi1} \\
&& \{ \eta_{p}^r, \eta_{k}^{s\dagger} \}
= \{ \tilde{\eta}_{p}^r, \tilde{\eta}_{k}^{s\dagger} \}
= (2\pi)^3 \delta^{(3)}(\Vec{p}-\Vec{k}) \delta^{rs}, 
\label{df_cr:xi2} 
\end{eqnarray}
with all other anticommutators vanish. 
The thermal vacuum, $|\theta\rangle$, 
is defined to be the state such that
\begin{eqnarray}
&& \xi_p^s |\theta\rangle = \tilde{\xi}_p^s |\theta\rangle
= \eta_p^s |\theta\rangle = \tilde{\eta}_p^s |\theta\rangle = 0, 
\label{df_vs1} \\
&& \langle \theta| \xi_p^{s\dagger} = \langle \theta| \tilde{\xi}_p^{s\dagger} 
=\langle \theta| \eta_p^{s\dagger} = \langle \theta| \tilde{\eta}_p^{s\dagger} =0.
\label{df_vs2}
\end{eqnarray}
The thermal expectation value is obtained by 
the expectation value under the thermal vacuum, $|\theta\rangle$.

Below we quantize the Dirac field in the interaction picture. 
To keep the time independence of the thermal vacuum, 
the Dirac field has to be decomposed as a Hermitian form 
with the transformed operators, 
\begin{eqnarray}
&&  \psi_{\xi}(x) \equiv \int \frac{d^3\Vec{p}}{(2\pi)^3}
 \frac{1}{\sqrt{2\omega_p}} \sum_s 
\left\{ \xi_{p}^{s}(t_x) u^s(p) 
 e^{ i \Vec{p} \cdot \Vec{x}} 
+ \eta_{p}^{s \dagger} (t_x) v^s(p)
 e^{ -i \Vec{p} \cdot \Vec{x}}\right\}, 
 \label{df_psixi1} \\
&&  \bar{\psi}_{\xi} (x) \equiv \int \frac{d^3\Vec{p}}{(2\pi)^3} 
\frac{1}{\sqrt{2\omega_p}} \sum_s
\left\{ \xi_{p}^{s \dagger}(t_x) \bar{u}^s(p)
 e^{ -i \Vec{p} \cdot \Vec{x}} 
+ \eta_{p}^{s} (t_x) \bar{v}^s(p)
 e^{ i \Vec{p} \cdot \Vec{x}}\right\}, 
\label{df_psixi2}
\end{eqnarray}
with 
\begin{eqnarray}
&&\xi_{p}^{s} (t_x)
 = \xi_{p}^{s} e^{-i \omega_{+,p}  t_x}, \quad
\xi_{p}^{s \dagger} (t_x)
 = \xi_{p}^{s\dagger} e^{i \omega_{+,p}  t_x}, 
\label{df_e1} \\
&& \eta_{p}^{s} (t_x)
 = \eta_{p}^{s} e^{-i \omega_{-,p}  t_x}, \quad
\eta_{p}^{s \dagger} (t_x)
 =\eta_{p}^{s \dagger} e^{i \omega_{-,p}  t_x},
\label{df_e2}
\end{eqnarray}
and 
\begin{eqnarray}
&& \omega_{+,p} \equiv \omega_p - \mu, \label{df_e_xi-1}\\
&& \omega_{-,p} \equiv \omega_p + \mu, \label{df_e_xi-2}
\end{eqnarray}
where $\omega_p$ is the relativistic energy eigenvalue. 
The chemical potential, $\mu$, 
is introduced according to the discussion in Appendix A. 
The functions, $u^s(p)$ and $v^s(p)$, are the solutions of the Dirac equation 
with the positive and the negative frequency parts, respectively. 
We write $\bar{u}^{s} \equiv u^{s \dagger} \gamma^0$ 
and $\bar{v}^s \equiv v^{s \dagger} \gamma^0$. 
The eigenfunctions obey 
\begin{eqnarray}
&&(\gamma^0 \omega_p - \Vec{\gamma}\cdot \Vec{p} -m) u^s (p)=0, 
\label{df_eigenf1} \\
&&(-\gamma^0 \omega_p + \Vec{\gamma} \cdot \Vec{p} -m) v^s (p)=0, 
\label{df_eigenf2} \\
&& \bar{u}^s (p) (\gamma^0 \omega_p - \Vec{\gamma}\cdot \Vec{p} -m )=0, 
\label{df_eigenf3} \\
&& \bar{v}^s (p) ( -\gamma^0 \omega_p + \Vec{\gamma}\cdot \Vec{p} -m )=0, 
\label{df_eigenf4} 
\end{eqnarray}
and
\begin{eqnarray}
&& \sum_s u^s(p) \bar{u}^s(p)
=\gamma^0 \omega_p - \Vec{\gamma}\cdot \Vec{p} + m, \label{df_eigenf5} \\
&& \sum_s v^s(p) \bar{v}^s(p)
=\gamma^0 \omega_p - \Vec{\gamma}\cdot \Vec{p} - m. \label{df_eigenf6}
\end{eqnarray}
The tilde conjugate for the Dirac field is given by
\begin{eqnarray}
\tilde{\psi}_{\xi}(x) &=& 
\int \frac{d^3\Vec{p}}{(2\pi)^3} \frac{1}{\sqrt{2\omega_p}} \sum_s 
\left\{ \tilde{\xi}_{p}^{s}(t_x) u^{s \ast} (p) 
 e^{ -i \Vec{p} \cdot \Vec{x}} 
+ \tilde{\eta}_{p}^{s \dagger} (t_x) v^{s \ast}(p)
 e^{ i \Vec{p} \cdot \Vec{x}} \right\},  \label{df_psi_tildexi1} \\
\tilde{\bar{\psi}}_{\xi} (x) &=& 
\int \frac{d^3\Vec{p}}{(2\pi)^3} \frac{1}{\sqrt{2\omega_p}} \sum_s
\left\{ \tilde{\xi}_{p}^{s \dagger}(t_x) \bar{u}^{s \ast} (p)
 e^{ i \Vec{p} \cdot \Vec{x}} 
+ \tilde{\eta}_{p}^{s} (t_x) \bar{v}^{s \ast} (p)
 e^{ -i \Vec{p} \cdot \Vec{x}}\right\}. \label{df_psi_tildexi2}
\end{eqnarray}

The time dependence of the non-tilde and tilde operators is combined 
into the thermal doublet notation, 
\begin{eqnarray}
&&\xi_{p}^{s, \alpha} (t_x)
 = \xi_{p}^{s, \alpha} e^{-i \omega_{+,p}  t_x},~~
\bar{\xi}_{p}^{s, \alpha} (t_x)
 =\bar{\xi}_{p}^{s, \alpha} e^{i \omega_{+,p}  t_x}, 
 \label{df_tdxi1} \\
&& \eta_{p}^{s, \alpha} (t_x)
 = \eta_{p}^{s, \alpha} e^{-i \omega_{-,p}  t_x},~~
\bar{\eta}_{p}^{s, \alpha} (t_x)
 =\bar{\eta}_{p}^{s, \alpha} e^{i \omega_{-,p}  t_x}.
 \label{df_tdeta1}
\end{eqnarray}
The Dirac field is rewritten in the thermal doublet form, 
\begin{eqnarray}
&& \psi_{\xi}^{\alpha}(x) 
 \equiv \int \frac{d^3\Vec{p}}{(2\pi)^3}
 \frac{1}{\sqrt{2\omega_p}} \sum_s 
\left\{ 
\left(
  \begin{array}{c}
	\xi_p^s (t_x) \\
	i \tilde{\xi}_p^{s \dagger} (t_x)
  \end{array}
\right)
u^s(p) e^{ i \Vec{p} \cdot \Vec{x}} 
+ \left(
  \begin{array}{c}
	\eta_p^{s \dagger}(t_x) \\
	i \tilde{\eta}_p^{s} (t_x)
  \end{array}
\right) 
v^s(p) e^{ -i \Vec{p} \cdot \Vec{x}} \right\} \nonumber \\
&& = \int \frac{d^3\Vec{p}}{(2\pi)^3}
 \frac{1}{\sqrt{2\omega_p}} \sum_s 
\left\{ \xi_{p}^{s, \alpha}(t_x) u^s(p) 
 e^{ i \Vec{p} \cdot \Vec{x}} 
+ (\tau_3 \bar{\eta}_{p}^s (t_x)^T)^\alpha v^s(p)
 e^{ -i \Vec{p} \cdot \Vec{x}} \right\}, 
 \label{df_psixi3} \\
&& \bar{\psi}_{\xi}^{\alpha}(x) \equiv \int \frac{d^3\Vec{p}}{(2\pi)^3} 
\frac{1}{\sqrt{2\omega_p}} \sum_s
\Bigl\{ \left(
  \begin{array}{cc}
	\xi_p^{s\dagger}(t_x) & -i \tilde{\xi}_{p}^{s} (t_x)
  \end{array}
\right)
\bar{u}^s(p) e^{ -i \Vec{p} \cdot \Vec{x}} \nonumber \\ 
&& \quad + \left(
  \begin{array}{cc}
	\eta_p^{s}(t_x) & -i \eta_p^{s\dagger} (t_x)
  \end{array}
\right) 
\bar{v}^s(p) e^{ i \Vec{p} \cdot \Vec{x}} \Bigr\} \nonumber \\
&&= \int \frac{d^3\Vec{p}}{(2\pi)^3} 
\frac{1}{\sqrt{2\omega_p}} \sum_s
\left\{ \bar{\xi}_{p}^{s, \alpha}(t_x) \bar{u}^s(p)
 e^{ -i \Vec{p} \cdot \Vec{x}} 
+ (\eta_{p}^{s} (t_x)^{T} \tau_3)^\alpha \bar{v}^s(p)
 e^{ i \Vec{p} \cdot \Vec{x}} \right\}. 
 \label{df_psixi4}
\end{eqnarray}
These fields satisfy the equal-time anticommutation relations, 
\begin{eqnarray}
&& \{ \psi_{\xi}^{\dagger \alpha}(t,\Vec{x}),\psi_{\xi}^\beta (t,\Vec{y})\}
 = \delta^{(3)}(\Vec{x}-\Vec{y}) \delta^{\alpha \beta},
\end{eqnarray}
where we write 
$\psi_{\xi}^{\dagger \alpha} \equiv \bar{\psi}_{\xi}^{\alpha}\gamma^0$. 

Next we decompose the Dirac field 
in terms of the original operators, $a_p^s$ and $b_p^s$. 
The time dependence of the original operators 
is fixed by the time evolution equations for the transformed operators. 
The time derivatives of Eq.~(\ref{df_tdxi1}) give
\begin{eqnarray}
&& \partial_{t_x} \xi_p^{s, \alpha} (t_x)
 = -i \omega_{+,p} \xi_p^{s, \alpha} (t_x), \label{df_eomxi1} \\
&&\partial_{t_x} \bar{\xi}_p^{s, \alpha} (t_x)
 = i  \omega_{+,p} \bar{\xi}^{s, \alpha} (t_x). \label{df_eomxi2} 
\end{eqnarray}
Applying the thermal Bogoliubov transformation (\ref{df_tbt1}) 
with Eqs.~(\ref{df_eomxi1}) and (\ref{df_eomxi2}), 
we obtain the time evolution equations for the original operators, 
$a_p^{s,\alpha}(t_x)$ and $\bar{a}_p^{s,\alpha}(t_x)$, 
\begin{eqnarray}
&&\partial_{t_x}a_{p}^{s,\alpha}(t_x)
= -i \{ \omega_{+,p}  + i \dot{n}_{+}(t_x;p)T_0 \}^{\alpha \beta} a_{p}^{s,\beta}(t_x), 
\label{df_eoma1} \\
&&\partial_{t_x}\bar{a}_{p}^{s,\alpha} (t_x) 
= \bar{a}_{p}^{s, \beta} (t_x)
 i \{ \omega_{+,p}  + i \dot{n}_{+}(t_x;p)T_0 \}^{\beta \alpha}, 
\label{df_eoma2}
\end{eqnarray}
where $T_0^{\alpha \beta}$ is given by Eq.~(\ref{t_0}). 
Eqs.~(\ref{df_eoma1}) and (\ref{df_eoma2}) 
indicate that the energy eigenvalue for the operators, 
$a_p^{s,\alpha}(t_x)$ and $\bar{a}_p^{s,\alpha}(t_x)$, 
depends on the time derivative of the thermal Bogoliubov parameter, $\dot{n}_{+}(t;p)$. 
The solutions for these equations are found to be 
\begin{eqnarray}
a_{p}^{s, \alpha} (t_x) &=& 
 {\rm exp} \left\{ -i\int_{-\infty}^{t_x} dt_s \Omega_{+,p}(t_s) 
 \right\}^{\alpha \beta}
 a_{p}^{s, \beta},  \label{df_a1} \\
\bar{a}_{p}^{s,\alpha} (t_x) &=& \bar{a}_{p}^{s, \beta}\ 
 {\rm exp} \left\{ i \int_{-\infty}^{t_x} dt_s \Omega_{+,p}(t_s)
  \right\}^{\beta \alpha}, 
 \label{df_a2} 
\end{eqnarray}
with
\begin{eqnarray}
&&\Omega_{+,p}^{\alpha \beta} (t_s)
 \equiv \omega_{+,p} \delta^{\alpha \beta}
  + i \dot{n}_{+}(t_s;p) T_0^{\alpha \beta}. \label{df_ea1} 
\end{eqnarray}

The time dependence of the operators, 
$b_p^{s,\alpha}(t_x)$ and $\bar{b}_p^{s,\alpha}(t_x)$, 
is found from Eq.~(\ref{df_tdeta1}). 
The equations of motion for 
$\eta_p^{s,\alpha}(t_x)$ and $\bar{\eta}_p^{s,\alpha}(t_x)$ are given by
\begin{eqnarray}
\partial_{t_x} \eta^{s, \alpha} (t_x)
 &=& -i \omega_{-,p}  \eta^{s, \alpha} (t_x), \label{df_eometa1} \\
\partial_{t_x} \bar{\eta}^{s, \alpha} (t_x)
 &=& i  \omega_{-,p} \bar{\eta}^{s, \alpha} (t_x). \label{df_eometa2}
\end{eqnarray}
Applying the thermal Bogoliubov transformation, 
we find the time evolution equations of the operators, 
$b_p^{s,\alpha}(t_x)$ and $\bar{b}_p^{s,\alpha}(t_x)$, 
\begin{eqnarray}
&&\partial_{t_x}b_{p}^{s, \alpha}(t_x)
=-i \{ \omega_{-,p} + i \dot{n}_{-}(t_x;p)T_0 \}^{\alpha \beta} b_{p}^{s,\beta}(t_x), 
 \label{df_eomb1} \\
&&\partial_{t_x}\bar{b}_{p}^{s,\alpha} (t_x) 
= \bar{b}_{p}^{s,\beta} (t_x)
 i \{ \omega_{-,p} + i \dot{n}_{-}(t_x;p)T_0 \}^{\beta \alpha}. 
 \label{df_eomb2}
\end{eqnarray}
The solutions for these equations are represented as
\begin{eqnarray}
&&b_{p}^{s, \alpha} (t_x) = 
 {\rm exp} \left\{ -i\int_{-\infty}^{t_x} dt_s \Omega_{-,p}(t_s) 
 \right\}^{\alpha \beta} b_{p}^{s,\beta}, 
 \label{df_b1} \\
&&\bar{b}_{p}^{s, \alpha} (t_x) = \bar{b}_{p}^{s, \beta} 
{\rm exp} \left\{ i \int_{-\infty}^{t_x} dt_s \Omega_{-,p}(t_s)
  \right\}^{\beta \alpha}, 
 \label{df_b2}
\end{eqnarray}
where 
\begin{eqnarray}
\Omega_{-,p}^{\alpha \beta} (t_s)
= \omega_{-,p} \delta^{\alpha \beta}
 + i \dot{n}_{-}(t_s;p) T_0^{\alpha \beta}. \label{df_eb1} 
\end{eqnarray}

Hence the Dirac field is decomposed in terms of the original operators, 
\begin{eqnarray}
&&  \psi_{a}^\alpha(x) \equiv \int \frac{d^3\Vec{p}}{(2\pi)^3}
 \frac{1}{\sqrt{2\omega_p}} \sum_s 
\left\{ 
a_{p}^{s, \alpha} (t_x) u^s (p) 
e^{ i \Vec{p} \cdot \Vec{x}} 
+ (\tau_3 \bar{b}_{p}^s (t_x)^T)^\alpha v^s (p)
e^{ -i \Vec{p} \cdot \Vec{x}} \right\},  \nonumber \\ 
 \label{df_psia1} \\
&&  \bar{\psi}_{a}^\alpha (x) \equiv \int \frac{d^3\Vec{p}}{(2\pi)^3} 
\frac{1}{\sqrt{2\omega_p}} \sum_s 
\left\{ \bar{a}_{p}^{s, \alpha} (t_x) \bar{u}^s(p) 
e^{ -i \Vec{p} \cdot \Vec{x}} 
+ ( b_{p}^s(t_x)^{T} \tau_3)^\alpha \bar{v}^s(p) 
e^{ i \Vec{p} \cdot \Vec{x}} \right\}. \nonumber \\ 
 \label{df_psia2}
\end{eqnarray}
It is easy to find that these fields satisfy 
the equal-time anticommutation relations, 
\begin{eqnarray}
\{ \psi_{a}^{\dagger \alpha}(t,\Vec{x}),\psi_{a}^{\beta}(t,\Vec{y})\}
 = \delta^{(3)}(\Vec{x}-\Vec{y}) \delta^{\alpha \beta}, \label{df_anticr1}
\end{eqnarray}
where we define $\psi_{a}^{\dagger \alpha} \equiv \bar{\psi}_{a}^{\alpha} \gamma^0$. 

Time evolution of the Dirac field is generated by the hat-Hamiltonian (\ref{hat-H}). 
The unperturbed hat-Hamiltonian, $\hat{H}_Q$, 
for the Dirac field, $\psi_a^{\alpha}$ and $\bar{\psi}_a^{\alpha}$, 
can be found from the equations of motion for the Dirac field. 
Differentiating Eqs.~(\ref{df_psia1}) and (\ref{df_psia2}) 
with respect to time and using Eqs.~(\ref{df_eigenf1})-(\ref{df_eigenf4}), 
we derive the equations of motion for the Dirac field, 
\begin{eqnarray}
&&i\gamma^0 \partial_{t_x}\psi_a^{\alpha}(x) 
=(-i \Vec{\gamma} \cdot \nabla_x + m 
- \mu \gamma^0)\psi_a^{\alpha}(x) \nonumber \\
&&\quad + i\gamma^0 \dot{n}_{+}(t_x;|\nabla_x|) T_0^{\alpha \beta} \psi_{a,+}^{\beta}(x)
-i \gamma^0 \dot{n}_{-}(t_x;|\nabla_x|) T_0^{\alpha\beta} \psi_{a,-}^{\beta}(x), 
 \label{df_eom_psi_a1} \\
&& -i \partial_{t_x} \bar{\psi}_a^{\alpha}(x) \gamma^0 
= \bar{\psi}_a^{\alpha}(x) 
(i \Vec{\gamma} \cdot \overleftarrow{\nabla}_x + m 
- \mu \gamma^0) \nonumber \\
&&\quad + i \dot{n}_{+}(t_x;|\nabla_x|) 
\bar{\psi}_{a,-}^{\beta}(x) \gamma^0  T_0^{\beta \alpha} 
-i \dot{n}_{-}(t_x;|\nabla_x|) \bar{\psi}_{a,+}^{\beta}(x) \gamma^0  T_0^{\beta \alpha},
 \label{df_eom_psi_a2}
\end{eqnarray}
where $\psi_{a,\pm}^{\alpha}$ and $\bar{\psi}_{a,\pm}^{\alpha}$ describe 
the positive and the negative frequency parts of the Dirac field (\ref{df_psia1}) 
and (\ref{df_psia2}), respectively.

The unperturbed hat-Hamiltonian in an equilibrium system 
for the Dirac field at finite chemical potential is given by
\begin{eqnarray}
\hat{H}_0 = \int d^3\Vec{x} 
 \bar{\psi}_a^{\alpha}(x)
 \left( -i \Vec{\gamma} \cdot \nabla_x +m 
-\mu \gamma^0 \right) \psi_{a}^{\alpha}(x). \label{df_Ha0}
\end{eqnarray}
We introduce the thermal counter term for the Dirac field 
in the following form, 
\begin{eqnarray}
&&\hat{Q} = \int d^3{\Vec{x}} \Biggl[ 
 -i \bar{\psi}_a^{\alpha}(x) 
\frac{ \dot{\bar{n}}_{d}(t_x;|\nabla_x|) }{\hat{\omega}_{\nabla_x}} 
(-i \Vec{\gamma} \cdot \nabla_x +m)
T_0^{\alpha \beta}  \psi_{a}^{\beta} (x) \nonumber \\
&&\quad - \frac{i}{2} \bar{\psi}_a^{\alpha}(x)
 \dot{\mu}_{d}(t_x;|\nabla_x|) 
 \gamma^0 T_0^{\alpha \beta}  \psi_{a}^{\beta} (x) \Biggr].	\label{df_Q1}
\end{eqnarray}
where $\dot{\bar{n}}_{d}(t_x;|\nabla_x|)$ and 
$\dot{\mu}_{d}(t_x;|\nabla_x|)$ are 
the coefficients consist with the Bogoliubov parameters, 
$n_{+}(t_x;|\nabla_x|)$ and $n_{-}(t_x;|\nabla_x|)$.  
The unperturbed hat-Hamiltonian for 
the non-equilibrium Dirac field, $\hat{H}_Q$, is defined by 
\begin{eqnarray}
&& \hat{H}_Q = \hat{H}_0 - \hat{Q} \nonumber \\
&&= \int d^3\Vec{x} \Biggl[ 
 \bar{\psi}_a^{\alpha}(x)
 \left( 1 + i \frac{ \dot{\bar{n}}_{d}(t_x;|\nabla_x|) }{\hat{\omega}_{\nabla_x}} T_0
 \right)^{\alpha \beta} 
(-i \Vec{\gamma} \cdot \nabla_x +m) \psi_{a}^{\beta}(x)  \nonumber \\
&&\quad + \bar{\psi}_a^{\alpha}(x) \gamma^0 
\left( -\mu 
+ \frac{i}{2} \dot{\mu}_{d} (t_x;|\nabla_x| ) T_0 \right)^{\alpha \beta} 
\psi_{a}^{\beta} (x) \Biggr].	\label{df_Ha1}
\end{eqnarray}
The Heisenberg equations should reproduce 
the equations of motion (\ref{df_eom_psi_a1}) and (\ref{df_eom_psi_a2}). 
This condition fixes two types of the coefficients, $\dot{\bar{n}}_{d}(t_x;|\nabla_x|)$ and 
$\dot{\mu}_{d}(t_x;|\nabla_x|)$, in the thermal counter term as 
\begin{eqnarray}
&&\dot{\bar{n}}_{d}(t_x;|\nabla_x|)=\frac{ \dot{n}_{+}(t_x;|\nabla_x|) 
+ \dot{n}_{-}(t_x;|\nabla_x|) }{2}, \\
&&\dot{\mu}_{d}(t_x;|\nabla_x|)=\dot{n}_{+}(t_x;|\nabla_x|)-\dot{n}_{-}(t_x;|\nabla_x|).
\end{eqnarray}
These equations have the same form with those 
for the charged scalar field (\ref{csf_nc_1}) and (\ref{csf_nu_1}).

\section{Self-Consistency Condition for the Dirac Field}

In this section we introduce a self-consistency renormalization condition 
for the Dirac field with an interaction, $\hat{H}_{int}$. 
As is shown in the previous section, 
the unperturbed hat-Hamiltonian (\ref{df_Ha1}) 
contains the thermal counter term in NETFD. 
In a similar manner with \S 3, 
the interaction hat-Hamiltonian, $\hat{H}_I$, 
is given by $\hat{H}_{int} + \hat{Q}$. 
The perturbative calculation should be developed under the unperturbed 
hat-Hamiltonian, $\hat{H}_Q$, with the interaction hat-Hamiltonian, $\hat{H}_I$. 

\subsection{The Bogoliubov Parameter and the Number Density}
Following the procedure in Ref.~\refcite{CQNS1}, 
we show that the Bogoliubov parameters for the Dirac field 
correspond to the number densities 
under the self-consistency renormalization conditions. 
In TFD the observable particle number densities are 
given by the vacuum expectation values of the number operators, 
\begin{eqnarray}
&&(2\pi)^3 \delta^{sr} \delta^{(3)}(\Vec{p}-\Vec{k}) n_{H,+}(t;p) 
= \langle \theta| a_{full,p}^{s \dagger}(t) a_{full,k}^r(t) |\theta \rangle, \label{df_app2_1} \\
&&(2\pi)^3 \delta^{sr} \delta^{(3)}(\Vec{p}-\Vec{k}) n_{H,-}(t;p) 
= \langle \theta| b_{full,p}^{s \dagger}(t) b_{full,k}^r(t) |\theta \rangle, \label{df_app2_2}
\end{eqnarray}
where the lower index "$full$" indicates the perturbed operators.

We define the Bogoliubov transformation for the perturbed operators, 
\begin{eqnarray}
&&\!\!\!\!\!\! \xi_{full,p}^{s,\alpha} (t) 
= B(n_{+}(t;p))^{\alpha \beta} a_{full,p}^{s,\beta} (t),~ 
\bar{\xi}_{full,p}^{s,\alpha} (t)
= \bar{a}_{full,p}^{s,\beta} (t) B^{-1}(n_{+}(t;p))^{\beta \alpha}, 
 \label{df_app2_3} \\
&&\!\!\!\!\!\! \eta_{full,p}^{s,\alpha} (t) 
= B(n_{-}(t;p))^{\alpha \beta} b_{full,p}^{s,\beta} (t),~ 
\bar{\eta}_{full,p}^{s,\alpha} (t)
= \bar{b}_{full,p}^{s,\beta} (t) B^{-1}(n_{-}(t;p))^{\beta \alpha},
 \label{df_app2_4}
\end{eqnarray}
with the Bogoliubov parameters, $n_{\pm}(t;p)$. 
From the condition, $\langle \theta|\hat{H}_I=0$, it is found that 
the transformed operators, $\xi_{full}^s(t)$ and $\xi_{full}^{s\dagger}(t)$, satisfy 
\begin{eqnarray}
\langle \theta | \xi_{full,p}^{s\dagger}(t) = \langle \theta | \tilde{\xi}_{full,p}^{s\dagger}(t)=0,
\qquad 
\xi_{full,p}^{s}(t)|\theta\rangle \ne 0,
\qquad  
\tilde{\xi}_{full,p}^{s}(t) |\theta \rangle \ne 0. 
\label{df_app2_5}
\end{eqnarray}
Substituting Eqs.~(\ref{df_app2_3}) and (\ref{df_app2_5}) into 
Eq.~(\ref{df_app2_1}), we obtain
\begin{eqnarray}
&&\langle \theta| a_{full,p}^{s\dagger}(t) a_{full,k}^{r}(t) |\theta\rangle \nonumber \\
&&=\langle \theta| \bigl\{ i \tilde{\xi}_{full,p}^s(t) 
+ (1-n_{+}(t;p)) \xi_{full,p}^{s\dagger}(t) \bigr\} 
\bigl\{ \xi_{full,k}^r(t) -i n_{+}(t;k) \tilde{\xi}_{full,k}^{r\dagger}(t) \bigr\}|\theta\rangle 
\nonumber \\ 
&&=i \langle \theta|\tilde{\xi}_{full,p}^s(t) \xi_{full,k}^r(t)|\theta \rangle 
+ n_{+}(t;k) \langle \theta|\tilde{\xi}_{full,p}^s(t) \tilde{\xi}_{full,k}^{r\dagger}(t)|\theta\rangle.
\label{df_app2_6}
\end{eqnarray}
This equation implies
\begin{eqnarray}
(2\pi)^3 \delta^{sr} \delta^{(3)}(\Vec{p}-\Vec{k}) \bigl\{ n_{H,+}(t;p) - n_{+}(t;p) \bigr\} 
= i \langle \theta| \tilde{\xi}_{full,p}^{s}(t) \xi_{full,k}^r(t) |\theta \rangle. 
\label{df_app2_7}
\end{eqnarray}
Thus the Bogoliubov parameter, $n_{+}(t;p)$, 
coincides with the number density, $n_{H,+}(t;p)$, 
under the condition, 
$\langle \theta| \tilde{\xi}_{full,p}^s(t) \xi_{full,k}^r(t)|\theta\rangle=0$.

We also derive the condition for the antiparticle number density (\ref{df_app2_2}). 
The operators, $\eta_{full,p}^{s}(t)$ and $\eta_{full,p}^{s,\dagger}(t)$, satisfy 
\begin{eqnarray}
\langle \theta | \eta_{full,p}^{s\dagger}(t) 
= \langle \theta | \tilde{\eta}_{full,p}^{s\dagger}(t)=0,
\qquad 
\eta_{full,p}^{s}(t)|\theta\rangle \ne 0,
\qquad  
\tilde{\eta}_{full,p}^{s}(t) |\theta \rangle \ne 0. 
\label{df_app2_8}
\end{eqnarray}
In a similar manner to Eq.~(\ref{df_app2_6}) we obtain
\begin{eqnarray}
(2\pi)^3 \delta^{sr} \delta^{(3)}(\Vec{p}-\Vec{k}) \bigl\{ n_{H,-}(t;p) - n_{-}(t;p) \bigr\} 
= i \langle \theta| \tilde{\eta}_{full,p}^{s}(t) \eta_{full,k}^r(t) |\theta \rangle. 
\label{df_app2_9}
\end{eqnarray}
Therefore the Bogoliubov parameter, $n_-(t;p)$, 
also coincides with the anti-particle number density, $n_{H,-}(t;p)$, 
under the condition, 
$\langle \theta|\tilde{\eta}_{full,p}^s (t) \eta_{full,k}^r (t)|\theta\rangle=0$. 
Thus we impose the self-consistency renormalization conditions for the Dirac field,
\begin{eqnarray}
&&\langle \theta| \tilde{\xi}_{full,p}^s(t) \xi_{full,k}^r(t)|\theta\rangle=0, \\
&&\langle \theta|\tilde{\eta}_{full,p}^s (t) \eta_{full,k}^r (t)|\theta\rangle=0.
\end{eqnarray}

\subsection{Time Evolution for the Bogoliubov Parameters}

The time evolution is derived from 
the self-consistency renormalization conditions. 
Here we begin with the perturbative expansion for 
the full Dirac propagator in NETFD. 
In the interaction picture the full Dirac propagator is written as
\begin{eqnarray}
S_{H}^{\alpha\beta}(t_x,t_y,\Vec{x}-\Vec{y})
= \langle \theta| T[\psi_a^{\alpha} (x) \bar{\psi}_a^{\beta} (y)
u(\infty,-\infty)]|\theta\rangle,
\label{df_pro:1}
\end{eqnarray}
where the operator $u(t,t^\prime)$ has the same form with Eq.~(\ref{u1}). 
Applying the thermal Bogoliubov transformation, 
the full propagator (\ref{df_pro:1}) is rewritten as
\begin{eqnarray}
&&S_H^{\alpha \beta}(t_{x}, t_{y},\Vec{x}-\Vec{y})  \nonumber \\
&& = B^{-1}(n_{+}(t_x;|\nabla_x|) ) 
 \left(
   \begin{array}{cc}
	s_1^{11}(x,y) & s_1^{12}(x,y) \\
	0 & s_1^{22}(x,y)
   \end{array}
\right) B(n_{+}(t_y;|\overleftarrow{\nabla}_y|) ) \nonumber \\
&& + B^{-1}(n_{+}(t_x;|\nabla_x|) ) 
 \left(
   \begin{array}{cc}
		s_2^{11}(x,y) & s_2^{12}(x,y) \\
		s_2^{21}(x,y) & 0
   \end{array}
\right) 
 B^{-1}(n_{-}(t_y;|\overleftarrow{\nabla}_y|) )^T \tau_3  \nonumber \\
&& + \tau_3 B(n_{-}(t_x;|\nabla_x|) )^T  
 \left(
   \begin{array}{cc}
	0 & s_3^{12}(x, y) \\
	s_3^{21}(x,y) & s_3^{22}(x,y)
   \end{array}
\right) 
 B(n_{+}(t_y;|\overleftarrow{\nabla}_y|) ) \nonumber \\
&& +  \tau_3 B(n_{-}(t_x;|\nabla_x|) )^T 
 \left(
   \begin{array}{cc}
	s_4^{11} (x,y) & 0 \\
	s_4^{21} (x,y) & s_4^{22}(x, y)
   \end{array}
\right) 
   B^{-1}(n_{-}(t_y;|\overleftarrow{\nabla}_y|) )^T \tau_3, 
\label{DF_sp:1-2}
\end{eqnarray}
with
\begin{eqnarray}
&& s_1^{\gamma_1 \gamma_2}(x,y) 
=  \theta(t_x-t_y) \langle \theta| \psi_{\xi,+}^{\gamma_1}(x)
 u(t_x,t_y) \bar{\psi}_{\xi,-}^{\gamma_2}(y)u(t_y,-\infty)|\theta\rangle  \nonumber \\
&&\quad - \theta (t_y-t_x) \langle \theta| \bar{\psi}_{\xi,-}^{\gamma_2}(y)
 u(t_y,t_x) \psi_{\xi,+}^{\gamma_1}(x)u(t_x,-\infty)|\theta\rangle, 
\label{s-1} \\
&& s_2^{\gamma_1 \gamma_2}(x,y)
=\theta (t_x-t_y) \langle \theta| \psi_{\xi,+}^{\gamma_1}(x)
  u(t_x,t_y) \{\bar{\psi}_{\xi,+}(y)\tau_3\}^{\gamma_2}
u(t_y,-\infty)|\theta\rangle	\nonumber \\
&&\quad - \theta (t_y-t_x) \langle \theta| \{ \bar{\psi}_{\xi,+}(y) \tau_3 \}^{\gamma_2} 
 u(t_y,t_x) \psi_{\xi,+}^{\gamma_1}(x) u(t_x,-\infty) |\theta \rangle, 
\label{s-2} \\
&& s_3^{\gamma_1 \gamma_2}(x,y)
=\theta (t_x-t_y) \langle \theta| \{ \tau_3 \psi_{\xi,-}(x) \}^{\gamma_1} 
 u(t_x,t_y) \bar{\psi}_{\xi,-}^{\gamma_2}(y) u(t_y,-\infty) |\theta \rangle \nonumber \\
&&\quad - \theta (t_y-t_x)\langle \theta| \bar{\psi}_{\xi,-}^{\gamma_2}(y)
  u(t_y,t_x) \{ \tau_3 \psi_{\xi,-}(x) \}^{\gamma_1}u(t_x,-\infty)|\theta\rangle, 
\label{s-3} \\
&& s_4^{\gamma_1 \gamma_2}(x,y)
=\theta (t_x-t_y) 
\langle \theta| \{ \tau_3 \psi_{\xi,-}(x)\}^{\gamma_1}
 u(t_x,t_y) \{ \bar{\psi}_{\xi,+}(y)\tau_3\}^{\gamma_2}
 u(t_y,-\infty) |\theta\rangle 
\nonumber \\
&&\quad - \theta (t_y-t_x)
\langle \theta| \{ \bar{\psi}_{\xi,+}(y) \tau_3 \}^{\gamma_2}
 u(t_y,t_x) \{ \tau_3 \psi_{\xi,-}(x) \}^{\gamma_1}
 u(t_x,-\infty) |\theta\rangle, 
\label{s-4}
\end{eqnarray}
where $\psi_{\xi,\pm}^{\alpha}$ and $\bar{\psi}_{\xi,\pm}^{\alpha}$ show 
the positive and the negative frequency parts of the Dirac field, 
(\ref{df_psixi3}) and (\ref{df_psixi4}), respectively. 

We expand the time evolution operator $u(t,t^\prime)$ in terms of 
the interaction hat-Hamiltonian (\ref{cs_HI1}) 
and evaluate the propagator in the leading order of 
the thermal counter term, $\hat{Q}$. 
After the Bogoliubov transformation 
the thermal counter term, $\hat{Q}$, is 
written in terms of the transformed operators, $\xi_p^s$ and $\eta_p^s$, 
\begin{eqnarray}
\hat{Q}= - \int \frac{d^3\Vec{p}}{(2\pi)^3} \sum_{s} 
\left\{ \dot{n}_{+}(t_x;p)
\xi_{p}^{s\dagger} \tilde{\xi}_{p}^{s\dagger} 
- \dot{n}_{-}(t_x;p) \tilde{\eta}_{p}^{s\dagger} \eta_{p}^{s\dagger}  \right\}.
\label{df_Q_xi}
\end{eqnarray}
Substituting Eq.~(\ref{df_Q_xi}) with Eq.~(\ref{u1}) into Eq.~(\ref{DF_sp:1-2}) 
and dropping the higher order corrections, we obtain
\begin{eqnarray}
&& \langle \theta| T[\psi_a^\alpha (x) \bar{\psi}_a^\beta (y)
u(\infty,-\infty)]|\theta\rangle 
-\langle \theta| T[\psi_a^\alpha (x) \bar{\psi}_a^\beta (y)]|\theta\rangle 
 \nonumber \\
&&= \int \frac{d^3 \Vec{p}}{(2\pi)^3} 
\frac{\omega_p \gamma^0 -\Vec{p} \cdot \Vec{\gamma} +m }{2 \omega_p}
 e^{-i\omega_{+,p} (t_x-t_y)} e^{i \Vec{p}\cdot (\Vec{x}-\Vec{y})} \nonumber \\
&&~ \times \Biggl( \theta (t_x-t_y) \int_{-\infty}^{t_y} dt_s 
  (-1) \dot{n}_{+} (t_s;p)  
 + \theta(t_y-t_x) \int_{-\infty}^{t_x} dt_s 
 (-1)\dot{n}_{+} (t_s;p) 
 \Biggr) \nonumber \\
&&~ \times B^{-1}(n_{+}(t_x;p)) 
\left(
   \begin{array}{cc}
	0 & 1 \\
	0 & 0 \\
   \end{array}
\right) B(n_{+}(t_y;p)) \nonumber \\
&& + \int \frac{d^3 \Vec{p}}{(2\pi)^3} 
\frac{\omega_p \gamma^0 + \Vec{p} \cdot \Vec{\gamma} -m}{2\omega_p} 
e^{ i \omega_{-,p} (t_x-t_y)} 
e^{-i \Vec{p}\cdot (\Vec{x}-\Vec{y})} \nonumber \\
&&~ \times \Biggl( \theta (t_x-t_y)\int_{-\infty}^{t_y} dt_s 
  \dot{n}_{-}(t_s;p)  
 + \theta (t_y-t_x) \int_{-\infty}^{t_x} dt_s 
  \dot{n}_{-}(t_s;p) 
 \Biggr) \nonumber \\
&& ~ \times  \tau_3 B(n_{-}(t_x;p))^T  
\left(
   \begin{array}{cc}
	0 & 0 \\
	1 & 0 \\
   \end{array}
\right) 
B^{-1}(n_{-}(t_y;p))^T \tau_3.
\label{df_sp:1}
\end{eqnarray}
A finite correction appears for $s_1^{12}$ and $s_4^{21}$ in Eq.~(\ref{DF_sp:1-2}). 

Next we consider the radiative correction from the interaction, $\hat{H}_{int}$. 
At the leading order it is represented as
\begin{eqnarray}
&& \int d^4 z_1 d^4 z_2 \ S_0^{\alpha \gamma_1}(t_x,t_{z_1},\Vec{x}-\Vec{z}_1)
(-i)\Sigma^{\gamma_1 \gamma_2}(t_{z_1},t_{z_2},\Vec{z}_1-\Vec{z}_2)
S_0^{\gamma_2\beta}(t_{z_2},t_{y},\Vec{z}_2-\Vec{y}) \nonumber \\
&&=  \int d^4 z_1 d^4 z_2 
\bigl[ B^{-1}(n_{+}(t_x;|\nabla_x|) )^{\alpha \gamma_1}
\delta \Sigma_{F,1}^{\gamma_1 \gamma_2}(x,z_1,z_2,y) 
B(n_{+}(t_y;|\overleftarrow{\nabla}_y|) )^{\gamma_2 \beta} \nonumber \\
&& + B^{-1}(n_{+}(t_x;|\nabla_x|) )^{\alpha \gamma_1} 
\delta \Sigma_{F,2}^{\gamma_1 \gamma_2}(x,z_1,z_2,y) 
\{ B^{-1}(n_{-}(t_y;|\overleftarrow{\nabla}_y|) )^T \tau_3\}^{\gamma_2 \beta} 
\nonumber \\
&& + \{ \tau_3 B (n_{-}(t_x;|\nabla_x|) )^T \}^{\alpha \gamma_1} 
\delta \Sigma_{F,3}^{\gamma_1 \gamma_2}(x,z_1,z_2,y) 
B(n_{+}(t_y;|\overleftarrow{\nabla}_y|) )^{\gamma_2 \beta} \nonumber \\
&& + \{ \tau_3 B(n_{-}(t_x;|\nabla_x|) )^T \}^{\alpha \gamma_1}
\delta \Sigma_{F,4}^{\gamma_1 \gamma_2}(x,z_1,z_2,y)
\{ B^{-1}(n_{-}(t_y;|\overleftarrow{\nabla}_y|) )^T \tau_3 \}^{\gamma_2 \beta} \bigr], 
\label{df_sp:2}
\end{eqnarray}
with
\begin{eqnarray}
&& \delta \Sigma_{F,1}^{\gamma_1 \gamma_2}(x,z_1,z_2,y)  \\
&& =\left(
   \begin{array}{c}
	S_{0,R}^{11}(x - z_1) (-i) \Sigma_R (t_{z_1}, t_{z_2},\Vec{z}_1-\Vec{z}_2)
 S_{0,R}^{11}(t_{z_2} - t_y) \quad
 \delta \Sigma_{F,1}^{12} ( x,z_1, z_2, y)  \\
  \qquad \qquad 0 \qquad \qquad 
 S_{0,R}^{22}(x - z_1) (-i) \Sigma_A (t_{z_1}, t_{z_2},\Vec{z}_1-\Vec{z}_2) 
 S_{0,R}^{22} ( z_2 - y)  \\
   \end{array}
\right),  \nonumber \\
&& \delta \Sigma_{F,2}^{\gamma_1 \gamma_2}(x,z_1,z_2,y)  \\
&& =\left(
   \begin{array}{c}
\delta \Sigma_{F,2}^{11} ( x, z_1, z_2, y) 
\quad S_{0,R}^{11}(x -  z_1) i \Sigma_R (t_{z_1}, t_{z_2},\Vec{z}_1-\Vec{z}_2) 
 S_{0,A}^{22}( z_2 - y) \\
 S_{0,R}^{22}( x - z_1) i \Sigma_A (t_{z_1}, t_{z_2},\Vec{z}_1-\Vec{z}_2) 
 S_{0,A}^{11} ( z_2 - y) 
\qquad \qquad 0 \qquad \qquad \\
   \end{array}
\right), \nonumber \\
&& \delta \Sigma_{F,3}^{\gamma_1 \gamma_2}(x,z_1,z_2,y) \\
&& =\left(
   \begin{array}{c}
\qquad \qquad 0 \qquad \qquad 
 S_{0,A}^{11}( x - z_1) i \Sigma_A(t_{z_1}, t_{z_2},\Vec{z}_1-\Vec{z}_2) 
 S_{0,R}^{22} ( z_2 - y)  \\
 S_{0,A}^{22}(x - z_1) i \Sigma_R (t_{z_1}, t_{z_2},\Vec{z}_1-\Vec{z}_2)
 S_{0,R}^{11}( z_2 - y) \quad \delta \Sigma_{F,3}^{22}( x, z_1, z_2, y)    \\
   \end{array}
\right), \nonumber \\
&& \delta \Sigma_{F,4}^{\gamma_1 \gamma_2}(x,z_1,z_2,y) \\
&& =\left(
   \begin{array}{c}
	S_{0,A}^{11}( x -  z_1) (-i) \Sigma_A (t_{z_1}, t_{z_2},\Vec{z}_1-\Vec{z}_2)
S_{0,A}^{11}( z_2 - y) 
\qquad \qquad 0 \qquad \qquad   \\
  \delta \Sigma_{F,4}^{21}( x, z_1, z_2, y) 
\quad S_{0,A}^{22}(x - z_1) (-i) \Sigma_A (t_{z_1}, t_{z_2},\Vec{z}_1-\Vec{z}_2)
 S_{0,A}^{22}( z_2 - y)  \\
   \end{array}
\right), \nonumber
\end{eqnarray}
where $S_0^{\alpha \beta}$ is the free propagator given by (\ref{df_app1_3}) and
\begin{eqnarray}
&& \delta \Sigma_{F,1}^{12}(x,z_1,z_2,y) \nonumber \\
&& = S_{0,R}^{11}( x - z_1)
 \bigl\{ -i \Sigma^{12} (t_{z_1}, t_{z_2},\Vec{z}_1-\Vec{z}_2)  
 - (-i) \Sigma_R (t_{z_1}, t_{z_2},\Vec{z}_1-\Vec{z}_2) 
n_{+}(t_{z_2};|\overleftarrow{\nabla}_{z_2}|) \nonumber \\ 
&&\quad + n_{+}(t_{z_1};|\nabla_{z_1}|) (-i)\Sigma_A (t_{z_1}, t_{z_2},\Vec{z}_1-\Vec{z}_2) 
\bigr\}S_{0,R}^{22}( z_2 - y), 
\label{df_SEND1} \\
&& \delta \Sigma_{F,2}^{11}( x, z_1, z_2, y) \nonumber \\
&& = S_{0,R}^{11}( x - z_1)
 \bigl\{ -i \Sigma^{11} (t_{z_1}, t_{z_2},\Vec{z}_1-\Vec{z}_2)  
 + i \Sigma_R (t_{z_1}, t_{z_2},\Vec{z}_1-\Vec{z}_2) 
 n_{-} (t_{z_2};|\overleftarrow{\nabla}_{z_2}|) \nonumber \\
&&\quad + n_{+}(t_{z_1};|\nabla_{z_1}|) i \Sigma_A (t_{z_1}, t_{z_2},\Vec{z}_1-\Vec{z}_2) \bigr\}
 S_{0,A}^{11}( z_2 - y),  
\label{df_SEND2}  \\
&& \delta \Sigma_{F,3}^{22}( x, z_1, z_2, y) \nonumber \\
&& = S_{0,A}^{22}( x - z_1)
 \bigl\{ i\Sigma^{22} (t_{z_1}, t_{z_2},\Vec{z}_1-\Vec{z}_2) 
  + (-i) \Sigma_R (t_{z_1}, t_{z_2},\Vec{z}_1-\Vec{z}_2) 
 n_{+} (t_{z_2};|\overleftarrow{\nabla}_{z_2}|) \nonumber \\
&&\quad + n_{-} (t_{z_1};|\nabla_{z_1}|) 
(-i) \Sigma_A (t_{z_1}, t_{z_2},\Vec{z}_1-\Vec{z}_2) \bigr\} S_{0,R}^{22}( z_2 - y), 
\label{df_SEND3} \\
&& \delta \Sigma_{F,4}^{21} (x, z_1, z_2, y) \nonumber \\
&&= S_{0,A}^{22} ( x - z_1)
 \bigl\{ i\Sigma^{21} (t_{z_1}, t_{z_2},\Vec{z}_1-\Vec{z}_2)  
  + (-i) \Sigma_R (t_{z_1}, t_{z_2},\Vec{z}_1-\Vec{z}_2) 
 n_{-} (t_{z_2};|\overleftarrow{\nabla}_{z_2}|) \nonumber \\
&&\quad 
- n_{-} (t_{z_1};|\nabla_{z_1}|) (-i) \Sigma_A (t_{z_1}, t_{z_2},\Vec{z}_1-\Vec{z}_2) \bigr\}
 S_{0,A}^{11}( z_2 - y).  \label{df_SEND4}
\end{eqnarray}
The retarded and the advanced parts, $\Sigma_R$ and $\Sigma_A$, are defined by 
\begin{eqnarray}
\Sigma_R \equiv \Sigma^{11}+\Sigma^{12} = \Sigma^{21}+\Sigma^{22},
\quad
\Sigma_A \equiv  \Sigma^{11}-\Sigma^{21} = \Sigma^{22}-\Sigma^{12}.
\label{df_SelfERA}
\end{eqnarray}

The self-consistency renormalization conditions, 
$\langle \theta|\tilde{\xi}_{full,p}^s(t_x) \xi_{full,k}^r(t_x)|\theta\rangle=0$ 
and $\langle \theta|\tilde{\eta}_{full,p}^s (t_x) \eta_{full,k}^r (t_x)|\theta\rangle=0$, 
can be satisfied by fixing the terms, 
$s_1^{12}$ and $s_4^{21}$ in Eq.~(\ref{DF_sp:1-2}) at the equal time limit.\footnote{
Substituting the fields (\ref{df_psixi3})-(\ref{df_psixi4}) and taking the equal time limit, 
we obtain
\begin{eqnarray*}
\lim_{t_x\rightarrow t_y}
s_1^{12}(x,y)&=& -i \int \frac{d^3 \Vec{p}}{(2\pi)^3} \frac{d^3 \Vec{k}}{(2\pi)^3}
\frac{1}{\sqrt[]{2\omega_p}} \frac{1}{\sqrt[]{2\omega_k}} 
\sum_{s} \sum_{r} u^s(p) \bar{u}^r(k)  
e^{ i \Vec{p} \cdot \Vec{x}} e^{ -i \Vec{k} \cdot \Vec{y}} \nonumber \\ 
&& \times \langle \theta|  T[\xi_p^s(t_x) \tilde{\xi}_k^r(t_x)u(\infty, -\infty)] |\theta\rangle,\\
\lim_{t_x\rightarrow t_y}
s_4^{21}(x,y)
&=& -i \int \frac{d^3 \Vec{p}}{(2\pi)^3} \frac{d^3 \Vec{k}}{(2\pi)^3}
\frac{1}{\sqrt[]{2\omega_p}} \frac{1}{\sqrt[]{2\omega_k}} 
\sum_{s} \sum_{r} v^s(p) \bar{v}^r(k)
e^{ -i \Vec{p} \cdot \Vec{x}} e^{ i \Vec{k} \cdot \Vec{y}} \nonumber \\
&& \times \langle \theta| T[ \tilde{\eta}_p^s (t_x) \eta_k^r (t_x)u(\infty, -\infty)]
 |\theta\rangle.
\end{eqnarray*}
}
From Eqs.~(\ref{df_sp:1}) and (\ref{df_sp:2}) we can find 
the self-consistency renormalization conditions for the Dirac field, 
\begin{eqnarray}
&&\int_{-\infty}^{t_x} dt_s \int \frac{d^3 \Vec{p}}{(2\pi)^3} 
\frac{\omega_p \gamma^0 -\Vec{p} \cdot \Vec{\gamma} +m }{2 \omega_p} 
e^{i\Vec{p}\cdot (\Vec{x}-\Vec{y})} 
(-1)\dot{n}_{+}(t_s;p) \nonumber \\
&&\quad +\lim_{t_x\rightarrow t_y} \int d^4 z_1d^4 z_2
 \delta \Sigma_{F,1}^{12} (x, z_1, z_2, y)=0,  \label{df_Self-C1} \\
&&\int_{-\infty}^{t_x} dt_s \int \frac{d^3 \Vec{p}}{(2\pi)^3}
\frac{\omega_p \gamma^0 + \Vec{p} \cdot \Vec{\gamma} -m}{2\omega_p} 
e^{-i\Vec{p}\cdot (\Vec{x}-\Vec{y})} 
\dot{n}_{-}(t_s;p) \nonumber \\
&&\quad +\lim_{t_x\rightarrow t_y} \int d^4 z_1 d^4 z_2
  \delta \Sigma_{F,4}^{21} (x, z_1, z_2, y)=0. \label{df_Self-C2}
\end{eqnarray}
The t-representation is more convenient in practical calculations. 
Performing the spatial Fourier transformation 
and acting the time differential operator, $\partial_{t_x}$, on 
Eqs.~(\ref{df_Self-C1}) and (\ref{df_Self-C2}), we obtain 
\begin{eqnarray}
&& \frac{\omega_p \gamma^0 -\Vec{p} \cdot \Vec{\gamma} +m }{2 \omega_p} 
\dot{n}_{+}(t_x;p) 
= \partial_{t_x} \lim_{t_x\rightarrow t_y}  \int dt_{z_1}dt_{z_2}
 \delta \Sigma_{F,1}^{12} (t_x,t_{z_1},t_{z_2},t_y;\Vec{p})=0, \nonumber \\
\label{df_Boltz_eq1} \\
&& \frac{\omega_p \gamma^0 + \Vec{p} \cdot \Vec{\gamma} -m}{2\omega_p} 
\dot{n}_{-}(t_x;p) 
= -\partial_{t_x} \lim_{t_x\rightarrow t_y} \int dt_{z_1}dt_{z_2}
  \delta \Sigma_{F,4}^{21} (t_x,t_{z_1},t_{z_2},t_y;\Vec{p})=0, \nonumber \\
\label{df_Boltz_eq2}
\end{eqnarray}
with
\begin{eqnarray}
&& \delta \Sigma_{F,1}^{12} (t_x,t_{z_1},t_{z_2},t_y;\Vec{p}) \nonumber \\
&& = S_{0,R}^{11}( t_x - t_{z_1};\Vec{p})
 \bigl\{ -i \Sigma^{12} (t_{z_1}, t_{z_2};\Vec{p})  
 - n_{+}(t_{z_2};p) (-i) \Sigma_R (t_{z_1}, t_{z_2};\Vec{p}) \nonumber \\ 
&&\quad + n_{+}(t_{z_1};p) (-i)\Sigma_A (t_{z_1}, t_{z_2};\Vec{p}) 
\bigr\}S_{0,R}^{22}( t_{z_2} - t_y;\Vec{p}), 
\label{df_Boltz_eq3} 
\end{eqnarray}
\begin{eqnarray}
&& \delta \Sigma_{F,4}^{21} (t_x, t_{z_1}, t_{z_2}, t_y;\Vec{p}) \nonumber \\
&&= S_{0,A}^{22} ( t_x - t_{z_1};\Vec{p})
 \bigl\{ i\Sigma^{21} (t_{z_1}, t_{z_2};\Vec{p})  
  + n_{-} (t_{z_2};p) (-i) \Sigma_R (t_{z_1}, t_{z_2};\Vec{p}) \nonumber \\
&&\quad - n_{-} (t_{z_1};p) 
(-i) \Sigma_A (t_{z_1}, t_{z_2};\Vec{p}) \bigr\}
 S_{0,A}^{11}( t_{z_2} - t_y;\Vec{p}). 
\label{df_Boltz_eq4}
\end{eqnarray}
As is shown in \S 6.1, the Bogoliubov parameters, $n_{\pm}(t_x;p)$, 
under the self-consistency renormalization conditions 
(\ref{df_Boltz_eq1}) and (\ref{df_Boltz_eq2}) 
correspond to the particle and antiparticle number densities. 
Thus we conclude that 
the time evolution for the particle and anti-particle number densities 
is described by Eqs.~(\ref{df_Boltz_eq1}) and (\ref{df_Boltz_eq2}). 

\section{Boltzmann Equation for the Scalar and the Dirac Fields}

Here we consider a Yukawa type interaction between 
neutral scalar and Dirac fields. 
To show the validity of the time evolution equations 
(\ref{df_Boltz_eq1}) and (\ref{df_Boltz_eq2}). 
The time evolution of the thermal Bogoliubov parameters 
is studied for the scalar and the Dirac fields. 
We start from the Hamiltonian, 
\begin{eqnarray}
H &=& \int d^3 \Vec{x} 
\Bigl[ \bar{\psi}_a(x)( -i \Vec{\gamma}\cdot \nabla_x + m_F 
- \mu \gamma^0 )\psi_a (x)
 \nonumber \\
&&+\frac{1}{2} \{  \pi_a (x)^2
+ \phi_a(x) (-\nabla_x^2 + m_{B}^2) \phi_a(x) \}
+ g\bar{\psi}_a(x)\psi_a(x) \phi_a (x)\Bigr], \label{Ykw_mdl1}
\end{eqnarray}
where $\phi_a$ describe the neutral scalar field 
and $g$ is the Yukawa coupling constant. 
In this section all the quantities with subscripts $B$ and $F$ represent 
ones for the scalar and the Dirac fields, respectively. 
In NETFD the Hamiltonian (\ref{Ykw_mdl1}) is extended to 
the hat-Hamiltonian (\ref{hat-H}), 
\begin{eqnarray}
\hat{H} 
&=& \int d^3 \Vec{x} 
\Bigl[ \bar{\psi}_a^{\alpha} (x) ( -i \Vec{\gamma}\cdot \nabla_x + m_F
 -\mu \gamma^0 ) 
\psi_a^{\alpha} (x) \nonumber \\
&&+ \frac{1}{2} \left\{ \bar{\pi}_a^{\alpha} (x) \pi_a^{\alpha} (x)
+ \bar{\phi}_a^\alpha (x) (-\nabla_x^2 + m_{B}^2) \phi_a^{\alpha} (x) \right\} \nonumber \\
&&+ \sum_{\gamma = 1}^{2} g^{\gamma} 
\{ \bar{\psi}_a (x) \tau_3 \}^{\gamma} \psi_a^{\gamma}(x) \phi_a^\gamma (x) \Bigr],
\end{eqnarray}
where the neutral scalar field in the thermal doublet notation, 
$\phi_a^{\alpha}$ and $\bar{\phi}_a^{\alpha}$, 
and their canonical conjugate, $\pi_a^{\alpha}$ and $\bar{\pi}_a^{\alpha}$, 
are given in Ref.~\refcite{CQNS1}.

In Feynman diagrams the thermal propagator, 
$S_0^{\alpha \beta}\tau_3^{\beta \gamma}$, 
is assigned to the internal fermion lines. (See Appendix C.)
The free thermal propagator for the neutral scalar field is given by\cite{CQNS1} 
\begin{eqnarray}
&& D_0^{\alpha\beta} (t_x, t_y;\Vec{p}) =
B^{-1}(n_{B}(t_x;p))^{\alpha\gamma_1} 
D_{0,R}^{\gamma_1 \gamma_2} (t_x-t_y;\Vec{p}) 
B(n_{B}(t_y;p))^{\gamma_2\beta}  \nonumber \\
&& +\{\tau_3 B(n_{B}(t_x;p))^T \}^{\alpha\gamma_1} 
 D_{0,A}^{\gamma_1 \gamma_2} (t_x-t_y;\Vec{p})
 \{ B^{-1}(n_{B}(t_y;p))^T\tau_3 \}^{\gamma_2 \beta},	\label{ns_prop1}
\end{eqnarray}
where we write 
\begin{eqnarray}
&& D_{0,R}^{11} (t_x-t_y;\Vec{p}) 
= \theta(t_x-t_y) \frac{1}{2\omega_p}  e^{-i \omega_p (t_x - t_y)}, 
\label{ns_prop_com1} \\
&& D_{0,R}^{22} (t_x-t_y;\Vec{p})
= - \theta(t_y-t_x) \frac{1}{2\omega_p} e^{-i \omega_p (t_x-t_y)}, 
\label{ns_prop_com2} \\
&& D_{0,A}^{11} (t_x - t_y;\Vec{p})
=\theta(t_y-t_x) \frac{1}{2\omega_p} e^{i \omega_p (t_x - t_y)}, 
\label{ns_prop_com3} \\
&& D_{0,A}^{22} (x-y)
= - \theta(t_x-t_y) \frac{1}{2\omega_p} e^{i \omega_p (t_x - t_y)}, 	
\label{ns_prop_com4}	\\
&&{\rm other \ components }=0. \nonumber
\end{eqnarray}
We assign $D_0^{\alpha\beta} \tau_3^{\beta\gamma}$ 
on the internal scalar lines. 
The Feynman rule for the vertex is given by, 
\begin{eqnarray}
(-i) g^{\gamma} = (-i) g
\left(
   \begin{array}{c}
	1 \\
	-1 \\
   \end{array}
\right ), \label{Ykw-C1}
\end{eqnarray}
where $\gamma$ denotes the thermal index. 

\begin{figure}[pb]
\centerline{\psfig{file=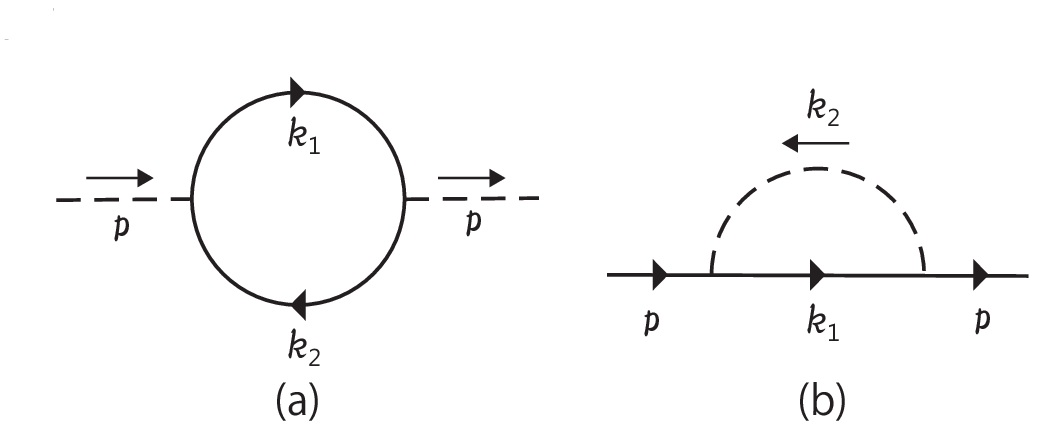}}
\vspace*{8pt}
	\caption{1-loop thermal self-energy derived three point scalar interaction model}
	\label{figTSE1}
\end{figure}
We evaluate the time evolution equation for 
the Bogoliubov parameters at the 1-loop level. 
The self-energy for the neutral scalar field is illustrated 
in Fig.~\ref{figTSE1}(a). It is calculated as
\begin{eqnarray}
&& i \Sigma_{B}^{\alpha\beta}(t_{z_1}, t_{z_2};\Vec{p}) 
=	\int \frac{d^3\Vec{k}_1}{(2\pi)^3} \frac{d^3\Vec{k}_2}{(2\pi)^3} 
(2\pi)^3 \delta^{(3)} (\Vec{p} - \Vec{k}_1 + \Vec{k}_2) \nonumber \\
&&~	\times (-1) \tau_3^{\alpha \gamma} 
{\rm tr} \Bigl[ (-i) g^{\gamma} 
 \{ S_0(t_{z_1}, t_{z_2}; \Vec{k}_1) \tau_3 \}^{\gamma \beta} 
 (-i)g^{\beta} 
 \{ S_0(t_{z_2}, t_{z_1}; \Vec{k}_2) \tau_3 \}^{\beta \gamma}
 \Bigr],
\end{eqnarray}
where ${\rm tr}[\cdots]$ shows trace manipulation on the spinor indices. 
Contracting the thermal index "$\gamma$", we obtain 
\begin{eqnarray}
&&\!\!\! i \Sigma_{B}^{\alpha\beta}(t_{z_1}, t_{z_2};\Vec{p}) 
= - g^2 \sum_{i_1=1}^2 \sum_{i_2=1}^2 \int \frac{d^3\Vec{k}_1}{ (2\pi )^3} 
\frac{d^3\Vec{k}_2}{ (2\pi )^3} 
(2\pi)^3 \delta^{(3)} (\Vec{p} - \Vec{k}_1 + \Vec{k}_2) \nonumber \\
&&\!\!\!  \times \frac{ E_{F,k_1,i_1} E_{F,k_2,i_2}
 - \Vec{k}_1 \cdot \Vec{k}_2 + m_F^2}
	{ E_{F,k_1,i_1} E_{F,k_2,i_2}}
\  e^{ -i ( E_{C,k_1,i_1} - E_{C,k_2,i_2}) (t_{z_1} - t_{z_2})}	\nonumber \\
&&\!\!\! \times \Biggl[ \theta (t_{z_1} - t_{z_2}) 
	\left(
   \begin{array}{cc}
	\bar{f}_{F,i_1}(t_{z_2};k_1) f_{F,i_2}(t_{z_2};k_2)	
	&	-f_{F,i_1}(t_{z_2};k_1) \bar{f}_{F,i_2}(t_{z_2};k_2) \\
	\bar{f}_{F,i_1}(t_{z_2};k_1) f_{F,i_2}(t_{z_2};k_2)
	&	- f_{F,i_1}(t_{z_2};k_1) \bar{f}_{F,i_2}(t_{z_2};k_2) \\
   \end{array}
\right)	\nonumber \\
&&\!\!\! + \theta (t_{z_2} - t_{z_1}) 
	\left(
   \begin{array}{cc}
	f_{F,i_1}(t_{z_1};k_1) \bar{f}_{F,i_2}(t_{z_1};k_2)	
	& - f_{F,i_1}(t_{z_1};k_1) \bar{f}_{F,i_2}(t_{z_1};k_2) \\
	\bar{f}_{F,i_1}(t_{z_1};k_1) f_{F,i_2}(t_{z_1};k_2)	
	&	- \bar{f}_{F,i_1}(t_{z_1};k_1) f_{F,i_2}(t_{z_1};k_2) \\
   \end{array}
\right) \Biggr],	\label{Ykw-SEB1}
\end{eqnarray}
where we set 
\begin{eqnarray}
&& E_{F,q,1}\equiv \omega_{F,q},\quad E_{F,q,2} \equiv -\omega_{F,q}, \label{df_se_n_1} \\
&& E_{C,q,1}\equiv \omega_{+,q}=\omega_{F,q}-\mu,\quad 
E_{C,q,2} \equiv -\omega_{-,q}=-\omega_{F,q}-\mu,  \label{df_se_n_2} \\
&& f_{F,1}(t;q) \equiv n_{+}(t;q),\quad f_{F,2}(t;q) \equiv 1 - n_{-}(t;q), \label{df_se_n_3} \\
&& \bar{f}_{F,1}(t;q) \equiv 1 - n_{+}(t;q),\quad \bar{f}_{F,2}(t;q) \equiv n_{-}(t;q). 
\label{df_se_n_4}
\end{eqnarray} 
It is found that the chemical potential is canceled out from the 
exponent in Eq.~(\ref{Ykw-SEB1}). 

The time evolution equation for the Bose particle 
is derived from the self-consistency renormalization condition 
for the neutral scalar field\cite{CQNS1}. 
It is given by 
\begin{eqnarray}
&&\!\!\! \dot{n}_{B}(t_x;p) =
-2 \omega_p \partial_{t_x} \Bigl[ \lim_{t_x\rightarrow t_y}\int dt_{z_1}dt_{z_2}
D_{0,R}^{11}(t_x-t_{z_1};\Vec{p})
 \Big\{ i\Sigma^{12} (t_{z_1},t_{z_2};\Vec{p})\nonumber \\
&&\!\!\! + n_{B} (t_{z_2};p) i \Sigma_R (t_{z_1},t_{z_2};\Vec{p}) 
 - n_{B}(t_{z_1};p) i \Sigma_A(t_{z_1},t_{z_2};\Vec{p}) \Bigr\} D_{0,R}^{22}(t_{z_2}-t_y;\Vec{p}) 
 \Bigr].  \label{ns_BoltzEq1} 
\end{eqnarray}
Inserting the components of 
the propagator (\ref{ns_prop_com1}), (\ref{ns_prop_com2}) 
and the 1-loop quantum correction (\ref{Ykw-SEB1}) into Eq.~(\ref{ns_BoltzEq1}), 
we obtain 
\begin{eqnarray}
&& \dot{n}_{B}(t_x;p) = - g^2 \sum_{i_1=1}^2 \sum_{i_2=1}^2 \int_{-\infty}^{t_x} dt_s
\int \frac{d^3\Vec{k}_1}{ (2\pi )^3} 
\frac{d^3\Vec{k}_2}{ (2\pi )^3} 
(2\pi)^3 \delta^{(3)} (\Vec{p} - \Vec{k}_1 + \Vec{k}_2)  \nonumber \\
&& \times \frac{ E_{F,k_1,i_1} E_{F,k_2,i_2}
 - \Vec{k}_1 \cdot \Vec{k}_2 + m_F^2} { \omega_{B,p} E_{F,k_1,i_1} E_{F,k_2,i_2}}
 \cos \left\{ (\omega_{B,p} - E_{F,k_1,i_1} + E_{F,k_2,i_2}) (t_x - t_s) \right\}
 \nonumber \\
&& \times \bigl\{ n_{B}(t_s;p) \bar{f}_{F,i_1}(t_s;k_1)f_{F,i_2}(t_s;k_2) 
 - (1+n_{B}(t_s;p))f_{F,i_1}(t_s;k_1) \bar{f}_{F,i_2}(t_s;k_2) \bigr\}. \nonumber \\
 \label{ns_Boltz_mdl1}
\end{eqnarray}
This equation has the consistent statistical structure 
with the quantum Boltzmann equation.

At 1-loop the thermal self-energy for the Dirac field, Fig.~\ref{figTSE1}(b), 
is given by
\begin{eqnarray}
&&-i\Sigma_{F}^{\alpha\beta}(t_{z_1}, t_{z_2};\Vec{p})=
	 \int \frac{d^3 \Vec{k}_1}{(2\pi)^3} \frac{d^3 \Vec{k}_2}{(2\pi)^3} 
	 (2\pi)^3 \delta^{(3)}(\Vec{p} - \Vec{k}_1 + \Vec{k}_2) \nonumber \\
&&~ \times \tau_3^{\alpha \gamma} (-i) g^{\gamma} 
 \{ S_0(t_{z_1}, t_{z_2};\Vec{k}_1) \tau_3 \}^{\gamma \beta}
 (-i)g^{\beta} \{ D_0(t_{z_2}, t_{z_1}; \Vec{k}_2) \tau_3 \}^{\beta \gamma}.
\end{eqnarray}
Substituting the free propagators (\ref{csf_app1_5}) and (\ref{df_app1_5}) and 
contracting the thermal index "$\gamma$", 
it is found to be
\begin{eqnarray}
&&-i\Sigma_{F}^{\alpha\beta}(t_{z_1}, t_{z_2};\Vec{p})
=-g^2 \sum_{i_1=1}^2 \sum_{i_2=1}^2 
\int \frac{d^3\Vec{k}_1}{ (2\pi )^3} \frac{d^3\Vec{k}_2}{ (2\pi )^3}
(2\pi)^3 \delta^{(3)} (\Vec{p}-\Vec{k}_1+\Vec{k}_2) \nonumber \\
&& \times \frac{ E_{F,k_1,i_1} \gamma^0
 - \Vec{k}_1 \cdot \Vec{\gamma} + m_F}
	{4 E_{F,k_1,i_1} \omega_{B,k_2}} 
\ e^{ -i ( E_{C,k_1,i_1} - E_{B,k_2,i_2}) (t_{z_1} - t_{z_2})}	\nonumber \\
&&\times \Biggl[ \theta (t_{z_1} - t_{z_2}) 
	\left(
   \begin{array}{cc}
	\bar{f}_{F,i_1}(t_{z_2};k_1) f_{B,i_2}(t_{z_2};k_2)
	&	f_{F,i_1}(t_{z_2};k_1) \bar{f}_{B,i_2}(t_{z_2};k_2) \\
	\bar{f}_{F,i_1}(t_{z_2}k_1) f_{B,i_2}(t_{z_2};k_2)
	&	f_{F,i_1}(t_{z_2};k_1) \bar{f}_{B,i_2}(t_{z_2};k_2) \\
   \end{array}
\right)	\nonumber \\
&&\ + \theta (t_{z_2} - t_{z_1}) 
	\left(
   \begin{array}{cc}
	- f_{F,i_1}(t_{z_1};k_1) \bar{f}_{B,i_2}(t_{z_1};k_2)
	&	f_{F,i_1}(t_{z_1};k_1) \bar{f}_{B,i_2}(t_{z_1};k_2) \\
	\bar{f}_{F,i_1}(t_{z_1};k_1) f_{B,i_2}(t_{z_1};k_2)
	&	- \bar{f}_{F,i_1}(t_{z_1};k_1) f_{B,i_2}(t_{z_1};k_2) \\
   \end{array}
\right) \Biggr],	 \nonumber \\ 
 \label{df_fermi_SE1}
\end{eqnarray}
where we write 
\begin{eqnarray}
&& E_{B,q,1} \equiv \omega_{B,q},\ \ E_{B,q,2} \equiv -\omega_{B,q}, \\
&& f_{B,1}(t;q) \equiv n_{B}(t;q),~ f_{B,2}(t;q) \equiv 1 + n_{B}(t;q), \\
&& \bar{f}_{B,1}(t;q) \equiv 1 + n_{B}(t;q),~ \bar{f}_{B,2}(t;q) \equiv n_{B}(t;q).
\end{eqnarray} 
We substitute the self-energy (\ref{df_fermi_SE1}) 
into Eqs.~(\ref{df_Boltz_eq1})-(\ref{df_Boltz_eq2}) 
and integrate over the time variables. 
After taking the equal time limit, we find 
\begin{eqnarray}
&& \lim_{t_x\rightarrow t_y} \int dt_{z_1}dt_{z_2} 
\delta \Sigma_{F,1}^{12} (t_x,t_{z_1},t_{z_2},t_y;\Vec{p}) \nonumber \\
&& = g^2 \sum_{i_1=1}^2 \sum_{i_2=1}^2               
\int_{-\infty}^{t_x} dt_s 
\int \frac{d^3\Vec{k}_1}{(2\pi)^3} \frac{d^3\Vec{k}_2}{(2\pi)^3}
(2\pi)^3 \delta^{(3)} (\Vec{p} - \Vec{k}_1 + \Vec{k}_2) 
\frac{\omega_{F,p}\gamma^0 - \Vec{p}\cdot\Vec{\gamma} + m_F}{ 2\omega_{F,p}}  
\nonumber \\ 
&& \times \frac{  \omega_{F,p}  E_{F,k_1,i_1}
 - \Vec{p} \cdot \Vec{k}_1 + m_F^2}
{2 \omega_{F,p} E_{F,k_1,i_1} \omega_{B,k_2}} \ 
 \frac{ {\sin}\{(\omega_{+,p} - E_{C,k_1,i_1} + E_{B,k_2,i_2})(t_x-t_s)\}}
{\omega_{+,p} - E_{C,k_1,i_1} + E_{B,k_2,i_2}} \nonumber \\
&& \times \bigl\{ (1-n_{+}(t_s;p))f_{F,i_1}(t_s; k_1) \bar{f}_{B,i_2}(t_s; k_2)
	- n_{+}(t_s;p)  \bar{f}_{F,i_1}(t_s;k_1) f_{B,i_2}(t_s;k_2) \bigr\}, \nonumber \\ 
\label{df_off_dia+} \\
&&\lim_{t_x\rightarrow t_y} \int dt_{z_1}dt_{z_2} 
\delta \Sigma_{F,4}^{21} (t_x,t_{z_1},t_{z_2},t_y;\Vec{p}) \nonumber \\
&&= -g^2 \sum_{i_1=1}^2 \sum_{i_2=1}^2               
\int_{-\infty}^{t_x} dt_s 
\int \frac{d^3\Vec{k}_1}{(2\pi)^3} \frac{d^3\Vec{k}_2}{(2\pi)^3} 
(2\pi)^3 \delta^{(3)} (\Vec{p} - \Vec{k}_1 + \Vec{k}_2)
 \frac{\omega_{F,p}\gamma^0 + \Vec{p} \cdot \Vec{\gamma} - m_F}{2\omega_{F,p}}  
\nonumber \\
&& \times \frac{ \omega_{F,p} E_{F,k_1,i_1} 
 + \Vec{p} \cdot \Vec{k}_1 - m_F^2}
{2 \omega_{F,p} E_{F,k_1,i_1} \omega_{B,k_2}} 
\ \frac{ {\sin}\{ (\omega_{-,p} + E_{C,k_1,i_1} - E_{B,k_2,i_2})(t_x-t_s)\}}
{\omega_{-,p} + E_{C,k_1,i_1} - E_{B,k_2,i_2}}	\nonumber \\
&&\times \bigl\{ n_{-}(t_s;p) f_{F,i_1}(t_s;k_1) \bar{f}_{B,i_2}(t_s;k_2)
	- ( 1- n_{-}(t_s;p))  \bar{f}_{F,i_1}(t_s;k_1) f_{B,i_2}(t_s;k_2) \bigr\}. 
\nonumber \\ \label{df_off_dia-}
\end{eqnarray}
Due to the relationship (\ref{df_se_n_2}), 
these equations are independent on $\mu$. 
We insert Eqs.~(\ref{df_off_dia+})-(\ref{df_off_dia-}) 
into Eqs.~(\ref{df_Boltz_eq1})-(\ref{df_Boltz_eq2}), 
and find the time evolution equations for the Fermi particle 
and anti-particle number densities, 
\begin{eqnarray}
&& \dot{n}_{+}(t_x;p) 
 = g^2 \sum_{i_1=1}^2 \sum_{i_2=1}^2               
\int_{-\infty}^{t_x} dt_s 
\int \frac{d^3\Vec{k}_1}{(2\pi)^3} \frac{d^3\Vec{k}_2}{(2\pi)^3}
(2\pi)^3 \delta^{(3)} (\Vec{p} - \Vec{k}_1 + \Vec{k}_2)  \nonumber \\ 
&& \times \frac{  \omega_{F,p}  E_{F,k_1,i_1}
 - \Vec{p} \cdot \Vec{k}_1 + m_F^2}
{2 \omega_{F,p} E_{F,k_1,i_1} \omega_{B,k_2}} \ 
 {\cos}\{(\omega_{F,p} - E_{F,k_1,i_1} + E_{B,k_2,i_2})(t_x-t_s)\} \nonumber \\
&& \times \bigl\{ (1-n_{+}(t_s;p))f_{F,i_1}(t_s;k_1) \bar{f}_{B,i_2}(t_s;k_2)
	- n_{+}(t_s;p)  \bar{f}_{F,i_1}(t_s;k_1) f_{B,i_2}(t_s;k_2) \bigr\}, \nonumber \\ 
\label{df_boltz+} \\
&& \dot{n}_{-}(t_x;p) 
= -g^2 \sum_{i_1=1}^2 \sum_{i_2=1}^2               
\int_{-\infty}^{t_x} dt_s 
\int \frac{d^3\Vec{k}_1}{(2\pi)^3} \frac{d^3\Vec{k}_2}{(2\pi)^3} 
(2\pi)^3 \delta^{(3)} (\Vec{p} - \Vec{k}_1 + \Vec{k}_2)  \nonumber \\
&& \times \frac{ \omega_{F,p} E_{F,k_1,i_1} 
 + \Vec{p} \cdot \Vec{k}_1 - m_F^2}
{2 \omega_{F,p} E_{F,k_1,i_1} \omega_{B,k_2}} 
\  {\cos}\{ (\omega_{F,p} + E_{F,k_1,i_1} - E_{B,k_2,i_2})(t_x-t_s)\} \nonumber \\
&&\times \bigl\{ n_{-}(t_s;p) f_{F,i_1}(t_s;k_1) \bar{f}_{B,i_2}(t_s;k_2)
	- ( 1- n_{-}(t_s;p))  \bar{f}_{F,i_1}(t_s;k_1) f_{B,i_2}(t_s;k_2) \bigr\}. \nonumber \\ 
\label{df_boltz-}
\end{eqnarray}
The last lines in Eqs.~(\ref{df_boltz+}) and (\ref{df_boltz-}) have the 
consistent forms with the collision term in the quantum Boltzmann equations. 

\section{Conclusion}

The relativistic complex scalar and Dirac fields have been 
discussed at finite chemical potential in NETFD. 
The thermal degree of freedom is introduced through 
the Bogoliubov transformation.
The fields are decomposed in terms of the Bogoliubov 
transformed operators, $\xi_p$ and $\eta_p$, with
Hermitian energy eigenvalues. 
The time dependent Bogoliubov transformation modifies 
the unperturbed hat-Hamiltonian and induces the thermal counter term. 
The self-consistency renormalization condition 
has been generalized for the complex scalar and the Dirac fields.  
Imposing the self-consistency renormalization condition, 
we obtain the quantum Boltzmann equations at the leading order. 
A contribution beyond the quantum Boltzmann equation 
may appear from the higher order. 

In Ref.~\refcite{NETFD_DF1} 
the construction of a relativistic Dirac field in NETFD 
has not been accomplished by straightforward extending for the non-relativistic fields. 
In our formalism the time dependence of 
the particle and the anti-particle is independently fixed. 
As is mentioned in Ref.~\refcite{NETFD_DF1}, 
the charge conjugation invariance can be broken. 
Hence we introduce the time dependent chemical potential, $\mu$, 
and successfully quantize.

There are some alternative methods to derive the quantum Boltzmann 
equation. In Ref.~\refcite{SD-2} the quantum Boltzmann equation 
for a relativistic neutral scalar field has been derived 
from the Schwinger-Dyson equation in NETFD. 
The thermal propagator in an equilibrium system 
has been simply extended to the one with the 
time dependent particle number density. 
The full propagator has been calculated based on 
the extension with the same diagram in an equilibrium system. 
In this paper the thermal propagators are 
derived as an expectation value of the composite operators 
constructed by two fields, (\ref{csf_app1_1}) and (\ref{df_app1_1}). 
This procedure based on the canonical quantization is more 
fundamental and essential in NETFD. 
It is now possible to approach high energy non-equilibrium phenomena in NETFD 
starting from the Schwinger-Dyson equation and the canonical quantization.

There are some remaining problems.
In this paper we assume that the energy eigenvalues 
for the Bogoliubov transformed free fields are independent on time variable. 
However, the time dependence for the eigenvalue is 
induced thought quantum corrections in an out of equilibrium system.
We have to generalize the procedure 
with the time dependent energy eigenvalue at higher order. 
Here the thermal Bogoliubov parameters for the Dirac field are 
represented as spin independent quantities. 
It is necessary to extend the parameters as a spinor 
dependent ones for a helicity dependent configuration.

A symmetry behavior is 
one of the most important problem in non-equilibrium phenomena 
of relativistic quark matter. 
It is investigated by observing an order parameter 
which is described as an expectation value of fields. 
Since we directly evaluate the expectation value of fields in NETFD, 
the procedure developed here expected to be useful for calculating 
the order parameter which determines the symmetry breaking.

It is also interesting to apply the procedure to 
the early stage evolution in heavy ion collisions. 
For this purpose we should consider a spatially inhomogeneous state. 
In Refs.~\refcite{inhomo1} and \refcite{inhomo2} 
spatially inhomogeneous states have been studied for the cold atom system according to 
the Bogoliubov-de Gennes (BdG) method in NETFD. 
The field is expanded by a complete set of wave functions 
under the spatially inhomogeneous potential. 
To apply the BdG method, 
it is necessary to impose the field obeys 
the canonical (anti-)commutation relation. 
It is expected that the procedure developed in this paper 
is applicable to the spatially inhomogeneous system. 

We hope to solve these problems and report the result in future. 

\section*{Acknowledgements}

The authors wold like to thank Y.~Yamanaka and Y.~Nakamura 
for fruitful discussions.

\appendix

\section{Finite Chemical Potential in TFD}

We discuss a chemical potential, $\mu$, 
for complex scalar and Dirac fields in TFD. 
At $\mu=0$ a free hat-Hamiltonian for complex scalar field is given by 
\begin{eqnarray}
\hat{H}_0|_{\mu=0} &=& 
 \int d^3\Vec{x} \bigl\{ \bar{\pi}_a^{\alpha}(x) \pi_a^{\alpha} (x)
 + \bar{\phi}_a^{\alpha}(x) (-\nabla_x^2 + m^2) \phi_a^{\alpha}(x) \bigr\}. 
\label{chm_app1_1}
\end{eqnarray}
The complex scalar field is decomposed as 
\begin{eqnarray}
&&  \phi_{a}^\alpha(x) = \int \frac{d^3\Vec{p}}{(2\pi)^3}
 \frac{1}{\sqrt{2\omega_p}} 
\left\{ 
a_{p}^\alpha (t_x) 
e^{ i \Vec{p} \cdot \Vec{x}} 
+ (\tau_3 \bar{b}_{p}(t_x)^T)^\alpha 
e^{ -i \Vec{p} \cdot \Vec{x}} \right\}, 
\label{chm_app1_2-1} \\
&&  \bar{\phi}_{a}^\alpha (x) = \int \frac{d^3\Vec{p}}{(2\pi)^3} 
\frac{1}{\sqrt{2\omega_p}} 
\left\{ \bar{a}_{p}^\alpha (t_x) 
e^{ -i \Vec{p} \cdot \Vec{x}} 
+ ( b_{p}(t_x)^{T} \tau_3)^\alpha 
e^{ i \Vec{p} \cdot \Vec{x}} \right\}, 
 \label{chm_app1_2-2}
\end{eqnarray}
and 
\begin{eqnarray}
&&  \pi_{a}^{\alpha} (x) = (-i) \int \frac{d^3\Vec{p}}{(2\pi)^3}
 \sqrt{\frac{\omega_p}{2}} 
\left\{ 
a_{p}^\alpha (t_x) 
e^{ i \Vec{p} \cdot \Vec{x}} 
- ( \tau_3 \bar{b}_{p}(t_x)^T )^\alpha 
e^{ -i \Vec{p} \cdot \Vec{x}} \right\}, 
\label{chm_app1_3-1}  \\
&&  \bar{\pi}_{a}^{\alpha}(x) = (-i) \int \frac{d^3\Vec{p}}{(2\pi)^3} 
\sqrt{\frac{\omega_p}{2}} 
\left\{ -\bar{a}_{p}^\alpha (t_x) 
e^{ -i \Vec{p} \cdot \Vec{x}} 
+ ( b_{p}(t_x)^T \tau_3)^\alpha 
e^{ i \Vec{p} \cdot \Vec{x}} \right\}, 
\label{chm_app1_3-2}
\end{eqnarray}
with the following time dependence, 
\begin{eqnarray}
&&a_{p}^\alpha (t_x)
 = a_{p}^\alpha e^{-i \omega_{p} t_x},~~
\bar{a}_{p}^\alpha (t_x)
 =\bar{a}_{p}^{\alpha} e^{i \omega_{p} t_x}, \label{chm_app1_4-1} \\
&& b_{p}^\alpha (t_x)
 = b_{p}^\alpha e^{-i \omega_{p} t_x},~~ 
\bar{b}_{p}^\alpha (t_x)
 =\bar{b}_{p}^{\alpha} e^{i \omega_{p} t_x}. \label{chm_app1_4-2}
\end{eqnarray}
The canonical commutation relations are given by 
\begin{eqnarray}
 [\phi_{a}^\alpha (t,\Vec{x}),\bar{\pi}_{a}^\beta (t,\Vec{y})]
 =  [ \bar{\phi}_{a}^\alpha (t,\Vec{x}),\pi_{a}^\beta (t,\Vec{y}) ]
 = i \delta^{(3)}(\Vec{x}-\Vec{y}) \delta^{\alpha \beta}. \label{chm_app1_4-3}
\end{eqnarray}

In TFD a conserved charge, $\hat{Q}_{consv}$, 
i.e. the zeroth component of the Noether current, is represented as
\begin{eqnarray}
\hat{Q}_{consv}= i \int d^{3}x \left\{ \bar{\phi}_{a}^{\alpha}(x)\pi_{a}^{\alpha}(x) 
- \bar{\pi}_{a}^{\alpha}(x) \phi_{a}^{\alpha}(x) \right\}. \label{chm_app1_5}
\end{eqnarray}
The chemical potential, $\mu$, can be introduced for this conserved charge. 
Thus the free hat-Hamiltonian (\ref{chm_app1_1}) is modified as 
\begin{eqnarray}
\hat{H}_0 &=& \hat{H}_0|_{\mu=0} - \mu \hat{Q}_{consv} \nonumber \\
 &=& \int d^3\Vec{x} \Bigl[ \bar{\pi}_a^{\alpha}(x) \pi_a^{\alpha} (x)
 + \bar{\phi}_a^{\alpha}(x) (-\nabla_x^2 + m^2) \phi_a^{\alpha}(x) \nonumber \\
&& -i \mu \left\{ \bar{\phi}_{a}^{\alpha}(x)\pi_{a}^{\alpha}(x) 
- \bar{\pi}_{a}^{\alpha}(x) \phi_{a}^{\alpha}(x) \right\}
 \Bigr]. \label{chm_app1_6}
\end{eqnarray}
It should be noticed that 
the conserved charge, $\hat{Q}_{consv}$, commutes 
with the hat-Hamiltonian, $\hat{H}|_{\mu=0}$. 
It means that time evolution by the finite chemical potential 
is independent on that by $\hat{H}|_{\mu=0}$. 
Then the time dependence of the ordinary operators 
for the complex scalar field (\ref{chm_app1_2-1})-(\ref{chm_app1_3-2}) 
is modified as 
\begin{eqnarray}
&&a_{p}^\alpha (t_x)
 = e^{i \hat{H}_0 t_x} a_{p}^{\alpha}(0) e^{-i \hat{H}_{0}t_x}
 = a_{p}^\alpha e^{-i \omega_{+,p} t_x}, \label{chm_app1_7-1} \\
&& \bar{a}_{p}^\alpha (t_x)
 = e^{i \hat{H}_0 t_x} \bar{a}_{p}^{\alpha}(0) e^{-i \hat{H}_{0}t_x}
 = \bar{a}_{p}^{\alpha} e^{i \omega_{+,p} t_x}, \label{chm_app1_7-2} \\
&& b_{p}^\alpha (t_x)
 = e^{i \hat{H}_0 t_x} b_{p}^{\alpha}(0) e^{-i \hat{H}_{0}t_x}
 = b_{p}^\alpha e^{-i \omega_{-,p} t_x}, \label{chm_app1_7-3} \\
&& \bar{b}_{p}^\alpha (t_x)
 = e^{i \hat{H}_0 t_x} \bar{b}_{p}^{\alpha}(0) e^{-i \hat{H}_{0}t_x}
 =\bar{b}_{p}^{\alpha} e^{i \omega_{-,p} t_x}, \label{chm_app1_7-4}
\end{eqnarray}
where we set 
\begin{eqnarray}
&& \omega_{+,p} = \omega_p - \mu, \label{chm_app1_8-1} \\
&& \omega_{-,p} = \omega_p + \mu. \label{chm_app1_8-2}
\end{eqnarray}
We replace the operators (\ref{chm_app1_4-1}) and (\ref{chm_app1_4-2}) 
to (\ref{chm_app1_7-1})-(\ref{chm_app1_7-4}), respectively. 
The complex scalar field (\ref{chm_app1_2-1})-(\ref{chm_app1_3-2}) 
with the operators (\ref{chm_app1_7-1})-(\ref{chm_app1_7-4}) 
satisfies the canonical commutation relations (\ref{chm_app1_4-3}). 
The time evolution equations of the fields 
(\ref{chm_app1_2-1})-(\ref{chm_app1_3-2}) 
are obtained by the Heisenberg equations. 

In the equilibrium system 
the thermal Bogoliubov transformations are defined by
\begin{eqnarray}
&&\xi_{p}^\alpha (t) 
= B(n_{+}(p))^{\alpha \beta} a_{p}^{\beta} (t), \quad 
\bar{\xi}_{p}^\alpha (t)
= \bar{a}_{p}^{\beta} (t) B^{-1}(n_{+}(p))^{\beta \alpha}, \label{chm_app1_9-1} \\
&&\eta_{p}^\alpha (t) 
= B(n_{-}(p))^{\alpha \beta} b_{p}^{\beta} (t), \quad 
\bar{\eta}_{p}^\alpha (t)
= \bar{b}_{p}^{\beta} (t) B^{-1}(n_{-}(p))^{\beta \alpha}, \label{chm_app1_9-2}
\end{eqnarray}
where the thermal Bogoliubov matrices,
$B$ and $B^{-1}$, are given by Eqs.~(\ref{cs_bm1}) and (\ref{cs_bm2}). 
In the equilibrium system 
the thermal Bogoliubov parameters, $n_{\pm}(p)$, for the scalar field 
represent the time independent particle and anti-particle number densities,
\cite{henning1,csf_henning1} 
\begin{eqnarray}
n_{+} (p)&=&\frac{1}{e^{ \beta \omega_{+,p} } -1}, \label{chm_app1_10-1} \\
n_{-} (p)&=&\frac{1}{e^{ \beta \omega_{-,p} } -1}, \label{chm_app1_10-2}
\end{eqnarray}
where $\beta$ denotes the inverse of the temperature. 

Since the thermal Bogoliubov transformations in the equilibrium system 
(\ref{chm_app1_9-1}) and (\ref{chm_app1_9-2}) are independent on time, 
the time dependence for the transformed operators, 
$\xi_p^\alpha(t)$ and $\eta_p^\alpha(t)$, 
coincides with the one for the ordinary oscillators, 
$a_p^{\alpha}(t)$ and $b_{p}^{\alpha}(t)$, 
\begin{eqnarray}
&&\xi_{p}^\alpha (t_x)
 = \xi_{p}^\alpha e^{-i \omega_{+,p} t_x}, \quad 
\bar{\xi}_{p}^\alpha (t_x)
 =\bar{\xi}_{p}^{\alpha} e^{i \omega_{+,p} t_x}, \label{chm_app1_11-1} \\
&& \eta_{p}^\alpha (t_x)
 = \eta_{p}^\alpha e^{-i \omega_{-,p} t_x}, \quad 
\bar{\eta}_{p}^\alpha (t_x)
 =\bar{\eta}_{p}^{\alpha} e^{i \omega_{-,p} t_x}. \label{chm_app1_11-2}
\end{eqnarray}
Thus the complex scalar field 
is decomposed in terms of the transformed operators 
(\ref{chm_app1_11-1}) and (\ref{chm_app1_11-2}) as 
\begin{eqnarray}
&&  \phi_{\xi}^{\alpha}(x) 
= \int \frac{d^3\Vec{p}}{(2\pi)^3}
 \frac{1}{\sqrt{2\omega_p}} 
\left\{ \xi_{p}^{\alpha}(t_x) 
 e^{ i \Vec{p} \cdot \Vec{x}} 
+ (\tau_3 \bar{\eta}_{p} (t_x)^T)^\alpha 
 e^{ -i \Vec{p} \cdot \Vec{x}} \right\}, \label{chm_app1_12-1} \\
&&  \bar{\phi}_{\xi}^{\alpha}(x) 
= \int \frac{d^3\Vec{p}}{(2\pi)^3} 
\frac{1}{\sqrt{2\omega_p}} 
\left\{ \bar{\xi}_{p}^{\alpha}(t_x) 
 e^{ -i \Vec{p} \cdot \Vec{x}} 
+ (\eta_{p} (t_x)^{T} \tau_3)^\alpha 
 e^{ i \Vec{p} \cdot \Vec{x}} \right\}, \label{chm_app1_12-2}
\end{eqnarray}
and
\begin{eqnarray}
&&  \pi_{\xi}^\alpha(x) = (-i)\int \frac{d^3\Vec{p}}{(2\pi)^3}
 \sqrt{\frac{\omega_p}{2}} 
\left\{ \xi_{p}^\alpha (t_x) 
 e^{ i \Vec{p} \cdot \Vec{x}} 
- (\tau_3 \bar{\eta}_{p}(t_x)^T)^\alpha 
 e^{ -i \Vec{p} \cdot \Vec{x}} \right\}, \label{chm_app1_13-1} \\
&&  \bar{\pi}_{\xi}^\alpha(x) = (-i)\int \frac{d^3\Vec{p}}{(2\pi)^3} 
 \sqrt{\frac{\omega_p}{2}} 
\left\{ -\bar{\xi}_{p}^\alpha (t_x) 
 e^{ -i \Vec{p} \cdot \Vec{x}} 
+ (\eta_{p} (t_x)^{T} \tau_3)^\alpha 
 e^{ i \Vec{p} \cdot \Vec{x}} \right\}. \label{chm_app1_13-2}
\end{eqnarray}
The complex scalar field 
and its canonical conjugate (\ref{chm_app1_12-1})-(\ref{chm_app1_13-2}) 
satisfy the equal-time canonical commutation relations, 
\begin{eqnarray}
 [\phi_{\xi}^\alpha (t,\Vec{x}),\bar{\pi}_{\xi}^\beta (t,\Vec{y})]
 =  [ \bar{\phi}_{\xi}^\alpha (t,\Vec{x}),\pi_{\xi}^\beta (t,\Vec{y}) ]
 = i \delta^{(3)}(\Vec{x}-\Vec{y}) \delta^{\alpha \beta}. \label{chm_app1_14}
\end{eqnarray}
Our definition of the thermal vacuum preserves 
the representations of the complex scalar field, 
(\ref{chm_app1_12-1})-(\ref{chm_app1_13-2}), in NETFD. 

In a similar manner the chemical potential can be introduced in the Dirac field. 
The conserved charge, $\hat{Q}_{consv}$, is given for the Dirac field as 
\begin{eqnarray}
\hat{Q}_{consv} = \int d^3\Vec{x} 
 \bar{\psi}_a^{\alpha}(x) \gamma^0  \psi_{a}^{\alpha}(x). \label{chm_app2_1}
\end{eqnarray}
In the equilibrium system the conserved charge 
introduces the chemical potential, $\mu$. 
Thus the free hat-Hamiltonian is modified as 
\begin{eqnarray}
\hat{H}_0 &=& \hat{H}_0|_{\mu=0} - \mu \hat{Q}_{consv} \nonumber \\
&=& \int d^3\Vec{x} 
 \bar{\psi}_a^{\alpha}(x)
 \left( -i \Vec{\gamma} \cdot \nabla_x +m 
-\mu \gamma^0 \right) \psi_{a}^{\alpha}(x). \label{chm_app2_2}
\end{eqnarray}
At finite chemical potential the Dirac field is decomposed as 
\begin{eqnarray}
&&  \psi_{a}^\alpha (x) = \int \frac{d^3\Vec{p}}{(2\pi)^3}
 \frac{1}{\sqrt{2\omega_p}} \sum_s 
\left\{ 
a_{p}^{s, \alpha} (t_x) u^s (p) 
e^{ i \Vec{p} \cdot \Vec{x}} 
+ (\tau_3 \bar{b}_{p}^s (t_x)^T)^\alpha v^s (p)
\ e^{ -i \Vec{p} \cdot \Vec{x}} \right\},  \nonumber \\ 
 \label{chm_app2_3-1} \\
&&  \bar{\psi}_{a}^\alpha (x) = \int \frac{d^3\Vec{p}}{(2\pi)^3} 
\frac{1}{\sqrt{2\omega_p}} \sum_s 
\left\{ \bar{a}_{p}^{s, \alpha} (t_x) \bar{u}^s(p) 
e^{ -i \Vec{p} \cdot \Vec{x}} 
+ ( b_{p}^s(t_x)^{T} \tau_3)^\alpha \bar{v}^s(p) 
e^{ i \Vec{p} \cdot \Vec{x}} \right\}, \nonumber \\ 
 \label{chm_app2_3-2}
\end{eqnarray}
with
\begin{eqnarray}
&& a_{p}^{s, \alpha} (t_x) = a_{p}^{s, \alpha} e^{-i\omega_{+,p} t_x},~~
\bar{a}_{p}^{s,\alpha} (t_x) = \bar{a}_{p}^{s, \alpha} e^{ i \omega_{+,p} t_x}, 
 \label{chm_app2_4-1} \\
&& b_{p}^{s, \alpha} (t_x) = b_{p}^{s,\alpha} e^{-i \omega_{-,p} t_x},~~
\bar{b}_{p}^{s, \alpha} (t_x) = \bar{b}_{p}^{s, \alpha}e^{i \omega_{-,p} t_x}, 
 \label{chm_app2_4-2}
\end{eqnarray}
where the energy eigenvalues, $\omega_{\pm,p}$, 
are given by Eqs.~(\ref{chm_app1_8-1}) and (\ref{chm_app1_8-2}). 
The eigenfunctions, $u^s(p)$ and $v^s(p)$, satisfy, 
\begin{eqnarray}
&&(\gamma^0 \omega_p - \Vec{\gamma}\cdot \Vec{p} -m) u^s (p)=0, 
\label{chm_app2_5-1} \\
&&(-\gamma^0 \omega_p + \Vec{\gamma} \cdot \Vec{p} -m) v^s (p)=0, 
\label{chm_app2_5-2} \\
&& \bar{u}^s (p) (\gamma^0 \omega_p - \Vec{\gamma}\cdot \Vec{p} -m )=0, 
\label{chm_app2_5-3} \\
&& \bar{v}^s (p) ( -\gamma^0 \omega_p + \Vec{\gamma}\cdot \Vec{p} -m )=0, 
\label{chm_app2_5-4} 
\end{eqnarray}
and
\begin{eqnarray}
&& \sum_s u^s(p) \bar{u}^s(p)
=\gamma^0 \omega_p - \Vec{\gamma}\cdot \Vec{p} + m, \label{chm_app2_6-1} \\
&& \sum_s v^s(p) \bar{v}^s(p)
=\gamma^0 \omega_p - \Vec{\gamma}\cdot \Vec{p} - m. \label{chm_app2_6-2}
\end{eqnarray}
We should notice that these conditions contain no chemical potential. 
The Dirac field with the chemical potential obeys the 
anticommutation relation, 
\begin{eqnarray}
&& \{ \psi_{a}^{\dagger \alpha}(t,\Vec{x}),\psi_{a}^{\beta}(t,\Vec{y})\}
 = \delta^{(3)}(\Vec{x}-\Vec{y}) \delta^{\alpha \beta}, \label{chm_app2_7}
\end{eqnarray}
where we write $\psi_{a}^{\dagger \alpha} \equiv \bar{\psi}_{a}^{\alpha} \gamma^0$.  
The time evolution equations for the fields 
(\ref{chm_app2_3-1}) and (\ref{chm_app2_3-2}) 
are reproduced by the Heisenberg equations 
and Eqs.~(\ref{chm_app2_5-1})-(\ref{chm_app2_5-4}). 

In the equilibrium system the thermal Bogoliubov transformation are defined by 
\begin{eqnarray}
&&\xi_{p}^{s,\alpha} (t) 
= B(n_{+}(p))^{\alpha \beta} a_{p}^{s,\beta} (t),~~ 
\bar{\xi}_{p}^{s,\alpha} (t)
= \bar{a}_{p}^{s,\beta} (t) B^{-1}(n_{+}(p))^{\beta \alpha},  \label{chm_app2_8-1} \\
&&\eta_{p}^{s,\alpha} (t) 
= B(n_{-}(p))^{\alpha \beta} b_{p}^{s,\beta} (t),~~ 
\bar{\eta}_{p}^{s,\alpha} (t)
= \bar{b}_{p}^{s,\beta} (t) B^{-1}(n_{-}(p))^{\beta \alpha}, \label{chm_app2_8-2}
\end{eqnarray}
where the thermal Bogoliubov matrices, $B(n_{\pm})$ and $B^{-1}(n_{\pm})$, 
are given by Eqs.~(\ref{df_tbm1}) and (\ref{df_tbm2}). 
The thermal Bogoliubov parameters in the equilibrium system 
are defined with the Fermi particle and anti-particle number densities,
\cite{henning1} 
\begin{eqnarray}
n_{+} (p)&=&\frac{1}{e^{ \beta \omega_{+,p} } +1}, \label{chm_app2_9-1} \\
n_{-} (p)&=&\frac{1}{e^{ \beta \omega_{-,p} } +1}. \label{chm_app2_9-2}
\end{eqnarray}

From the thermal Bogoliubov transformations 
(\ref{chm_app2_8-1}) and (\ref{chm_app2_8-2}) 
we obtain the time dependence for the transformed operators, 
$\xi_p^{s,\alpha}(t)$ and $\eta_p^{s,\alpha}(t)$, 
at finite chemical potential, 
\begin{eqnarray}
&&\xi_{p}^{s,\alpha} (t_x)
 = \xi_{p}^{s,\alpha} e^{-i \omega_{+,p} t_x},~~
\bar{\xi}_{p}^{s,\alpha} (t_x)
 = \bar{\xi}_{p}^{s,\alpha} e^{i \omega_{+,p} t_x}, \label{chm_app2_10-1} \\
&& \eta_{p}^{s,\alpha} (t_x)
 = \eta_{p}^{s,\alpha} e^{-i \omega_{-,p} t_x},~~
\bar{\eta}_{p}^{s,\alpha} (t_x)
 =\bar{\eta}_{p}^{s,\alpha} e^{i \omega_{-,p} t_x}. \label{chm_app2_10-2}
\end{eqnarray}
With the transformed operators (\ref{chm_app2_10-1}) and (\ref{chm_app2_10-2}) 
we can construct the Dirac field at finite chemical potential as
\begin{eqnarray}
&&  \psi_{\xi}^{\alpha} (x) = \int \frac{d^3\Vec{p}}{(2\pi)^3}
 \frac{1}{\sqrt{2\omega_p}} \sum_s 
\left\{ \xi_{p}^{s,\alpha}(t_x) u^s(p) 
 e^{ i \Vec{p} \cdot \Vec{x}} 
+ ( \tau_3 \bar{\eta}_{p}^{s} (t_x)^T)^{\alpha} v^s(p)
 e^{ -i \Vec{p} \cdot \Vec{x}} \right\},  \nonumber \\ 
\label{chm_app2_11-1} \\
&&  \bar{\psi}_{\xi}^{\alpha} (x) = \int \frac{d^3\Vec{p}}{(2\pi)^3} 
\frac{1}{\sqrt{2\omega_p}} \sum_s
\left\{ \bar{\xi}_{p}^{s,\alpha}(t_x) \bar{u}^s(p)
 e^{ -i \Vec{p} \cdot \Vec{x}} 
+ (\eta_{p}^{s} (t_x)^T \tau_3)^{\alpha} \bar{v}^s(p)
 e^{ i \Vec{p} \cdot \Vec{x}} \right\}. \nonumber \\ 
\label{chm_app2_11-2}
\end{eqnarray}
The Dirac field satisfies the canonical anticommutation relation,
\begin{eqnarray}
&& \{ \psi_{\xi}^{\dagger \alpha}(t,\Vec{x}),\psi_{\xi}^{\beta}(t,\Vec{y})\}
 = \delta^{(3)}(\Vec{x}-\Vec{y}) \delta^{\alpha \beta}. \label{chm_app2_12}
\end{eqnarray}
In our definition of NETFD the Dirac field has the same expressions 
in Eqs.~(\ref{chm_app2_11-1}) and (\ref{chm_app2_11-2}). 

\section{Propagator for a Free Complex Scalar Field}
In TFD the Feynman propagator for a free complex scalar 
field is given by the expectation value of the time 
ordered product of a composite operator constructed by 
two scalar fields. It has the $2\times 2$ matrix 
form in the thermal doublet notation, 
\begin{eqnarray}
D_{0}^{\alpha\beta}(t_x, t_y,\Vec{x}-\Vec{y})\equiv\langle \theta|
  T[\phi_a^{\alpha}(x)\bar{\phi}_a^{\beta}(y)]|\theta\rangle. \label{csf_app1_1}
\end{eqnarray}
After the thermal Bogoliubov transformation the propagator
(\ref{csf_app1_1}) reads
\begin{eqnarray}
&& D_{0}^{\alpha\beta}(t_x, t_y,\Vec{x}-\Vec{y}) \nonumber \\
&& = \theta(t_x-t_y)
\Bigl[ B^{-1}(n_{+}(t_x;|\nabla_x|))^{\alpha\gamma_1}\langle \theta|
 \phi_{\xi,+}^{\gamma_1} (x)\bar{\phi}_{\xi,-}^{\gamma_2}(y)|\theta\rangle
 B(n_{+} (t_y;|\overleftarrow{\nabla}_y|))^{\gamma_2 \beta} \nonumber \\
&& +\{\tau_3 B(n_{-}(t_x;|\nabla_x|) )^{T}\}^{\alpha\gamma_1} \nonumber \\
&&\quad \times \langle \theta| \{ \tau_3 \phi_{\xi,-}(x)\}^{\gamma_1} 
\{ \bar{\phi}_{\xi,+}(y) \tau_3\}^{\gamma_2} |\theta\rangle 
\{ B^{-1}(n_{-} (t_y;|\overleftarrow{\nabla}_y|))^T \tau_3 \}^{\gamma_2\beta} 
\Bigr]\nonumber \\
&& + \theta (t_y-t_x)
\Bigl[ B^{-1}(n_{+}(t_x;|\nabla_x|))^{\alpha\gamma_1}\langle \theta|
 \bar{\phi}_{\xi,-}^{\gamma_2} (y) \phi_{\xi,+}^{\gamma_1}(x)|\theta\rangle
 B(n_{+} (t_y;|\overleftarrow{\nabla}_y|))^{\gamma_2 \beta} \nonumber \\
&& +\{\tau_3 B(n_{-}(t_x;|\nabla_x|))^{T}\}^{\alpha\gamma_1} \nonumber \\
&&\quad \times
\langle \theta| \{ \bar{\phi}_{\xi,+}(y) \tau_3\}^{\gamma_2} 
\{ \tau_3 \phi_{\xi,-}(x)\}^{\gamma_1} |\theta\rangle 
\{ B^{-1}(n_{-} (t_y;|\overleftarrow{\nabla}_y|))^T \tau_3 \}^{\gamma_2\beta}
\Bigr],  \label{csf_app1_2}
\end{eqnarray}
where the fields, $\phi_{\xi,\pm}^{\alpha}$ and $\bar{\phi}_{\xi,\pm}^{\alpha}$, 
represent the positive and the negative frequency parts in 
Eqs.~(\ref{csf_phixi1}) and (\ref{csf_phixi2}), respectively. 

Due to the definition of the thermal vacuum (\ref{csf_vs1}) and (\ref{csf_vs2}), 
the thermal propagator, $D_{0}^{\alpha\beta}$, reduces to
\begin{eqnarray}
&& D_{0}^{\alpha\beta} (t_x, t_y,\Vec{x}-\Vec{y}) = 
B^{-1}(n_{+}(t_x;|\nabla_x|))^{\alpha\gamma_1} 
D_{0,R}^{\gamma_1 \gamma_2} (x-y) 
B(n_{+}(t_y;|\overleftarrow{\nabla}_y|))^{\gamma_2\beta}  \nonumber \\
&& +\bigl\{ \tau_3 B(n_{-}(t_x;|\nabla_x|))^T \bigr\}^{\alpha\gamma_1} 
 D_{0,A}^{\gamma_1 \gamma_2} (x-y)
 \bigl\{ B^{-1}(n_{-}(t_y;|\overleftarrow{\nabla}_y|))^T\tau_3 \bigr\}^{\gamma_2 \beta},	
\label{csf_app1_3}
\end{eqnarray}
where $ D_{0,R}^{\gamma_1 \gamma_2} (x-y)$ and 
$D_{0,A}^{\gamma_1 \gamma_2} (x-y)$ are given by
\begin{eqnarray}
&& D_{0,R}^{11} (x-y)=\int \frac{d^3\Vec{p}}{(2\pi)^3}
\theta (t_x-t_y) \frac{1}{2\omega_p} e^{-i \omega_{+,p} (t_x - t_y)}
 e^{i \Vec{p} \cdot (\Vec{x} - \Vec{y})}, \label{csf_app1_4-1} \\
&& D_{0,R}^{22} (x-y)=-\int \frac{d^3\Vec{p}}{(2\pi)^3}
\theta (t_y-t_x) \frac{1}{2\omega_p} e^{-i \omega_{+,p} (t_x - t_y)}
 e^{i \Vec{p} \cdot (\Vec{x} - \Vec{y})}, \label{csf_app1_4-2} \\
&& D_{0,A}^{11} (x-y)=\int \frac{d^3\Vec{p}}{(2\pi)^3}
\theta (t_y-t_x) \frac{1}{2\omega_p} e^{i \omega_{-,p} (t_x - t_y)}
 e^{-i \Vec{p} \cdot (\Vec{x} - \Vec{y})}, \label{csf_app1_4-3} \\
&& D_{0,A}^{22} (x-y)=-\int \frac{d^3\Vec{p}}{(2\pi)^3}
\theta(t_x-t_y) \frac{1}{2\omega_p} e^{i \omega_{-,p} (t_x-t_y)} 
 e^{-i \Vec{p} \cdot (\Vec{x} - \Vec{y})}, 	\label{csf_app1_4-4} \\
&&{\rm other \ components }=0. \nonumber
\end{eqnarray}

In a homogeneous non-equilibrium system 
the t-representation is convenient for practical calculations. 
Performing the spatial Fourier transformation, we rewrite the 
thermal propagator in the t-representation, 
\begin{eqnarray}
&& D_{0}^{\alpha\beta} (t_x,t_y;\Vec{p}) = 
B^{-1}(n_{+}(t_x;p))^{\alpha\gamma_1} 
D_{0,R}^{\gamma_1 \gamma_2} (t_x-t_y;\Vec{p}) 
B(n_{+}(t_y;p))^{\gamma_2\beta}  \nonumber \\
&& +\{\tau_3 B(n_{-}(t_x;p))^T \}^{\alpha\gamma_1} 
 D_{0,A}^{\gamma_1 \gamma_2} (t_x-t_y;\Vec{p})
 \{ B^{-1}(n_{-}(t_y;p))^T\tau_3 \}^{\gamma_2 \beta}, \label{csf_app1_5}
\end{eqnarray}
where we define 
\begin{eqnarray}
&& D_{0,R}^{11} (t_x-t_y;\Vec{p})=
 \theta (t_x-t_y) \frac{1}{2\omega_p}  e^{-i \omega_{+,p} (t_x - t_y)}, 
\label{csf_app1_6-1} \\
&& D_{0,R}^{22} (t_x-t_y;\Vec{p})=
-\theta (t_y-t_x) \frac{1}{2\omega_p} e^{-i \omega_{+,p} (t_x - t_y)}, 
 \label{csf_app1_6-2} \\
&& D_{0,A}^{11} (t_x-t_y;\Vec{p})=
\theta (t_y-t_x) \frac{1}{2\omega_p} e^{i \omega_{-,p} (t_x - t_y)}, 
 \label{csf_app1_6-3} \\
&& D_{0,A}^{22} (t_x-t_y;\Vec{p})=
-\theta(t_x-t_y) \frac{1}{2\omega_p} e^{i \omega_{-,p} (t_x-t_y)}, 
 \label{csf_app1_6-4} \\
&&{\rm other \ components }=0. \nonumber
\end{eqnarray}
These expressions are used to evaluate Feynman diagrams for the
complex scalar field. 

\section{Propagator for a Free Dirac Field}

We evaluate the thermal propagator for a free Dirac field in NETFD. 
The thermal propagator has the $2\times 2$ 
matrix form in the thermal doublet notation. 
It is defined by 
\begin{eqnarray}
S_0^{\alpha\beta}(t_x, t_y,\Vec{x}-\Vec{y}) \equiv\langle \theta|
  T[\psi_a^{\alpha}(x)\bar{\psi}_a^{\beta}(y)]|\theta\rangle. \label{df_app1_1}
\end{eqnarray}
Applying the thermal Bogoliubov transformation, we obtain 
\begin{eqnarray}
&& S_0^{\alpha\beta}(t_x, t_y,\Vec{x}-\Vec{y}) \nonumber \\
&& = \theta(t_x-t_y)
\Bigl[ B^{-1}(n_{+}(t_x;|\nabla_x|))^{\alpha\gamma_1}\langle \theta|
 \psi_{\xi,+}^{\gamma_1} (x)\bar{\psi}_{\xi,-}^{\gamma_2}(y)|\theta\rangle
 B(n_{+} (t_y;|\overleftarrow{\nabla}_y|))^{\gamma_2 \beta} \nonumber \\
&& +\{\tau_3 B(n_{-}(t_x;|\nabla_x|))^{T}\}^{\alpha\gamma_1} \nonumber \\
&&\quad \times \langle \theta| \{ \tau_3 \psi_{\xi,-}(x)\}^{\gamma_1} 
\{ \bar{\psi}_{\xi,+}(y) \tau_3\}^{\gamma_2}
|\theta\rangle \{ B^{-1}(n_{-} (t_y;|\overleftarrow{\nabla}_y|))^T \tau_3
\}^{\gamma_2\beta} \Bigr]\nonumber \\
&& - \theta (t_y-t_x)
\Bigl[ B^{-1}(n_{+}(t_x;|\nabla_x|))^{\alpha\gamma_1}\langle \theta|
 \bar{\psi}_{\xi,-}^{\gamma_2} (y) \psi_{\xi,+}^{\gamma_1}(x)|\theta\rangle
 B(n_{+} (t_y;|\overleftarrow{\nabla}_y|))^{\gamma_2 \beta} \nonumber \\
&& + \{\tau_3 B(n_{-}(t_x;|\nabla_x|))^{T}\}^{\alpha\gamma_1} \nonumber \\
&& \quad \times \langle \theta| \{ \bar{\psi}_{\xi,+}(y) \tau_3\}^{\gamma_2} 
\{ \tau_3 \psi_{\xi,-}(x) \}^{\gamma_1}
|\theta\rangle \{ B^{-1}(n_{-} (t_y;|\overleftarrow{\nabla}_y|))^T \tau_3 
\}^{\gamma_2\beta} \Bigr], \label{df_app1_2}
\end{eqnarray}
where the fields, $\psi_{\xi,\pm}^{\alpha}$ and $\bar{\psi}_{\xi,\pm}^{\alpha}$, show 
the positive and the negative frequency parts of the Dirac field, 
(\ref{df_psixi3}) and (\ref{df_psixi4}). 

Since the transformed operators annihilate the thermal vacuum, 
the thermal propagator reads 
\begin{eqnarray}
&& S_0^{\alpha\beta} (t_x, t_y,\Vec{x}-\Vec{y}) = 
B^{-1}(n_{+}(t_x;|\nabla_x|) )^{\alpha\gamma_1} 
S_{0,R}^{\gamma_1 \gamma_2} (x-y) 
B(n_{+}(t_y;|\overleftarrow{\nabla}_y|))^{\gamma_2\beta}   \nonumber \\
&& +\{\tau_3 B(n_{-}(t_x;|\nabla_x|))^T \}^{\alpha\gamma_1} 
 S_{0,A}^{\gamma_1 \gamma_2} (x-y)
 \{ B^{-1}(n_{-}(t_y;|\overleftarrow{\nabla}_y|))^T\tau_3 \}^{\gamma_2 \beta}, 
\label{df_app1_3}
\end{eqnarray}
where $ S_{0,R}^{\gamma_1 \gamma_2} (x-y)$ and 
$S_{0,A}^{\gamma_1 \gamma_2} (x-y)$ are written as 
\begin{eqnarray}
&&\!\!\!\!\!\!\!\!\! 
S_{0,R}^{11} (x-y) = \int \frac{d^3\Vec{p}}{(2\pi)^3} \theta (t_x-t_y) 
\frac{\omega_p \gamma^0 - \Vec{p}\cdot \Vec{\gamma} + m}{2\omega_p} 
e^{-i \omega_{+,p}  (t_x - t_y)} 
e^{i \Vec{p} \cdot (\Vec{x} - \Vec{y})}, \\
&&\!\!\!\!\!\!\!\!\! 
S_{0,R}^{22} (x-y) = - \int \frac{d^3\Vec{p}}{(2\pi)^3} \theta (t_y-t_x) 
\frac{ \omega_p \gamma^0 - \Vec{p}\cdot \Vec{\gamma} + m }{2\omega_p}
e^{-i \omega_{+,p}  (t_x - t_y)} 
e^{i \Vec{p} \cdot (\Vec{x} - \Vec{y})}, \\
&&\!\!\!\!\!\!\!\!\! 
S_{0,A}^{11} (x-y) = - \int \frac{d^3\Vec{p}}{(2\pi)^3} \theta (t_y-t_x) 
\frac{ \omega_p \gamma^0 - \Vec{p}\cdot \Vec{\gamma} - m }{2\omega_p}
e^{i \omega_{-,p}  (t_x - t_y)} 
e^{-i \Vec{p} \cdot (\Vec{x} - \Vec{y})}, \\
&&\!\!\!\!\!\!\!\!\! 
S_{0,A}^{22} (x-y) = \int \frac{d^3\Vec{p}}{(2\pi)^3} \theta (t_x-t_y) 
\frac{ \omega_p \gamma^0 - \Vec{p}\cdot \Vec{\gamma} - m }{2\omega_p}
e^{i \omega_{-,p}  (t_x - t_y)} 
e^{-i \Vec{p} \cdot (\Vec{x} - \Vec{y})}, 
\label{df_app1_4}	\\
&&\!\!\!\!\!\!\!\!\! 
{\rm other \ components }=0. \nonumber
\end{eqnarray}

The thermal propagator (\ref{df_app1_3}) 
is expressed in the t-representation as 
\begin{eqnarray}
S_0^{\alpha\beta} (t_x,t_y;\Vec{p}) = 
B^{-1}(n_{+}(t_x;p))^{\alpha\gamma_1} 
S_{0,R}^{\gamma_1 \gamma_2} (t_x-t_y;\Vec{p}) 
B(n_{+}(t_y;p))^{\gamma_2\beta}  && \nonumber \\
  +\{\tau_3 B(n_{-}(t_x;p))^T \}^{\alpha\gamma_1} 
 S_{0,A}^{\gamma_1 \gamma_2} (t_x-t_y;\Vec{p})
 \{ B^{-1}(n_{-}(t_y;p))^T\tau_3 \}^{\gamma_2 \beta}, \label{df_app1_5}
\end{eqnarray}
where
\begin{eqnarray}
&& S_{0,R}^{11} (t_x-t_y;\Vec{p}) = \theta (t_x-t_y) 
\frac{\omega_p \gamma^0 - \Vec{p}\cdot \Vec{\gamma} + m}{2\omega_p} 
e^{-i \omega_{+,p}  (t_x - t_y)}, \\
&& S_{0,R}^{22} (t_x-t_y;\Vec{p}) = - \theta (t_y-t_x) 
\frac{ \omega_p \gamma^0 - \Vec{p}\cdot \Vec{\gamma} + m }{2\omega_p}
e^{-i \omega_{+,p}  (t_x - t_y)}, \\
&& S_{0,A}^{11} (t_x-t_y;\Vec{p}) = - \theta (t_y-t_x) 
\frac{ \omega_p \gamma^0 + \Vec{p}\cdot \Vec{\gamma} - m }{2\omega_p}
e^{i \omega_{-,p}  (t_x - t_y)}, \\
&& S_{0,A}^{22} (t_x-t_y;\Vec{p}) = \theta (t_x-t_y) 
\frac{ \omega_p \gamma^0 + \Vec{p}\cdot \Vec{\gamma} - m }{2\omega_p}
e^{i \omega_{-,p}  (t_x - t_y)}, 
\label{df_app1_6}	\\
&&{\rm other \ components }=0. \nonumber
\end{eqnarray}
We use these expressions in the calculations of the 1-loop self-energy.


\end{document}